\documentclass[12pt]{article}
\pdfoutput=1

\usepackage[a4paper,text={16.8cm,22.4cm}]{geometry}
\usepackage{amsmath,amsfonts,slashed,amssymb,tikz,bm,psfrag,graphicx,color,dsfont}
\usepackage{multicol,multirow}
\usepackage{float}
\usepackage{hyperref}     

\RequirePackage[sort&compress,square,comma,numbers]{natbib}
\allowdisplaybreaks
\renewcommand{\arraystretch}{1.2}

\hypersetup{
  colorlinks   = true,    
  urlcolor     = blue,    
  linkcolor    = blue,    
  citecolor    = red      
}

\begin{document}

\begin{titlepage}

\begin{flushright}
\normalsize
 TUM-HEP-1332/21 \\
November 23, 2021
\end{flushright}

\vspace{0.1cm}
\begin{center}
\Large\bf
QCD factorization for the four-body leptonic \\
$B$-meson decays
\end{center}

\vspace{0.5cm}
\begin{center}
{\bf  Chao Wang$^{a, \, b}$ \footnote{chaowang@nankai.edu.cn},
Yu-Ming Wang$^{a}$ \footnote{correspondence author: wangyuming@nankai.edu.cn},
Yan-Bing Wei$^{a, \, c}$ \footnote{correspondence author: weiyb@nankai.edu.cn}}\\
\vspace{0.7cm}
{\sl
${}^a$  School of Physics, Nankai University,
\\
Weijin Road 94, 300071 Tianjin, China \\
${}^b$  Department of Mathematics and Physics, \\
Meicheng East Road 1, Huaiyin Institute of Technology, \\
223200 Huaian, Jiangsu, China \\
${}^c$  Physik Department T31,\\
James-Franck-Stra\ss e~1,
Technische Universit\"at M\"unchen,\\
D--85748 Garching, Germany
}
\end{center}

\vspace{0.2cm}
\begin{abstract}

Employing the QCD factorization formalism  we compute
$B_{u}^{-} \to \gamma^{\ast} \, \ell \, \bar \nu_{\ell}$  form factors
with an off-shell photon state possessing  the virtuality of
order $m_b \, \Lambda_{\rm QCD}$ and $m_b^2$, respectively,
at next-to-leading order in QCD.
Perturbative resummation for the enhanced logarithms of $m_b / \Lambda_{\rm QCD}$
in the resulting factorization formulae is subsequently accomplished at next-to-leading
logarithmic accuracy with the renormalization-group technique in momentum space.
In particular, we derive the soft-collinear factorization formulae for a variety of 
the subleading power corrections to the exclusive radiative
$B_{u}^{-} \to \gamma^{\ast} \, W^{\ast}$  form factors with a hard-collinear photon 
at ${\cal O}(\alpha_s^{0})$.
We further construct the complete set of the angular observables governing the
full differential distribution of the four-body leptonic decays
$B_{u}^{-} \to  \ell^{\prime} \, \bar \ell^{\prime} \, \ell \, \bar \nu_{\ell}$
with $\ell, \, \ell^{\prime} \in \{ e, \, \mu \}$
and then perform an exploratory phenomenological analysis for a number of
decay observables accessible at the LHCb and Belle II experiments,
with an emphasis on the systematic uncertainties arising from the shape parameters
of the leading-twist $B$-meson light-cone distribution amplitude
in heavy quark effective theory.

\end{abstract}

\vfil

\end{titlepage}

\tableofcontents

\newpage

\section{Introduction}

The exclusive leptonic  decays of the charged $B_{u}^{-}$-meson are of paramount importance
for exploring the complex quark-flavour dynamics in the Standard Model (SM)
and for probing the nonstandard flavour-changing mechanisms beyond the electroweak scale.
However, the helicity suppression of the two-body leptonic $B_{u}^{-} \to \mu \,  \bar \nu_{\mu}$
decay process  is expected to yield the tiny branching fraction of ${\cal O}  (10^{-7})$,
which prevents the decisive measurements at the BaBar and Belle experiments
with more than $5 \, \sigma$ significance (see  \cite{Aubert:2009ar,Yook:2014kga,Sibidanov:2017vph,Prim:2019gtj}
for the available  searches at the $e^{+} e^{-}$ colliders).
On the other hand, the radiative leptonic $B_{u}^{-} \to \gamma \ell \bar \nu_{\ell}$ decays
with an energetic photon will evidently lift such unwanted  helicity suppression
\cite{Beneke:2011nf,Wang:2016qii,Wang:2018wfj,Beneke:2018wjp},
at the price of introducing the  additional suppression from the electromagnetic coupling constant
and from the Lorenz invariant three-body phase space factor.
Reconstructing the $B$-meson decay vertex with just a single charged particle
will unfortunately bring about the tremendous challenges  for searching
$B_{u}^{-} \to \mu \,  \bar \nu_{\mu}$ and  $B_{u}^{-} \to \gamma \ell \bar \nu_{\ell}$
at the LHCb experiment. It is therefore advantageous to investigate the four-body leptonic decays
$B_{u}^{-} \to  \ell^{\prime} \, \bar \ell^{\prime} \, \ell \, \bar \nu_{\ell}$
with $\ell, \, \ell^{\prime} \in \{ e, \, \mu \}$ with three charged tracks
for the sake of facilitating the experimental reconstruction of the bottom-meson candidates
in the hadronic collision environment and  circumventing the helicity suppression mechanism
applied to the two-body leptonic decays simultaneously.
From the QCD perspective, the rare leptonic decays
$B_{u}^{-} \to  \ell^{\prime} \, \bar \ell^{\prime} \, \ell \, \bar \nu_{\ell}$
with the invariant mass of the lepton pair $(\ell^{\prime} \, \bar \ell^{\prime})$
of order $m_b \, \Lambda_{\rm QCD}$ will further provide us with the valuable information
on the inverse moment  of the twist-two $B$-meson distribution amplitude
in heavy quark effective theory (HQET), which serves as an indispensable ingredient
for the theory description of a variety of the exclusive $B$-meson decay observables
\cite{Beneke:2000wa,Beneke:2001at,Beneke:2020fot,Shen:2020hfq,Wang:2015vgv,Wang:2017jow,Lu:2018cfc,Gao:2019lta}
based upon the heavy quark expansion as well as the dispersion technique.
Moreover, the exclusive four-body  decays
$B_{u}^{-} \to \gamma^{\ast} (\to \ell^{\prime} \, \bar \ell^{\prime} )\, \ell \, \bar \nu_{\ell}$
with the four-momentum ($p_{\mu}$) of the intermediate photon state satisfying $p^2 \sim {\cal O}(m_b^2)$
are apparently of interest for addressing the ``notorious" open issue of the systematic uncertainty
due to the analytical continuation of the (local) operator product expansion (OPE) from the Euclidean to
the Minkowskian  domain in the practical applications \cite{Blok:1997hs,Shifman:2000jv,Bigi:2001ys}.

In analogy to the exclusive electromagnetic penguin decays $B \to K^{(\ast)} \ell \bar \ell$
\cite{Khodjamirian:2010vf,Khodjamirian:2012rm},
the presence of the vector-meson resonances (e.g.,  $\rho$, $\omega$, etc)
in the $(\ell^{\prime} \, \bar \ell^{\prime})$  invariant mass spectrum
of $B_{u}^{-} \to \ell^{\prime} \, \bar \ell^{\prime} \, \ell \, \bar \nu_{\ell}$
will invalidate the applicability of the perturbative factorization approach for evaluating the resulting
hadronic tensor in the collinear regime of $p^2 \sim {\cal O}(\Lambda_{\rm QCD}^2)$.
As a consequence, we will focus on the kinematical region of the virtuality
of the photon state appearing in $B \to \gamma^{\ast} \,  \ell \, \bar \nu_{\ell}$
above the light vector-meson threshold, in contrast to the previous phenomenological
explorations in the entire kinematically allowed regions \cite{Danilina:2018uzr,Danilina:2019dji}
by employing   the vector-meson-dominance (VMD) ansatz,
which permits us to apply the appropriate OPE techniques for disentangling the strong interaction
dynamics at the separated distance scales.
In particular, QCD calculations of the four-body leptonic  $B_{u}^{-}$-meson decays
at leading power in an expansion in terms of $\Lambda_{\rm QCD}/m_b$
with the hard-collinear dilepton system
resemble constructing the soft-collinear factorization formula
for the vacuum-to-bottom-meson correlation function $T_{7B}^{\mu \nu}$
entering in the radiative leptonic  $B_{d, s} \to \gamma \ell \bar \ell$ amplitude
\cite{Beneke:2020fot} generated by the $B$-type insertion of the effective weak operators
\footnote{The hard-collinear matching coefficient entering the perturbative
factorization formulae of the radiative $B \to \gamma^{\ast}$ form factors
were originally computed at ${\cal O}(\alpha_s)$ in \cite{Wang:2016qii}
with the strategy of regions.}.
In addition, the nonperturbative hadronic dynamics imbedded in the timelike
$B_{u}^{-}  \to \gamma^{\ast} \ell \bar \nu_{\ell}$ form factors with
an off-shell photon  carrying the hard momentum $p_{\mu} \sim {\cal O}(m_b)$
will be characterized by the bottom-meson decay constant $f_{B_u}$,
which has been determined from the lattice-QCD simulation at $N_f=2+1+1$ with the relative precision
of approximately $0.68 \, \%$ \cite{Aoki:2019cca}.
As the power counting scheme for the virtuality of the photon state
dictates factorization properties of the non-hadronic radiative
$B_{u}^{-}  \to \gamma^{\ast} \ell \bar \nu_{\ell}$ decay form factors,
the non-local hadronic matrix element defined by the time-ordered product of
the weak transition current $\bar u \, \gamma_{\mu} \, (1-\gamma_5)  \, b $
and the bottom-quark electromagnetic current will result in an unsuppressed
contribution at $p^2 \sim {\cal O}(m_b^2)$ in the heavy quark expansion.

In contrast to the radiative leptonic $B_{u}^{-} \to \gamma \ell \bar \nu_{\ell}$ decays
with an on-shell photon state,  it necessitates the introduction of three independent hadronic form factors to parameterize
the non-local matrix element encoding the QCD effects for the four-body leptonic decays
$B_{u}^{-} \to \gamma^{\ast} (\to \ell^{\prime} \, \bar \ell^{\prime} )\, \ell \, \bar \nu_{\ell}$
by implementing the two nontrivial constraints from the Ward-Takahashi identity
\footnote{Alternatively, this observation can be understood from the Lotentz decompositions of
the (axial)-vector current matrix elements governing the exclusive  semileptonic
$B \to V \ell \bar \nu_{\ell}$ decays \cite{Beneke:2000wa},
where $V$ stands for a light vector meson.}.
Consequently, one of the major technical objectives of the present paper consists in computing
the leading-power contributions to the generalized  form factors  of
$B_{u}^{-}(p_B) \to \gamma^{\ast}(p)  \,  \ell(q_1) \, \bar \nu_{\ell}(q_2)$ in the heavy quark expansion
based upon the soft-collinear effective theory (SCET) approach and the local OPE technique
at  $p^2 \sim {\cal O}(m_b \, \Lambda_{\rm QCD})$ and  $p^2 \sim {\cal O}(m_b^2)$ respectively,
including the next-to-leading logarithmic (NLL) resummation for the parametrically large logarithms
of $m_b / \Lambda_{\rm QCD}$ in the obtained factorization expression
with the renormalization-group (RG)  formalism.
The yielding  formulae for  the radiative $B_{u}^{-} \to \gamma^{\ast}$  transition form factors
with an off-shell  hard-collinear photon can be further evaluated by postulating the complete momentum dependence
of the leading-twist  $B$-meson distribution amplitude rather than by introducing the inverse moment $\lambda_{B}$ and
the first two  inverse-logarithmic moments $\sigma_{B}^{(n)}$ at the NLL accuracy
as in the case of computing the on-shell $B_{u}^{-} \to \gamma$ form factors
\cite{Beneke:2011nf,Wang:2016qii,Wang:2018wfj,Beneke:2018wjp}.
Furthermore, we will endeavour to carry out the factorization analysis for
various subleading-power corrections to the exclusive rare $B_{u}^{-} \to \gamma^{\ast} \, W^{\ast}$ decay form factors
in the hard-collinear $p^2$-regime with the aid of the two-particle and three-particle higher-twist HQET
distribution amplitudes.
In addition, the primary phenomenological new ingredient of our analysis consists in
the comprehensive investigation of the full angular decay distribution for
$B_{u}^{-} \to \gamma^{\ast} (\to \ell^{\prime} \, \bar \ell^{\prime} )\, W^{\ast} (\to \ell \, \bar \nu_{\ell})$
in terms of five independent kinematical variables for both $\ell = \ell^{\prime}$
and $\ell \neq \ell^{\prime}$:
the invariant  masses of the dilepton system ($p^2$) and of the lepton-neutrino pair ($q^2$),
the three angles $\theta_1$, $\theta_2$ and $\phi$ defined in Appendix \ref{appendix: Kinematics},
which allows for the systematic construction of a numbers of observables
accessible at the LHCb and Belle II experiments.

The outline of this presentation is as follows.
We will set up the computational framework in Section  \ref{section: Preliminaries}
by establishing the general parametrization  of  the four-body leptonic
$B$-meson decay amplitude to the lowest non-vanishing order in the electromagnetic
interaction and by exploiting the interesting implications of the ${\rm U_{em} (1)}$ gauge symmetry
on the emerged $B_{u}^{-} \to \gamma^{\ast} \,  W^{\ast}$ decay form factors.
The matching procedure ${\rm QCD \rightarrow SCET_{I} \rightarrow SCET_{II}}$
for the appearing $B$-meson-to-vacuum correlation function defined by the flavour-changing $b \to u$ weak current
and the electromagnetic quark current carrying a hard-collinear momentum $p$ will be performed
at leading power in $\Lambda_{\rm QCD}/m_b$
in Section \ref{section: QCDF with a hard-collinear photon} with the accustomed perturbative factorization technique,
where the subleading power contribution from the virtual photon radiation off the heavy bottom-quark field
will be also derived at leading order (LO) in the strong coupling constant  with the OPE technique.
In particular, the Ward-Takahashi relations of the generalized radiative $B$-meson decay form factors will be
demonstrated explicitly at one loop.
The non-local power suppressed corrections from a number of distinct sources
(parametrized by the higher-twist bottom-meson distribution amplitudes)
will be further addressed here by employing the HQET equations of motion at tree level.
In Section \ref{section: QCDF with a hard photon} we will proceed to carry out the ${\rm QCD \rightarrow HQET}$
matching programme for the aforementioned $B$-meson-to-vacuum correlation function  at ${\cal O}(\alpha_s)$,
where the factorization-scale independence of the achieved expressions for
all the  $B_{u}^{-} \to \gamma^{\ast} \,  W^{\ast}$ form factors will be further verified
at next-to-leading order (NLO) in QCD
taking advantage of the RG evolution equation for the effective decay constant $\tilde{f}_B(\mu)$.
Having at our disposal the factorized expressions for these hadronic transition form factors,
we will turn to investigate their numerical implications with the three-parameter ansatz of
the HQET $B$-meson distribution amplitude as proposed in \cite{Beneke:2018wjp}
in Section \ref{section: numerical results}, where the phenomenological aspects of
the four-body leptonic $B$-meson decays will be subsequently explored with circumstances
on the basis of the corresponding full differential distribution
described by the  five independent kinematical variables as previously mentioned.
The concluding remarks and theory perspectives on the future  improvements
will be presented in Section \ref{section:summary}.
We collect in Appendix \ref{appendix: Kinematics} the kinematics of the exclusive reaction
$B_{u}^{-} \to  \ell^{\prime} \, \bar \ell^{\prime} \, \ell \, \bar \nu_{\ell}$
with $\ell = \ell^{\prime}$ and $\ell \neq \ell^{\prime}$
and then present in Appendix \ref{appendix: interference angular function} the explicit expressions
of the angular coefficient functions entering the interference term of the full differential
distribution for the four-body leptonic  decay with identical lepton flavours.

\section{Preliminaries}
\label{section: Preliminaries}

By analogy with the detailed discussions on the $B$-meson radiative leptonic decays
\cite{Beneke:2011nf,Wang:2016qii,Wang:2018wfj,Beneke:2018wjp},
the four-body leptonic $B_{u}^{-} \to  \ell^{\prime} \, \bar \ell^{\prime} \, \ell \, \bar \nu_{\ell}$
decay amplitude can be expressed as
\begin{eqnarray}
&& {\cal A} (B_{u}^{-} \to  \ell^{\prime} \, \bar \ell^{\prime} \, \ell \, \bar \nu_{\ell})
\nonumber \\
&& = {G_F \, V_{ub} \over \sqrt{2}}  \,
\langle \ell^{\prime}(p_1) \, \bar \ell^{\prime}(p_2) \, \ell(q_1) \, \bar \nu_{\ell}(q_2)
| \left [ \bar \ell \gamma^{\mu} (1- \gamma_5) \nu_{\ell} \right ] \,
\left [ \bar u \gamma_{\mu} (1- \gamma_5) b  \right ] |  B_{u}^{-}(p_B) \rangle \,,
\end{eqnarray}
where $p_B = m_B \, v = p + q$ is the four-momentum of the $B$-meson momentum
with $v$ being its velocity,
$p=p_1 + p_2$ and $q=q_1 + q_2$ denote the outgoing momenta carried by
the off-shell photon and the $W^{\ast}$ boson in the cascade decay process
$B_{u}^{-} \to \gamma^{\ast} (\to \ell^{\prime} \, \bar \ell^{\prime} )\, W^{\ast} (\to \ell \, \bar \nu_{\ell})$,
respectively.  It further proves convenient to work in the rest frame of the $B$-meson
and to introduce the two light-cone vectors $n_{\alpha}$ and $\bar n_{\alpha}$ fulfilling the general relations
$n^2 = \bar n^2 = 0$ and $n \cdot \bar n=2$ such that
\begin{eqnarray}
p_{\alpha} = {n \cdot p \over 2} \, \bar n_{\alpha} + {\bar n \cdot p \over 2} \,  n_{\alpha} \,,
\qquad
q_{\alpha} = {n \cdot q \over 2} \, \bar n_{\alpha} + {\bar n \cdot q \over 2} \,  n_{\alpha} \,,
\qquad
v_{\alpha} = {n_{\alpha} + \bar n_{\alpha} \over 2} \,.
\end{eqnarray}
Keeping the first-order contribution to the decay amplitude
${\cal A} (B_{u}^{-} \to  \ell^{\prime} \, \bar \ell^{\prime} \, \ell \, \bar \nu_{\ell})$
in the electromagnetic interaction gives rise to following expression
\begin{eqnarray}
{\cal A} (B_{u}^{-} \to  \ell^{\prime} \, \bar \ell^{\prime} \, \ell \, \bar \nu_{\ell})
&=&  {G_F \, V_{ub} \over \sqrt{2}}  \, {i g_{\rm em}^2 \,Q_{\ell^{\prime}} \over p^2 + i 0} \,
\left [ \bar \ell^{\prime}(p_1) \, \gamma^{\nu}  \, \ell^{\prime}(p_2) \right ]  \,
\nonumber \\
&& \times \,  \{ T_{\nu \mu}(p_B, p) \, \left [ \bar \ell (q_1) \gamma^{\mu} (1- \gamma_5) \nu_\ell(q_2) \right ]
 - \, i \, f_{B} \, p_{B}^{\mu} \,\,  L_{\nu \mu}(p_B, p)  \}  \,,
 \label{general amplitude: original form}
\end{eqnarray}
where the hadronic matrix element $T_{\nu \mu}$ and the leptonic rank-two tensor $L_{\nu \mu}$
are defined by
\begin{eqnarray}
T_{\nu \mu} (p_B, p) &=& \int d^4 x \, e^{i p \cdot x} \,
\langle 0 | {\rm T}  \{ j_{\nu}^{\rm em}(x),  \bar u(0) \gamma_{\mu} (1- \gamma_5) b(0) \}
| B_{u}^{-} (p_B)  \rangle \,,
\label{definition: hadronic tensor}
\\
L_{\nu \mu} (p_B, p) &=& \int d^4 x \, e^{i p \cdot x} \,
\langle  \ell(q_1) \, \bar \nu_{\ell}(q_2)
| {\rm T}  \{ j_{\nu}^{\rm em}(x),  \bar \ell (0) \gamma_{\mu} (1- \gamma_5) \nu_{\ell}(0) \}
| 0  \rangle  \,.
\end{eqnarray}
The explicit form of the  fermion electromagnetic current is given by
\begin{eqnarray}
j_{\nu}^{\rm em}(x)  =  \sum_q Q_q \, \bar q(x) \gamma_{\nu} q(x)
+ \sum_{\ell} \,  Q_{\ell} \,  \bar \ell(x)  \gamma_{\nu}  \ell(x)   \,,
\end{eqnarray}
and the $B$-meson  decay constant in QCD is defined by the local
axial-vector matrix element
\begin{eqnarray}
\langle 0 | \bar u \, \gamma^{\mu} \, \gamma_5 \,  b |  B_{u}^{-} (p_B)\rangle
= i \, f_B \, p_B^{\mu} \,.
\end{eqnarray}
Taking advantage of the general decomposition of the hadronic tensor \cite{Wang:2016qii}
\begin{eqnarray}
T_{\nu \mu}(p, q) &=& i \, \epsilon_{\mu \nu \rho \sigma} \, p^{\rho} \, v^{\sigma} \, F_V(p^2, n \cdot p)
+ v \cdot p \, g_{\mu \nu} \, \hat{F}_A(p^2, n \cdot p) + v_{\nu} \, p_{\mu} \, F_1(p^2, n \cdot p)  \nonumber \\
&& + v_{\mu} \, p_{\nu} \, F_2(p^2, n \cdot p) + v \cdot p \,\, v_{\mu} \, v_{\nu} \, F_3(p^2, n \cdot p)
+\frac{p_{\mu} \, p_{\nu}}{v \cdot p } \, F_4(p^2, n \cdot p)   \,,
\label{Original Lorentz decomposition}
\end{eqnarray}
and employing  the two relations due to the conservation of the electromagnetic current \cite{Grinstein:2000pc,Khodjamirian:2001ga}
\begin{eqnarray}
F_1(p^2, n \cdot p)  &=& - \hat{F}_A(p^2, n \cdot p)  - \frac{p^2}{(v \cdot p)^2} \, F_4(p^2, n \cdot p)   \,,
\nonumber \\
F_3(p^2, n \cdot p) &=&  - \frac{(Q_b-Q_u) \, f_B \, m_B}{(v \cdot p)^2}
- \frac{p^2}{(v \cdot p)^2} \, F_2(p^2, n \cdot p)  \,,
\label{Ward indentities: form I}
\end{eqnarray}
we can readily derive the following expression for the
$B_{u}^{-} \to  \ell^{\prime} \, \bar \ell^{\prime} \, \ell \, \bar \nu_{\ell}$ amplitude
\begin{eqnarray}
{\cal A} (B_{u}^{-} \to  \ell^{\prime} \, \bar \ell^{\prime} \, \ell \, \bar \nu_{\ell})
&=& {G_F \, V_{ub} \over \sqrt{2}}  \, {i g_{\rm em}^2 \,Q_{\ell^{\prime}} \over p^2 + i 0} \,
\left [ \bar \ell^{\prime}(p_1) \, \gamma^{\nu} \,  \ell^{\prime}(p_2) \right ] \,
\left [ \bar \ell (q_1) \gamma^{\mu} (1- \gamma_5) \nu_\ell(q_2) \right ]
\nonumber \\
&& \bigg  \{  i \, \epsilon_{\mu \nu p v} \, F_V(p^2,  n \cdot p)
+ v \cdot p \, g_{\mu \nu} \,  F_A(p^2,  n \cdot p)
\nonumber \\
&& \hspace{0.2 cm}  + \, p_{\mu} v_{\nu} \, \left [  F_1(p^2,  n \cdot p)
+ { v \cdot p \over m_B} \, F_3(p^2,  n \cdot p)   \right ]  \bigg  \}   \,,
\label{general decay amplitude}
\end{eqnarray}
with $\epsilon_{0123} = -1$ and the redefinition prescription
of the axial-vector form factor \cite{Beneke:2011nf}
\begin{eqnarray}
F_A(p^2,  n \cdot p)  = \hat{F}_A(p^2,  n \cdot p) + \frac{Q_{\ell} \, f_B}{v \cdot p}
\end{eqnarray}
to account for the second term in (\ref{general amplitude: original form})
due to the  virtual photon radiation off the lepton field.
Apparently, constructions of the perturbative factorization formulae for
the generalized  $B_{u}^{-} \to \gamma^{\ast}$ transition form factors
$F_V$, $\hat{F}_A$, $F_1$ and $F_3$ constitutes the primary task
in predicting the full differential distributions of the four-body leptonic bottom-meson decays.
To this end, it proves more convenient to introduce an alternative parametrization
of the nonlocal matrix element $T_{\nu \mu}(p, q)$ for facilitating the practical QCD calculations
\begin{eqnarray}
T_{\nu \mu}(p, q) &=& i \, \epsilon_{\mu \nu \rho \sigma} \, p^{\rho} \, v^{\sigma} \, F_V(p^2, n \cdot p)
+ v \cdot p \, g_{\mu \nu}^{\perp} \, \hat{F}_A(p^2, n \cdot p) + v_{\nu} \, p_{\mu} \, \hat{F}_1(p^2, n \cdot p)  \nonumber \\
&& + v_{\mu} \, p_{\nu} \, \hat{F}_2(p^2, n \cdot p) + v \cdot p \,\, v_{\mu} \, v_{\nu} \, \hat{F}_3(p^2, n \cdot p)
+\frac{p_{\mu} \, p_{\nu}}{v \cdot p } \, \hat{F}_4(p^2, n \cdot p)   \,,
\label{New Lorentz decomposition}
\end{eqnarray}
by separating the Lorentz structure $g_{\mu \nu}$ into the longitudinal and transverse components
\begin{eqnarray}
g_{\mu \nu} = g_{\mu \nu}^{\|}  + g_{\mu \nu}^{\perp}  \,,
\qquad
g_{\mu \nu}^{\|} = {n_{\mu} \bar n_{\nu} +  \bar n_{\mu} n_{\nu} \over 2}  \,.
\end{eqnarray}
It is then straightforward  to establish the transformation rules between $F_i$ and $\hat{F}_i$
($i=1,...,4$)
\begin{eqnarray}
F_1 = \hat{F}_1 - {r_1^2  \over r_1^2 - 4 \, r_2} \, \hat{F}_A \,,
\qquad
F_2 = \hat{F}_2 - {r_1^2  \over r_1^2 - 4 \, r_2} \, \hat{F}_A \,,
\nonumber \\
F_3  = \hat{F}_3 +  {4 \, r_2  \over r_1^2 - 4 \, r_2} \, \hat{F}_A \,,
\qquad
F_4 = \hat{F}_4  +  {r_1^2  \over r_1^2 - 4 \, r_2} \, \hat{F}_A  \,,
\label{relations between Fi and Fihat}
\end{eqnarray}
where we have introduced the two dimensionless kinematic variables
\begin{eqnarray}
r_1 = {2 v \cdot p \over \overline{m}_b}  \,,
\qquad
r_2 = {p^2  \over \overline{m}_b^2} \,,
\qquad
({\rm with}
\hspace{0.2 cm}  r_1^2 - 4 r_2 > 0)  \,,
\end{eqnarray}
allowing for an  equivalent formulation  of the Ward-Takahashi identities
(\ref{Ward indentities: form I})
\begin{eqnarray}
\hat{F}_1 = - {4 r_2 \over r_1^2} \,  \hat{F}_4 \,,
\qquad
\hat{F}_3 &=&  - \frac{4 \, (Q_b-Q_u) \, f_B \, m_B}{r_1^2 \,\, \overline{m}_b^2}
- {4 r_2 \over r_1^2}   \, \hat{F}_2    \,.
\label{New WT relations}
\end{eqnarray}
It is important to stress that these relations hold to all orders in the perturbative expansion
and to all orders in the heavy quark expansion, irrespective of the power-counting behaviour
of the off-shell photon momentum.

\section{QCD factorization for $B_{u}^{-} \to \gamma^{\ast} \, \ell \, \bar \nu_{\ell}$ with a hard-collinear photon}
\label{section: QCDF with a hard-collinear photon}

\subsection{The $B$-meson decay form factors at leading power}

We will proceed to derive the factorized expressions for the radiative leptonic
$B_{u}^{-}(p_B) \to \gamma^{\ast}(p)  \,  \ell(q_1) \, \bar \nu_{\ell}(q_2)$
decay form factors in the kinematic region of  $p^2 \sim {\cal O}(m_b \, \Lambda_{\rm QCD})$
by implementing the perturbative matching program ${\rm QCD} \to {\rm SCET_{I}} \to {\rm SCET_{II}}$
for the hadronic matrix element $T_{\nu \mu}$.
Integrating out the hard fluctuation modes with virtualities of order $m_b^2$
for the  $B$-meson-to-vacuum correlation function (\ref{definition: hadronic tensor})
results in the ${\rm SCET_{I}}$ representation
\footnote{For definiteness, here we employ the power counting scheme
of the two external momenta
\begin{eqnarray}
n \cdot p \sim {\cal O}(m_b) \,,
\qquad
\bar n \cdot p \sim {\cal O}(\Lambda_{\rm QCD}) \sim {\cal O}(\lambda^2) \, n \cdot p \,,
\qquad
n \cdot q \sim \bar n \cdot q \sim {\cal O}(m_b) \,,
\nonumber
\end{eqnarray}
where the scaling parameter $\lambda$ is defined to be of order $(\Lambda_{\rm QCD}/m_b)^{1/2}$.}
\begin{eqnarray}
{\cal T}_{\nu \mu}^{\|}(p, q) &=& Q_u \, \left[ C_V^{\rm (A0) \, 2}(n \cdot p, \mu)  \, v_{\mu}
+ \left (  C_V^{\rm (A0) \, 1}(n \cdot p, \mu) + C_V^{\rm (A0) \, 3}(n \cdot p, \mu)   \right ) \bar n_{\mu} \right ]
\nonumber \\
&&  \times \, \bigg \{   \int d^4 x \, e^{i p \cdot x} \,  \langle 0 |
{\rm T}  \{j_{\xi q_s, \,  \nu}^{(2), \, \|}(x),   (\bar \xi W_c)(0) \, (1 + \gamma_5) \, h_v(0)  \} |  B^{-}_{v}\rangle
\nonumber \\
&& \hspace{0.4 cm} +  \int d^4 x \, e^{i p \cdot x} \, \int d^4 y  \, \langle 0 |
{\rm T}  \{j_{\xi \xi, \,  \nu}^{(0), \, \|}(x), \,  i \,  {\cal L}_{\xi q_s}^{(2)}(y), \,
(\bar \xi W_c)(0) \, (1 + \gamma_5) \,  h_v(0)  \} |  B^{-}_{v}\rangle  \bigg \}
\nonumber \\
&& + \, {Q_u \over m_b} \, \int_0^1 d \tau \,  \left[ C_V^{\rm (B1) \, 1}(n \cdot p, \tau, \mu)  \, v_{\mu}
+  C_V^{\rm (B1) \, 2}(n \cdot p, \tau,  \mu)  \,  \bar n_{\mu} \right ]
\nonumber \\
&& \times \, \bigg \{ {n \cdot p \over 2 \pi} \, \int d^4 x \, e^{i p \cdot x} \,  \int d^4 y
 \int d r \, e^{-i n \cdot p \, \tau \, r}  \,
\nonumber \\
&& \hspace{0.5 cm}   \langle 0 | {\rm T}  \{j_{\xi \xi, \,  \nu}^{(0), \, \|}(x), \,  i \,  {\cal L}_{\xi q_s}^{(1)}(y), \,
(\bar \xi W_c)(0) \, (1 + \gamma_5) \,  (W_c^{\dagger} i \not \! \!  D_{\perp c} W_c)(r n) \,  h_v(0)  \} |  B^{-}_{v} \rangle  \bigg \}   \,,
\label{correlator L: SCET-I}
\hspace{1.0 cm}
\end{eqnarray}
for the longitudinal indices $\mu$ and $\nu$, and
\begin{eqnarray}
{\cal T}_{\nu \mu}^{\perp}(p, q) &=& Q_u \, C_V^{\rm (A0) \, 1}(n \cdot p, \mu) \,
\bigg \{   \int d^4 x \, e^{i p \cdot x} \,  \langle 0 |
{\rm T}  \{j_{\xi q_s, \,  \nu}^{(2), \, \perp}(x),
(\bar \xi W_c)(0) \, \gamma_{\mu}^{\perp} \, (1 - \gamma_5) \, h_v(0) \} |  B^{-}_{v}\rangle
\nonumber \\
&& +  \int d^4 x \, e^{i p \cdot x} \, \int d^4 y  \, \langle 0 |
{\rm T}  \{j_{\xi \xi, \,  \nu}^{(1), \, \perp}(x), \,  i \,  {\cal L}_{\xi q_s}^{(1)}(y), \,
(\bar \xi W_c)(0) \, \gamma_{\mu}^{\perp} \, (1 - \gamma_5)  \,  h_v(0)  \} |  B^{-}_{v}\rangle  \bigg \},
\nonumber \\
\label{correlator T: SCET-I}
\end{eqnarray}
for the transverse indices $\mu$ and $\nu$.
The explicit expressions of the effective electromagnetic currents
up to the ${\cal O}(\lambda^2)$ accuracy can be written as
\begin{eqnarray}
j_{\xi \xi, \,  \nu}^{(0), \, \|} &=& \bar \xi \, {\not \! n \over 2} \, \xi \,\, \bar n_{\nu}  \,,
\nonumber  \\
j_{\xi \xi, \,  \nu}^{(1), \, \perp} &=& \bar \xi \, \gamma_{\nu \perp} \,
{1 \over i \, n \cdot D_c} \, i \not \! \! D_{c \perp}\,  {\not \! n \over 2} \, \xi
+ \bar \xi \, i \not \! \! D_{c \perp}\,
{1 \over i \, n \cdot D_c} \, \gamma_{\nu \perp} \,  {\not \! n \over 2} \, \xi   \,,
\nonumber \\
j_{\xi q_s, \,  \nu}^{(2), \, \|}   &=& \left ( \bar \xi \, W_c \,  {\not \! n \over 2} \, Y_s^{\dagger} \, q_s
+ \bar q_s \, Y_s \,  {\not \! n \over 2} \,  W_c^{\dagger} \, \xi \right ) \,\, \bar n_{\nu} \,,
\nonumber \\
j_{\xi q_s, \,  \nu}^{(2), \, \perp}   &=&  \bar \xi \, W_c \,   \gamma_{\perp \nu}  \, Y_s^{\dagger}\, q_s
+ \bar q_s   \,   Y_s \,  \gamma_{\perp \nu} \, W_c^{\dagger} \, \xi \,.
\end{eqnarray}
The multipole expanded ${\rm SCET}$ Lagrangian with the homogenous power counting
in the expansion parameter $\lambda$ appearing in (\ref{correlator L: SCET-I})
and (\ref{correlator T: SCET-I}) reads  \cite{Beneke:2002ni}
(see \cite{Pirjol:2002km} also for an independent derivation in the hybrid momentum-position space)
\begin{eqnarray}
{\cal L}_{\xi q_s}^{(1)} &=& \bar q_s \,  W_c^{\dag} \,\,  i \not \! \! D_{\perp c}\,\, \xi
- \bar \xi \,\,   i \not \! \! \overleftarrow{D}_{\perp c} \,\,  W_c \, q_s,   \nonumber   \\
{\cal L}_{\xi q_s}^{(2)} &=& \bar q_s \,  W_c^{\dag} \,\,   \left ( i \, \bar n \cdot D
+ i \not \! \! D_{\perp c} \,\, {1 \over i \, n \cdot D_c} \, i \not \! \! D_{\perp c} \right ) \,\,
{\not \! n \over 2 }  \,\, \xi   \nonumber \\
&& -  \, \bar \xi \, {\not \! n \over 2 } \,\,  \left ( i \, \bar n \cdot \overleftarrow{D}
+ \, i \not \! \! \overleftarrow{D}_{\perp c} \,\, {1 \over i \, n \cdot \overleftarrow{D}_c} \,
i \not \! \! \overleftarrow{D}_{\perp c} \right ) \,\,
W_c \,\, q_s   \nonumber \\
&& + \,  \bar q_s  \, \overleftarrow{D}_s^{\mu} \, x_{\perp \mu} \,  W_c^{\dag} \,\,  i \not \! \! D_{\perp c}\,\, \xi
- \bar \xi  \,  i \not \! \! \! \overleftarrow{ D}_{\perp c} \,\,  W_c \, x_{\perp \mu}  \, D_s^{\mu} \,   q_s \,.
\end{eqnarray}
In addition, the perturbative matching coefficients of the ${\rm A0}$-type and ${\rm B1}$-type
${\rm SCET_{I}}$ operators at the  required accuracy are further given by
\cite{Bauer:2000yr,Beneke:2004rc,Hill:2004if,Beneke:2005gs}
\begin{eqnarray}
C_V^{\rm (A0) \, 1}(n \cdot p, \mu) &=&
1 + {\alpha_s \, C_F \over 4 \, \pi}  \,
\bigg \{ -2 \, \ln^2 \left ({n \cdot p \over \mu} \right )
+ 5 \, \ln \left  ({n \cdot p \over \mu} \right )- 2 \, {\rm Li}_2 (1-r)
- {3 - 2 \, r \over 1 - r} \, \ln r \nonumber \\
&&  - {\pi^2 \over 12} - 6 \bigg \} + {\cal O}(\alpha_s^2) \,,
\nonumber \\
C_V^{\rm (A0) \, 2}(n \cdot p, \mu) &=&  {\alpha_s \, C_F \over 4 \, \pi}  \,
\left \{  {2 \, r \over (1 - r)^2} \, \ln r + {2 \over 1-r}  \right \}
+ {\cal O}(\alpha_s^2) \,,
\nonumber \\
C_V^{\rm (A0) \, 3}(n \cdot p, \mu) &=& - {\alpha_s \, C_F \over 4 \, \pi}  \,
\left \{  \left [  {r^2 \over (1 - r)^2} - {1 \over 1-r} +1 \right ]  \, \ln r
+ {r \over 1-r}  \right \} + {\cal O}(\alpha_s^2) \,,
\nonumber \\
C_V^{\rm (B1) \, 1}(n \cdot p, \tau, \mu) &=& - {2 \over r} + {\cal O}(\alpha_s)\,,
\qquad
C_V^{\rm (B1) \, 2}(n \cdot p, \tau, \mu) =  \left ( -1 + {1 \over r}  \right )
+ {\cal O}(\alpha_s) \,,
\label{one-loop hard functions in SCET}
\end{eqnarray}
with the two abbreviations $r= n \cdot p / m_b$ and $\alpha_s=\alpha_s(\mu)$.
It is interesting to remark that the three-body ${\rm B1}$-type effective operators
cannot generate the leading-power contribution to the ${\rm SCET_{I}}$ correlation
function ${\cal T}_{\nu \mu}^{\perp}$ in comparison with
the hard-collinear factorization formula for the light-ray matrix element ${\cal T}_{\nu \mu}^{\|}$.

Implementing the ${\rm SCET_{I}} \to {\rm SCET_{II}}$  matching procedure for
(\ref{correlator L: SCET-I}) and  (\ref{correlator T: SCET-I})
by integrating out the hard-collinear fluctuation at the short-distance scale
$(m_b \, \Lambda_{\rm QCD})^{1/2}$ subsequently yields
\begin{eqnarray}
{\cal T}_{\nu \mu}^{\|}(p, q) &=&
Q_u \, \left[ C_V^{\rm (A0) \, 2}(n \cdot p, \mu)  \, v_{\mu}
+ \left (  C_V^{\rm (A0) \, 1}(n \cdot p, \mu) + C_V^{\rm (A0) \, 3}(n \cdot p, \mu)   \right ) \bar n_{\mu} \right ]
\, \bar n_{\nu}
\nonumber \\
&& \times \, {\tilde{f}_B(\mu) \, m_B \over 2} \, \int_0^{\infty} d \omega \,
{\phi_B^{-}(\omega, \mu) \over \bar n \cdot p - \omega + i 0}  \,
{\cal J}_{\|}^{\rm (A0)}(n \cdot p, \bar n \cdot p, \omega, \mu)
\nonumber \\
&& + \, {Q_u \over m_b} \, \int_0^1 d \tau \,  \left[ C_V^{\rm (B1) \, 1}(n \cdot p, \tau, \mu)  \, v_{\mu}
+  C_V^{\rm (B1) \, 2}(n \cdot p, \tau, \mu)  \,  \bar n_{\mu} \right ]
\, \bar n_{\nu}
\nonumber \\
&& \times \, {\tilde{f}_B(\mu) \, m_B \over 2} \, \int_0^{\infty} d \omega \,
\phi_B^{+}(\omega, \mu)  \,
{\cal J}_{\|}^{\rm (B1)}(n \cdot p, \tau, \bar n \cdot p, \omega, \mu)  \,,
\label{SCET factorization: L}
\end{eqnarray}
which can be taken from the analytical expressions of the soft-collinear factorization formulae for
the $B$-meson-to-vacuum correlation functions $\Pi_{\nu, \, \|}$ and $\tilde{\Pi}_{\nu, \, \|}$
obtained  in \cite{Gao:2019lta,DeFazio:2007hw},
and
\begin{eqnarray}
{\cal T}_{\nu \mu}^{\perp}(p, q) &=&  - Q_u  \, C_V^{\rm (A0) \, 1}(n \cdot p, \mu)  \,
\left ( g_{\mu \nu}^{\perp} - i \, \epsilon_{\mu \nu n v} \right ) \,
\nonumber \\
&& \times \, {\tilde{f}_B(\mu) \, m_B \over 2} \, \int_0^{\infty} d \omega \,
{\phi_B^{+}(\omega, \mu) \over \bar n \cdot p - \omega + i 0}  \,
{\cal J}_{\perp}^{\rm (A0)}(n \cdot p, \bar n \cdot p, \omega, \mu)  \,,
\label{SCET factorization: T}
\end{eqnarray}
which allows us to determine the two generalized $B_{u}^{-} \to \gamma^{\ast}$ form factors
with a transversely polarized virtual photon state.
The renormalized jet functions entering (\ref{SCET factorization: L})
and (\ref{SCET factorization: T}) have been worked out in \cite{Gao:2019lta,Wang:2016qii}
up to the ${\cal O}(\alpha_s)$ order
\begin{eqnarray}
{\cal J}_{\|}^{\rm (A0)}(n \cdot p, \bar n \cdot p, \omega, \mu)
&=& 1 + {\alpha_s \, C_F \over 4 \pi} \,
\bigg \{ \ln^2 {\mu^2 \over n \cdot p \, (\omega- \bar n \cdot p)}
- 2 \, \ln {\mu^2 \over n \cdot p \, (\omega-  \bar n \cdot p)}  \, \ln \left( 1 - {\omega \over \bar n \cdot p} \right)
\nonumber \\
&& - \ln^2  \left( 1 - {\omega \over \bar n \cdot p} \right)
- \left ( {2 \, \bar n \cdot p \over \omega} + 1 \right ) \, \ln \left( 1 - {\omega \over \bar n \cdot p} \right)
- {\pi^2 \over 6}  - 1  \bigg \} \,, \nonumber \\
{\cal J}_{\|}^{\rm (B1)}(n \cdot p, \tau, \bar n \cdot p, \omega, \mu)
&=&  {\alpha_s \, C_F \over 2 \pi} \, {n \cdot p \over \omega}  \,
\ln \left ( 1 - {\omega \over \bar n \cdot p} \right ) \,
(1- \tau) \, \theta(\tau) \,\theta(1-\tau) \,, \nonumber \\
{\cal J}_{\perp}^{\rm (A0)}(n \cdot p, \bar n \cdot p, \omega, \mu)
&=&  1 + {\alpha_s \, C_F \over 4 \pi} \,
\bigg \{ \ln^2 {\mu^2 \over n \cdot p \, (\omega- \bar  n \cdot p)}
- {\pi^2 \over 6} -1
\nonumber \\
&& - { \bar n \cdot p \over \omega} \, \ln {\bar n \cdot p - \omega \over \bar n \cdot p} \,
\left [  \ln {\mu^2 \over -p^2}  + \ln {\mu^2 \over n \cdot p \, (\omega- \bar n \cdot p)} + 3 \right ]
 \bigg \} \,.
\end{eqnarray}

Matching the achieved SCET representations  (\ref{SCET factorization: L})
and (\ref{SCET factorization: T}) for ${\cal T}_{\nu \mu}^{\|, \, \perp}$
onto the general decomposition (\ref{New Lorentz decomposition})
of the nonlocal matrix element $T_{\nu \mu}$ with the requirement
$T_{\nu \mu} = {\cal T}_{\nu \mu}^{\|} + {\cal T}_{\nu \mu}^{\perp}$
leads to the desired expressions of the $B_{u}^{-} \to \gamma^{\ast} \, \ell \, \bar \nu_{\ell}$
form factors with a hard-collinear photon
\begin{eqnarray}
F_{V, \, \rm LP} &=&  \hat{F}_{A, \, \rm LP} = -  {Q_u \, \tilde{f}_B(\mu) \, m_B \over n \cdot p} \,
C_V^{\rm (A0) \, 1}(n \cdot p, \mu)  \,
\int_0^{\infty} d \omega \,
{\phi_B^{+}(\omega, \mu) \over \bar n \cdot p - \omega + i 0}  \,
{\cal J}_{\perp}^{\rm (A0)}(n \cdot p, \bar n \cdot p, \omega, \mu)  \,,
\nonumber
\label{LP SCET factorization formula for FV} \\
\\
\hat{F}_{1, \, \rm LP} &=&  \left ( - {4 r_2 \over r_1^2} \right ) \,  \hat{F}_{4, \, \rm LP}  \,,
\label{LP SCET factorization formula for F1hat}  \\
\hat{F}_{2, \, \rm LP} &=&  {Q_u \, \tilde{f}_B(\mu) \, m_B \over n \cdot p} \,
\bigg \{  C_V^{\rm (A0) \, 2}(n \cdot p, \mu)  \,
\int_0^{\infty} d \omega \,
{\phi_B^{-}(\omega, \mu) \over \bar n \cdot p - \omega + i 0}  \,
{\cal J}_{\|}^{\rm (A0)}(n \cdot p, \bar n \cdot p, \omega, \mu)
\nonumber \\
&& + {1 \over m_b} \, \int_0^1 d \tau \,  C_V^{\rm (B1) \, 1}(n \cdot p, \tau, \mu)  \,
\int_0^{\infty} d \omega \, \phi_B^{+}(\omega, \mu)  \,
{\cal J}_{\|}^{\rm (B1)}(n \cdot p, \tau, \bar n \cdot p, \omega, \mu) \bigg \} \,,
\label{LP SCET factorization formula for F2hat}
 \\
\hat{F}_{3, \, \rm LP} &=& \left ( - {4 r_2 \over r_1^2}  \right )  \, \hat{F}_{2, \, \rm LP}
+  \frac{4 \, Q_u \, \tilde{f}_B(\mu)}{r_1^2 \, m_B} \, K(\mu) \,,
\label{LP SCET factorization formula for F3hat} \\
\hat{F}_{4, \, \rm LP} &=&  {Q_u \, \tilde{f}_B(\mu) \, m_B \over n \cdot p} \,
\bigg \{ \left [  C_V^{\rm (A0) \, 1}(n \cdot p, \mu) +  C_V^{\rm (A0) \, 3}(n \cdot p, \mu) \right ]  \,
\int_0^{\infty} d \omega \,
{\phi_B^{-}(\omega, \mu) \over \bar n \cdot p - \omega + i 0}  \,
\nonumber \\
&& \hspace{3 cm} \times \, {\cal J}_{\|}^{\rm (A0)}(n \cdot p, \bar n \cdot p, \omega, \mu)
\nonumber \\
&& + {1 \over m_b} \, \int_0^1 d \tau \,  C_V^{\rm (B1) \, 2}(n \cdot p, \tau, \mu)  \,
\int_0^{\infty} d \omega \, \phi_B^{+}(\omega, \mu)  \,
{\cal J}_{\|}^{\rm (B1)}(n \cdot p, \tau, \bar n \cdot p, \omega, \mu) \bigg \} \,,
\qquad
\label{SCET factorization of form factors: hard-colinear region}
\end{eqnarray}
where the soft-collinear factorization formulae of the subleading power form factors $\hat{F}_1$ and $\hat{F}_3$
are obtained by applying the two constraints (\ref{New WT relations}) from
the ${\rm U_{em} (1)}$ gauge symmetry  of the electromagnetic interaction.
The perturbative function $K(\mu)$ arises from expressing the QCD decay constant $f_B$
in terms of the static decay constant $\tilde{f}_B(\mu)$ \cite{Beneke:2011nf}
\begin{eqnarray}
f_B = \tilde{f}_B(\mu) \, K(\mu)
= \tilde{f}_B(\mu) \, \left [ 1 - {\alpha_s(\mu) \, C_F \over 2 \,  \pi} \,
\left ( {3 \over 4} \, \ln {\mu^2 \over m_b^2}  + 1 \right )
+  {\cal O} (\alpha_s^2) \right ]  \,.
\label{matching relation of fB}
\end{eqnarray}

Inspecting the obtained factorization formulae
(\ref{LP SCET factorization formula for FV}),
(\ref{LP SCET factorization formula for F1hat}),
(\ref{LP SCET factorization formula for F2hat}),
(\ref{LP SCET factorization formula for F3hat}),
(\ref{SCET factorization of form factors: hard-colinear region})
for the $B$-meson radiative decay form factors indicates that it is inevitable to generate the parametrically
enhanced logarithms of $m_b/\Lambda_{\rm QCD}$ by employing a universal value of the factorization scale $\mu$,
which warrant an all-order summation in perturbation theory at the desired accuracy.
To this end, we will set the factorization scale $\mu$ of order $\sqrt{\Lambda_{\rm QCD} \, m_b}$
and take advantage of the RG evolution equations for
the  hard matching coefficient $ C_V^{({\rm A0}), \, 1}$,
the conversion function $K$,
and the leading-twist $B$-meson distribution amplitude $\phi_B^{+}$ in momentum space
\cite{Hill:2004if,Beneke:2005gs,Lange:2003ff,Liu:2020ydl}
\begin{eqnarray}
&& {d \over d \ln \mu} \, C_V^{({\rm A0}), \, 1}(n \cdot p, \mu) =
\left [- \Gamma_{\rm cusp}(\alpha_s) \,
\ln \left ( {\mu \over n \cdot p} \right ) + \gamma^{(\rm A0)}(\alpha_s)\right ] \,
C_V^{({\rm  A0}), \, 1}(n \cdot p, \mu) \,,
\nonumber \\
&& {d \over d \ln \mu} \,  K^{-1}(\mu) =
\gamma_{K}(\alpha_s) \, K^{-1}(\mu)   \,,
\nonumber \\
&& {d \phi_B^{+}(\omega, \mu)  \over d \ln \mu} =
\left [ \Gamma_{\rm cusp}(\alpha_s) \, \ln {\omega \over \mu}
- \gamma_{\eta}(\alpha_s)  \right ] \, \phi_B^{+}(\omega, \mu)
+ \Gamma_{\rm cusp}(\alpha_s) \, \int_0^{\infty}  d x \,
\Gamma(1, x) \, \phi_B^{+}\left({\omega \over x}, \mu \right)
\nonumber \\
&& \hspace{2.5 cm} + \, {\cal O}(\alpha_s^2) \,.
\end{eqnarray}
The  perturbative expansions for the various anomalous dimensions read
\begin{eqnarray}
\Gamma_{\rm cusp}(\alpha_s) &=& \sum_{n=0}^{\infty}  \left ( {\alpha_s \over 4 \pi} \right )^{n+1}  \,
\Gamma_{\rm cusp}^{(n)} \,,
\qquad
\gamma^{(\rm A0)}(\alpha_s) = \sum_{n=0}^{\infty}  \left ( {\alpha_s \over 4 \pi} \right )^{n+1}  \,
\gamma^{{(\rm A0)}, \, (n)}  \,,
\nonumber \\
\gamma_{K}(\alpha_s) &=& \sum_{n=0}^{\infty}  \left ( {\alpha_s \over 4 \pi} \right )^{n+1}  \,
\gamma_{K}^{(n)} \,,
\qquad
\hspace{0.8 cm}
\gamma_{\eta}(\alpha_s) = \sum_{n=0}^{\infty}  \left ( {\alpha_s \over 4 \pi} \right )^{n+1}  \,
\gamma_{\eta}^{(n)} \,,
\end{eqnarray}
where the series coefficients of our interest are given by
\begin{eqnarray}
&& \Gamma_{\rm cusp}^{(0)}  = 4 \, C_F \,,
\qquad
\Gamma_{\rm cusp}^{(1)} =  C_F \left [ {268 \over 3} - 4 \, \pi^2  - {40 \over 9} \, n_f \right ]  \,,
\nonumber \\
&& \Gamma_{\rm cusp}^{(2)} =  C_F \left \{ 1470 - {536 \pi^2 \over 3} + {44 \pi^4 \over 5}
+ 264 \, \zeta(3)  + \left [-{1276 \over 9} + {80 \pi^4 \over 9}  -{208 \over 3} \, \zeta(3) \right ] \, n_f
- {16 \over 27} \, n_f^2   \right \}  \,, \nonumber \\
&& \gamma^{{(\rm A0)}, \, (0)}   =  - 5  \, C_F \,,
\qquad
\gamma^{{(\rm A0)}, \, (1)}   =   C_F \left [ - {1585 \over 18} - {5\, \pi^2 \over 6 }
+ \left ( {125 \over 27} + {\pi^2 \over 3} \right ) \, n_f \right ] \,,
\nonumber \\
&& \gamma_{K}^{(0)} = 3 \, C_F   \,,
\qquad
\gamma_{K}^{(1)} = C_F  \, \left [  {127 \over 6} +   {14 \pi^2 \over 9} - {5 \over 3} \, n_f  \right ] \,,
\qquad
\gamma_{\eta}^{(0)} = -2 \, C_F   \,,
 \\
&& \gamma_{\eta}^{(1)} =  C_F \, \left \{ C_F \,
\left [ - 4  +  {14 \pi^4 \over 3}  - 24 \,  \zeta(3) \right ]
+  \left [ {254 \over 9}  -  {55 \pi^4 \over 6}   - 18 \,  \zeta(3)  \right ]
+ \left [ - {32 \over 27}  +  {5 \pi^2 \over 9}   \right ] \, n_f  \right \}   \,.
\nonumber
\end{eqnarray}
The general solutions to these evolution equations can be further written as follows
\cite{Beneke:2011nf,Lee:2005gza}
\begin{eqnarray}
&& C_V^{({\rm A0}), \, 1}(n \cdot p, \mu) = U_1(n \cdot p, \mu_{h1}, \mu) \,
C_V^{({\rm A0}), \, 1}(n \cdot p, \mu_{h1})  \,,
\nonumber \\
&& K^{-1}(\mu) = U_2(\mu_{h2}, \mu) \, K^{-1}(\mu_{h2})   \,,
\nonumber \\
&& \phi_{B}^{+}(\omega, \mu) =  e^{V - 2 \, \gamma_E \, g} \,
{\Gamma(2 -g) \over \Gamma(g)} \, \int_0^{\infty} {d \eta \over \eta} \,
\phi_B^{+}(\eta, \mu_0) \, \left [ { {\rm max}(\omega, \eta) \over \mu_0} \right ]^g
\nonumber \\
&& \hspace{2.0 cm} \times \, \left [ { {\rm min}(\omega, \eta) \over {\rm max}(\omega, \eta) } \right ] \,
{}_2F_1 \left ( 1-g, 2-g, 2, { {\rm min}(\omega, \eta) \over {\rm max}(\omega, \eta) } \right )  \,,
\label{Lee-Neubert solution}
\end{eqnarray}
where the explicit expression of the RG functions  $U_1$ and $U_2$ at the NLL accuracy
can be found in the Appendix of \cite{Beneke:2011nf}
and the perturbative kernels $V$ and $g$ take the following forms \cite{Lee:2005gza,Bosch:2003fc,Bell:2013tfa}
\begin{eqnarray}
V \equiv V(\mu, \mu_0) &=& - \int_{\alpha_s(\mu_0)}^{\alpha_s(\mu)}
{d \alpha \over \beta(\alpha)} \,
\left [ \Gamma_{\rm cusp}(\alpha)  \, \int_{\alpha_s(\mu_0)}^{\alpha} \,
{d \alpha^{\prime} \over \beta(\alpha^{\prime})}  + \gamma_{\eta}(\alpha) \right ] \,,
\nonumber \\
g \equiv g(\mu, \mu_0) &=& \int_{\alpha_s(\mu_0)}^{\alpha_s(\mu)}  \,
d \alpha \, { \Gamma_{\rm cusp}(\alpha)  \over \beta(\alpha)}
\approx - {2 \, C_F \over \beta_0} \, \ln {\alpha_s(\mu) \over \alpha_s(\mu_0)}  \,.
\end{eqnarray}

\subsection{The $B$-meson decay form factors beyond leading power}

We now turn to evaluate the power suppressed contributions to the radiative
$B$-meson decay form factors from a number of distinct sources on the basis of
the perturbative QCD factorization technique:
\begin{itemize}

\item{The subleading correction to the hard-collinear quark propagator at ${\cal O}(\alpha_s^0)$
from the off-shell photon radiation off the light-flavour constituent of the bottom-meson.}

\item{The  two-particle and three-particle higher-twist corrections of the HQET $B$-meson
distribution amplitudes on the light-cone from the non-vanishing transverse motion of quarks
in the leading partonic configuration  and from the non-minimal Fock state
with an additional soft-gluon field. }

\item{The ``kinematic" power correction from the subleading component of the hard-collinear
photon momentum $\bar n \cdot p$ in the hadronic representation of the non-local matrix element
$T_{\nu \mu}$ as presented in (\ref{New Lorentz decomposition}).}

\item{The power suppressed local contribution from the energetic photon emission off
the heavy bottom-quark field at tree level.}

\end{itemize}

\begin{figure}
\begin{center}
\includegraphics[width=0.8 \columnwidth]{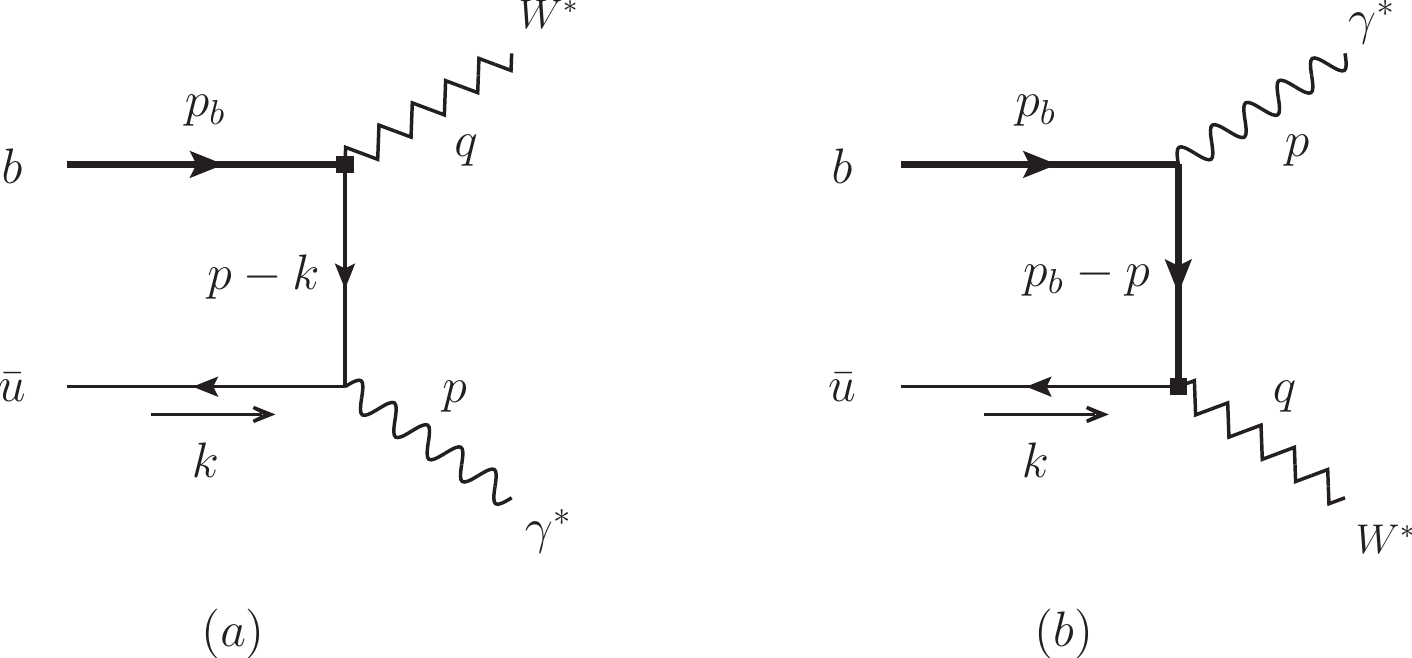}
\vspace*{-0.2 cm}
\caption{Diagrammatical representation of the QCD correlation function
$T_{\nu \mu}(p_B, q)$ at LO in the strong coupling constant. }
\label{fig: tree-level Feynman diagrams}
\end{center}
\end{figure}

Following the computational strategy detailed in \cite{Shen:2020hfq},
we start with the tree-level contribution to the QCD correlation function
(\ref{definition: hadronic tensor}) from the diagram \ref{fig: tree-level Feynman diagrams}(a)
\begin{eqnarray}
T_{\nu \mu} (p, q) & \supset & i \, Q_u \, \int d^4 x \, \int {d^4 k \over (2 \, \pi)^4} \,
{\rm exp}  \left ( i \, k \cdot x \right )  \,
{1 \over (p-k)^2 + i 0}
\nonumber \\
&& \times \, \left \langle 0 \left |\bar u(x) \, \gamma_{\nu} \,
(\not \! p - \not \! k)  \, \gamma_{\mu} \, (1- \gamma_5) \, h_v(0) \right | B^{-}_{u}(v) \right \rangle \,.
\label{tree-level amplitude of Qu}
\end{eqnarray}
Expanding the hard-collinear quark propagator appeared in (\ref{tree-level amplitude of Qu})
at the next-to-leading-power (NLP) accuracy leads to
\begin{eqnarray}
{\not \! p \, - \not \! k \over (p-k)^2}
&=& \underbrace{  {1 \over \bar n \cdot (p-k)} \, {\not \! {\bar n} \over 2} }  \,\,
+ \, \underbrace{\left \{  {n \cdot k \,\,  \bar n \cdot k \over n \cdot p \, \left [ \bar n \cdot (p-k) \right ]^2 } \,
{\not \! {\bar n} \over 2} \,
+  {1 \over n \cdot p} \, {\not \! n \over 2} \,
- {\not \! k_{\perp}  \over n \cdot p \, \bar n \cdot (p-k) }  \right \}}
+ ... \,, \hspace{1.0 cm}
\label{expansion of the hc propagator}
\\
&& \hspace{1.0 cm} {\rm LP}  \hspace{5.8 cm} {\rm NLP}
\nonumber
\end{eqnarray}
where the abbreviation ``${\rm LP}$" stands for the leading power term in the heavy quark expansion.
The yielding NLP correction from the first non-local term in curly brackets
can be computed with the well-known operator identity \cite{Kawamura:2001jm,Kawamura:2001bp}
\begin{eqnarray}
v_{\mu} \,  {\partial \over \partial x_{\mu}} \,
\left [  \bar q(x) \,\Gamma \,  h_v(0) \right ]
= i \, \int_0^1 d u \, \bar u \, \bar q(x) \,g_s \, G_{\alpha \beta}(u x) \,
x^{\alpha} \, v^{\beta} \, \Gamma \,  h_v(0)
+ (v \cdot \partial) \,  \left [  \bar q(x) \,\Gamma \,  h_v(0) \right ],
\end{eqnarray}
due to the HQET equations of motion at the classical level
(see \cite{Braun:2003wx,Bell:2008er} for further discussions).
Moreover, it proves necessary to implement the improved parametrization
of the vacuum-to-$B$-meson  matrix element of the three-body quark-gluon operator
\cite{Braun:2017liq}
\begin{eqnarray}
&& \langle 0 | \bar q_{\alpha}(\tau_1 \, n) \, g_s \, G_{\mu \nu}(\tau_2 \, n) \,
h_{v \, \beta}(0) | \bar B_q (v) \rangle \nonumber \\
&& = {\tilde{f}_{B_q}(\mu) \, m_{B_q} \over 4} \,
\bigg [ (1 + \slashed v) \, \bigg \{ (v_{\mu} \gamma_{\nu} - v_{\nu} \gamma_{\mu})  \,
\left [\Psi_A(\tau_1, \tau_2, \mu) - \Psi_V(\tau_1, \tau_2, \mu) \right ]
- i \, \sigma_{\mu \nu} \, \Psi_V(\tau_1, \tau_2, \mu) \nonumber  \\
&& \hspace{0.4 cm}
- (n_{\mu} \, v_{\nu} - n_{\nu} \, v_{\mu} ) \, X_A(\tau_1, \tau_2, \mu)
+ (n_{\mu} \, \gamma_{\nu} - n_{\nu} \, \gamma_{\mu} ) \,
\left [ W(\tau_1, \tau_2, \mu)  + Y_A(\tau_1, \tau_2, \mu)   \right ] \nonumber \\
&& \hspace{0.4 cm} + \, i \, \epsilon_{\mu \nu \alpha \beta} \,
n^{\alpha} \, v^{\beta}  \, \gamma_5 \, \tilde{X}_A(\tau_1, \tau_2, \mu)
- \, i \, \epsilon_{\mu \nu \alpha \beta} \,
n^{\alpha} \, \gamma^{\beta}  \, \gamma_5 \, \tilde{Y}_A(\tau_1, \tau_2, \mu)  \nonumber \\
&&  \hspace{0.4 cm}  - \, (n_{\mu} \, v_{\nu} - n_{\nu} \, v_{\mu} ) \,
\slashed  n \, W(\tau_1, \tau_2, \mu)
+ \, (n_{\mu} \, \gamma_{\nu} -  n_{\nu} \, \gamma_{\mu} ) \,
\slashed  n \, Z(\tau_1, \tau_2, \mu)   \bigg \}  \,
\gamma_5 \bigg ]_{\beta \, \alpha}  \,.
\label{definition of 3P B-meson LCDA}
\end{eqnarray}
Apparently, the relevant momentum-space distribution amplitudes can be obtained by performing
the Fourier transformation with respect to the light-cone variables $\tau_1$ and $\tau_2$
\begin{eqnarray}
&& \Psi_{X}  (\tau_1, \, \tau_2, \, \mu)
= \int_0^{\infty} \, d \omega_1  \, \int_0^{\infty} \, d \omega_2  \,
e^{-i \,(\omega_1 \, \tau_1 + \omega_2 \, \tau_2)} \,\,
\psi_X(\omega_1, \omega_2,  \mu)  \,,
\nonumber  \\
&& \Psi_{X}  \in  \left \{ \Psi_V, \, \Psi_A, \, X_A, \, Y_A, \,
\tilde{X}_A, \, \tilde{Y}_A, \,  W,  \, Z \right \} \,.
\end{eqnarray}
To facilitate the construction of the desired soft-collinear factorization formulae,
we express the eight invariant functions $\Psi_{X}$ in terms of the more suitable
distribution amplitudes with the definite collinear twist
(see \cite{Geyer:2005fb} for an alternative proposal of geometric twist)
\begin{eqnarray}
\Phi_3 &=& \Phi_A - \Phi_V \,,
\qquad  \hspace{3.8 cm}
\Phi_4 = \Phi_A + \Phi_V \,,
\nonumber \\
\hat{\Psi}_4  &=& \Psi_A + X_A \,,
\qquad  \hspace{3.8 cm}
\tilde{\Psi}_4  = \Psi_V - \tilde{X}_A \,,
\nonumber \\
\tilde{\Phi}_5 &=&  \Psi_A + \Psi_V + 2 \, Y_A - 2 \, \tilde{Y}_A + 2 \, W \,,
\qquad
\Psi_5 =  - \Psi_A + X_A - 2 \, Y_A  \,,
\nonumber \\
\tilde{\Psi}_5 &=&  - \Psi_V -  \tilde{X}_A + 2 \, \tilde{Y}_A  \,,
\qquad  \hspace{2.2 cm}
\Phi_6 =  \Phi_A - \Phi_V + 2 \, Y_A  + 2 \, W
+ 2 \, \tilde{Y}_A - 4 \, Z \,.
\hspace{1.0 cm}
\end{eqnarray}
We can then readily obtain the factorized expression for such NLP contribution
\begin{eqnarray}
T_{\nu \mu, \, {\rm NLP}}^{\rm hc, \, (I)} (p, q)
= - \frac{Q_{u} \, \tilde{f}_{B}(\mu) \, m_{B}}{n\cdot p} \,
\left [\left ( g^{\perp}_{\mu\nu} - i \, \epsilon_{\mu \nu n v} \right ) \,
\mathcal{G}_{\rm NLP, \, L}^{\rm hc, \, (I)}
+\bar{n}_{\mu}\bar{n}_{\nu}\,
\mathcal{G}_{{\rm NLP}, \, \bar n \, \bar n}^{\rm hc, \, (I)} \right ] \,,
\end{eqnarray}
where the newly defined ``form factors"  $\mathcal{G}_{X}^{\rm hc, \, (I)}$ are given by
\begin{eqnarray}
\mathcal{G}_{\rm NLP, \, L}^{\rm hc, \, (I)}
&=& \int^{\infty}_{0} d \omega_1 \,
\int^{\infty}_{0} d \omega_2  \, \int^{1}_{0} du \,\,
\frac{\bar{u}\, (\bar{n}\cdot p + \omega_1 + u \, \omega_2)}
{(\bar{n}\cdot p- \omega_1 -u \, \omega_2)^{3}}\,
\psi_{4}(\omega_1, \omega_2,  \mu) \nonumber \\
&& + \int^{\infty}_{0} d\omega\,
\frac{\omega}{(\bar{n}\cdot p-\omega)^{2}} \,
\left (\bar{\Lambda} - \frac{\omega}{2} \right )\,
\phi_B^{+}(\omega, \mu) \,,
\label{results of hc correction I: A}
\\
\mathcal{G}_{{\rm NLP}, \, \bar n \, \bar n}^{\rm hc, \, (I)}
&=& \int^{\infty}_{0} d \omega_1 \,
\int^{\infty}_{0} d \omega_2  \, \int^{1}_{0} du \,\,
\frac{\bar{u}\, (\bar{n}\cdot p + \omega_1 + u \, \omega_2)}
{(\bar{n}\cdot p- \omega_1 -u \, \omega_2)^{3}}\,
\psi_{5}(\omega_1, \omega_2,  \mu) \nonumber \\
&& + \int^{\infty}_{0} d\omega\,
\frac{\omega}{(\bar{n}\cdot p-\omega)^{2}} \,
\left (\bar{\Lambda} - \frac{\omega}{2} \right )\,
\phi_B^{-}(\omega, \mu) \,.
\label{results of hc correction I: B}
\end{eqnarray}
The hadronic parameter $\bar \Lambda$ entering (\ref{results of hc correction I: A})
and (\ref{results of hc correction I: B})can be defined in a manifestly covariant and
gauge invariant manner \cite{Neubert:1993mb}
\begin{eqnarray}
\bar \Lambda = \frac{\langle 0 | \bar q \,\,  i \, v \cdot \overleftarrow{D} \, \Gamma \,   h_v | \bar B_q (v) \rangle}
{\langle 0 | \bar q \,  \Gamma \,   h_v | \bar B_q (v) \rangle} \,.
\end{eqnarray}
The subleading power contribution from the second local term in curly brackets
of (\ref{expansion of the hc propagator}) can be evidently expressed by
the $B$-meson decay constant
\begin{eqnarray}
T_{\nu \mu, \, {\rm NLP}}^{\rm hc, \, (II)} (p, q)
= - \frac{Q_{u} \, \tilde{f}_{B}(\mu) \, m_{B}}{2 \, n\cdot p} \,
\left [\left ( g^{\perp}_{\mu\nu} + i \, \epsilon_{\mu \nu n v} \right ) \,
- n_{\mu} n_{\nu}  \right ] \,.
\end{eqnarray}
Applying an additional HQET operator identity from the equations of motion
\begin{eqnarray}
{\partial \over \partial x_{\rho}} \bar q(x) \, \gamma_{\rho} \, \Gamma \,   h_v(0)
&=& - i \, \int_0^1 d u \, u \,\, \bar q(x) \,g_s \, G^{\lambda \rho}(u x) \,
x_{\lambda} \, \gamma_{\rho} \, \Gamma \,  h_v(0)  \,,
\end{eqnarray}
we can proceed to construct the tree-level factorization formula
for the third non-local term in curly brackets
of (\ref{expansion of the hc propagator})
\begin{eqnarray}
T_{\nu \mu, \, {\rm NLP}}^{\rm hc, \, (III)} (p, q)
= \frac{Q_{u} \, \tilde{f}_{B}(\mu) \, m_{B}}{n\cdot p} \,
\left [{\bar n}_{\mu} n_{\nu}\,
\mathcal{G}_{{\rm NLP}, \,  \bar n \, n}^{\rm hc, \, (III)}
+ {n}_{\mu} \bar{n}_{\nu}\,
\mathcal{G}_{{\rm NLP}, \,  n \, \bar n}^{\rm hc, \, (III)} \right ]  \,,
\end{eqnarray}
where
\begin{eqnarray}
\mathcal{G}_{{\rm NLP}, \,  \bar n \, n}^{\rm hc, \, (III)}
&=& -  \int^{\infty}_{0} d\omega\,
\frac{\omega}{\bar{n}\cdot p-\omega} \,
{\phi_B^{-}(\omega, \mu)  \over 2}
- \int^{\infty}_{0} d \omega_1 \,
\int^{\infty}_{0} d \omega_2  \, \int^{1}_{0} du \,\,
\frac{u \, \phi_{3}(\omega_1, \omega_2,  \mu) }
{(\bar{n}\cdot p- \omega_1 -u \, \omega_2)^{2}}  \,,
\hspace{0.8 cm} \\
\mathcal{G}_{{\rm NLP}, \,  n \, \bar n}^{\rm hc, \, (III)}
&=&  - \int^{\infty}_{0} d \omega_1 \,
\int^{\infty}_{0} d \omega_2  \, \int^{1}_{0} du \,\,
\frac{1} {(\bar{n}\cdot p- \omega_1 -u \, \omega_2)^{2}} \,
\left [ u \, \phi_{4}(\omega_1, \omega_2,  \mu) + \psi_{4}(\omega_1, \omega_2,  \mu) \right ] \nonumber \\
&& -  \int^{\infty}_{0} d\omega\,
\frac{1}{\bar{n}\cdot p-\omega} \,
\left( \bar \Lambda - { \omega \over 2} \right) \,
\phi_B^{+}(\omega, \mu)  \,.
\end{eqnarray}
Adding the different pieces together, the ``dynamical" power corrections
to the  exclusive $B_{u}^{-} \to \gamma^{\ast} \, \ell \, \bar \nu_{\ell}$ form factors
due to the energetic photon emission from the light quark can be summarized in the following
\begin{eqnarray}
F_{V, \,  {\rm NLP} }^{\rm hc, \, dyn}
&=& - \frac{2 \, Q_{u} \, \tilde{f}_{B}(\mu) \, m_{B}}{(n\cdot p)^2} \,
\left (\mathcal{G}_{\rm NLP, \, L}^{\rm hc, \, (I)} - {1 \over 2} \right )
+ {\cal O} \left(\alpha_s, \, {\Lambda / m_b} \right)   \,,
\nonumber \\
\hat{F}_{A, \,  {\rm NLP} }^{\rm hc, \, dyn}
&=& - \frac{2 \, Q_{u} \, \tilde{f}_{B}(\mu) \, m_{B}}{(n\cdot p)^2} \,
\left (\mathcal{G}_{\rm NLP, \, L}^{\rm hc, \, (I)} + {1 \over 2} \right )
+ {\cal O} \left(\alpha_s, \, {\Lambda / m_b} \right)  \,,
\nonumber \\
\hat{F}_{1, \,  {\rm NLP} }^{\rm hc, \, dyn}
&=&  \frac{4 \, Q_{u} \, \tilde{f}_{B}(\mu) \, m_{B}}{(n\cdot p)^2} \,
\left ( \mathcal{G}_{{\rm NLP}, \,  \bar n \, n}^{\rm hc, \, (III)}  - {1 \over 2}\right  )
+ {\cal O} \left(\alpha_s, \, {\Lambda / m_b} \right)  \,,
\nonumber \\
\hat{F}_{2, \,  {\rm NLP} }^{\rm hc, \, dyn}
&=&  \frac{4 \, Q_{u} \, \tilde{f}_{B}(\mu) \, m_{B}}{(n\cdot p)^2} \,
\left ( \mathcal{G}_{{\rm NLP}, \,  n \, \bar  n}^{\rm hc, \, (III)}  - {1 \over 2}\right  )
+ {\cal O} \left(\alpha_s, \, {\Lambda / m_b} \right)    \,,
\nonumber \\
\hat{F}_{3, \,  {\rm NLP} }^{\rm hc, \, dyn}
&=&  \frac{4 \, Q_{u} \, \tilde{f}_{B}(\mu) \, m_{B}}{(n\cdot p)^2}
+ {\cal O} \left(\alpha_s, \, {\Lambda / m_b} \right)   \,,
\nonumber \\
\hat{F}_{4, \,  {\rm NLP} }^{\rm hc, \, dyn}
&=&  \frac{2 \, Q_{u} \, \tilde{f}_{B}(\mu) \, m_{B}}{(n\cdot p)^2} \,
\left ( \mathcal{G}_{{\rm NLP}, \, \bar n \, \bar n}^{\rm hc, \, (I)}
-  \mathcal{G}_{{\rm NLP}, \,  \bar n \, n}^{\rm hc, \, (III)}
- \mathcal{G}_{{\rm NLP}, \,  n \, \bar n}^{\rm hc, \, (III)}
+ {1 \over 2} \right )
+ {\cal O} \left(\alpha_s, \, {\Lambda / m_b} \right)  \,.
\label{dynamical NLP results}
\end{eqnarray}

We are now in a position to compute the subleading power corrections
to the radiative $B_{u}^{-} \to \gamma^{\ast} \, \ell \, \bar \nu_{\ell}$ form factors
from both the two-particle and three-particle $B$-meson distribution amplitudes
at tree level by employing the perturbative factorization technique.
Implementing the light-cone expansion of the hard-collinear quark propagator
in the background soft-gluon field \cite{Balitsky:1987bk}
(see \cite{Rusov:2017chr} for an improved discussion on the massive quark propagator)
\begin{eqnarray}
\langle 0 | {\rm T} \, \{\bar q (x), q(0) \} | 0\rangle
\supset   i \, g_s \, \int_0^{\infty} \,\, {d^4 \ell \over (2 \pi)^4} \,
{e^{- i \, \ell \cdot x} \over \ell^2 - m_q^2} \,
\int_0^1 \, d u \, \left  [ u \, x_{\mu} \, \gamma_{\nu}
 - \frac{(\slashed \ell + m_q ) \, \sigma_{\mu \nu}}{2 \, \left (\ell^2 - m_q^2 \right )}  \right ]
\, G^{\mu \nu}(u \, x) \,, \hspace{0.6 cm}
\end{eqnarray}
with the gluon-field strength tensor
$G^{\mu \nu} =  G_{\mu \nu}^{a} \, T^a = D_{\mu} \, A_{\nu} -  D_{\nu} \, A_{\mu}$,
and taking advantage of the general parametrization of the three-body HQET matrix element
(\ref{definition of 3P B-meson LCDA}),
we can immediately establish the soft-collinear factorization formulae
for the three-particle higher twist corrections
\begin{eqnarray}
F_{V, \,  {\rm NLP} }^{\rm 3PHT}
&=& \hat{F}_{A, \,  {\rm NLP} }^{\rm 3PHT}
= \frac{Q_{u} \, \tilde{f}_{B}(\mu) \, m_{B}}{(n\cdot p)^2} \,
\int^{\infty}_{0} \, d\omega_1  \int^{\infty}_{0} \, d \omega_2 \,
\int^{1}_{0} du\, \frac{1} {(\bar{n}\cdot p - \omega_1 -u \, \omega_2)^{2}}
\nonumber \\
&& \times \, \left [ (2u-1) \, \psi_{4} (\omega_1, \omega_2, \mu)-\tilde{\psi}_{4}(\omega_1, \omega_2, \mu) \right ]
+ {\cal O} \left(\alpha_s, \, {\Lambda / m_b} \right)  \,,
\nonumber  \\
\hat{F}_{1, \,  {\rm NLP} }^{\rm 3PHT}
&=& \frac{4 \, Q_{u} \, \tilde{f}_{B}(\mu) \, m_{B}}{(n\cdot p)^2} \,
\int^{\infty}_{0} \, d\omega_1  \int^{\infty}_{0} \, d \omega_2 \,
\int^{1}_{0} du\, \frac{u \, \phi_{3} (\omega_1, \omega_2, \mu)}
{(\bar{n}\cdot p - \omega_1 -u \, \omega_2)^{2}}
+ {\cal O} \left(\alpha_s, \, {\Lambda / m_b} \right) \,,
\nonumber  \\
\hat{F}_{2, \,  {\rm NLP} }^{\rm 3PHT}
&=& - \frac{4 \, Q_{u} \, \tilde{f}_{B}(\mu) \, m_{B}}{(n\cdot p)^2} \,
\int^{\infty}_{0} \, d\omega_1  \int^{\infty}_{0} \, d \omega_2 \,
\int^{1}_{0} du\, \frac{(1-u) \, \phi_{4} (\omega_1, \omega_2, \mu)}
{(\bar{n}\cdot p - \omega_1 -u \, \omega_2)^{2}}
+ {\cal O} \left(\alpha_s, \, {\Lambda / m_b} \right) \,,
\nonumber  \\
\hat{F}_{3, \,  {\rm NLP} }^{\rm 3PHT}
&=&  {\cal O} \left(\alpha_s, \, {\Lambda / m_b} \right)  \,,
\nonumber \\
\hat{F}_{4, \,  {\rm NLP} }^{\rm 3PHT}
&=&  \frac{2 \, Q_{u} \, \tilde{f}_{B}(\mu) \, m_{B}}{(n\cdot p)^2} \,
 \,
\int^{\infty}_{0} \, d\omega_1  \int^{\infty}_{0} \, d \omega_2 \,
\int^{1}_{0} du\, \frac{1} {(\bar{n}\cdot p - \omega_1 -u \, \omega_2)^{2}} \,
\Big [ (1-2u) \, \psi_{5} (\omega_1, \omega_2, \mu)
\nonumber \\
&&  + \, \tilde{\psi}_{5}(\omega_1, \omega_2, \mu)
+ (1-u) \, \phi_{4} (\omega_1, \omega_2, \mu)
- u \,   \phi_{3} (\omega_1, \omega_2, \mu) \Big ]
+ {\cal O} \left(\alpha_s, \, {\Lambda / m_b} \right)  \,,
\label{3PHT NLP results}
\end{eqnarray}
where the achieved expressions for the two form factors
$F_{V, \,  {\rm NLP} }^{\rm 3PHT}$ and
$F_{A, \,  {\rm NLP} }^{\rm 3PHT}$ are in accordance with
the analogous NLP contributions to the double radiative bottom-meson decays
in the kinematic limit $\bar n \cdot p \to 0$
as displayed in Eq. (4.37) in \cite{Shen:2020hfq}.

For the purpose of evaluating the two-particle higher-twist effects,
we will introduce the generalized decomposition of the two-body
non-local $B$-meson-to-vacuum matrix element with the off-light-cone corrections
up to the ${\cal O}(x^2)$ accuracy \cite{Braun:2017liq}
\begin{eqnarray}
&& \langle 0 | (\bar q_s Y_s)_{\beta}(x) \,\,\,  (Y_s^{\dag} h_v)_{\alpha}(0) |   \bar B_q \rangle
\nonumber \\
&& = -  {i \, \tilde{f}_{B_q}(\mu) \,  m_{B_q} \over 4} \,
\int_0^{\infty} d \omega \, e^{-i \omega v \cdot x} \, \bigg [  {1+ \slashed v \over 2}   \,
\bigg \{  2 \, \left [ \phi_B^{+}(\omega, \mu)  + x^2 \, g_B^{+}(\omega, \mu)  \right ]
\nonumber \\
&& \hspace{0.5 cm} \, -  {\slashed {x}   \over v \cdot x} \,
\left [ (\phi_B^{+}(\omega, \mu)-\phi_B^{-}(\omega, \mu))
+ x^2 \, ( g_B^{+}(\omega, \mu)-g_B^{-}(\omega, \mu))  \right ] \,
\bigg  \} \, \gamma_5  \bigg  ]_{\alpha \beta} \,.
\label{definition of the 2P LCDA with x2 correction}
\end{eqnarray}
It is then straightforward to write down the resulting factorized expression
\begin{eqnarray}
T_{\nu \mu}^{\rm 2PHT}(p, q) &=& - \frac{2 \, Q_{u} \, \tilde{f}_{B}(\mu) \, m_{B}}{n\cdot p} \,
\bigg \{  \left ( g_{\mu \nu}^{\perp} -  i \, \epsilon_{\mu \nu \rho \sigma} \, n^{\rho} \, v^{\sigma} \right ) \,
\int^{\infty}_{0} \, d \omega \, \frac{ g_B^{+}(\omega, \mu)}{(\bar{n}\cdot p-\omega)^{2}}  \,
\nonumber \\
&& - \, \bar n_{\mu} \, \bar n_{\nu}  \,
\int^{\infty}_{0} d \omega \, \frac{g_B^{-}(\omega, \mu)}{(\bar{n}\cdot p-\omega)^{2}}   \bigg \}  \,.
\label{2PHT factorization formula: original}
\end{eqnarray}
Applying the two non-trivial constraints on the subleading twist HQET distribution amplitudes
in momentum space \cite{Wang:2017jow,Gao:2019lta}
(see also \cite{Kawamura:2001jm,Kawamura:2001bp,Braun:2017liq} for the coordinate-space identities)
\begin{eqnarray}
-2 \, {d^2 \over d \omega^2} \, g_B^{+}(\omega, \mu) &=&
\left [ {3 \over 2} + (\omega - \bar \Lambda) \, {d \over d \omega}   \right ] \, \phi_B^{+}(\omega, \mu)
- {1 \over 2}  \, \phi_B^{-}(\omega, \mu)
+ \int_0^{\infty} \, {d \omega_2 \over \omega_2 } \, {d \over d \omega} \, \psi_4(\omega, \omega_2, \mu) \nonumber \\
&& - \int_0^{\infty} \, {d \omega_2 \over \omega_2^2 } \, \psi_4(\omega, \omega_2, \mu)
+ \int_0^{\omega} \, {d \omega_2 \over \omega_2^2 } \, \psi_4(\omega-\omega_2, \omega_2, \mu) \,,
\label{the EOM of gBplus }  \\
-2 \, {d^2 \over d \omega^2} \, g_B^{-}(\omega, \mu) &=&
\left [ {3 \over 2} + (\omega - \bar \Lambda) \, {d \over d \omega}   \right ] \, \phi_B^{-}(\omega, \mu)
- {1 \over 2}  \, \phi_B^{+}(\omega, \mu)
+ \int_0^{\infty} \, {d \omega_2 \over \omega_2 } \, {d \over d \omega} \, \psi_5(\omega, \omega_2, \mu) \nonumber \\
&& - \int_0^{\infty} \, {d \omega_2 \over \omega_2^2 } \, \psi_5(\omega, \omega_2, \mu)
+ \int_0^{\omega} \, {d \omega_2 \over \omega_2^2 } \, \psi_5(\omega-\omega_2, \omega_2, \mu) \,,
\label{the EOM of gBminus}
\end{eqnarray}
the twist-four and twist-five $B$-meson distribution amplitudes $g_B^{\pm}(\omega, \mu)$
can be  decomposed into the  Wandzura-Wilczek  contributions \cite{Wandzura:1977qf}
calculable from the lower-twist two-particle  distribution amplitudes  $\phi_B^{\pm}(\omega, \mu)$
and  the ``genuine" three-particle distribution amplitudes  of the same collinear twists
\begin{eqnarray}
g_B^{+}(\omega, \mu) &=& \hat{g}_B^{+}(\omega, \mu)
- {1 \over 2} \, \int_0^{\omega} \, d \omega_1 \, \int_0^1 d u \,
{\bar u \over u} \,\, \psi_4 \left (\omega, {\omega - \omega_1 \over u}, \mu \right ) \,,
\nonumber \\
g_B^{-}(\omega, \mu) &=& \hat{g}_B^{-}(\omega, \mu)
- {1 \over 2} \, \int_0^{\omega} \, d \omega_1 \, \int_0^1 d u \,
{\bar u \over u} \,\, \psi_5 \left (\omega, {\omega - \omega_1 \over u}, \mu \right ) \,,
\label{results of gBpm}
\end{eqnarray}
where the manifest expressions of the Wandzura-Wilczek terms are given by
\begin{eqnarray}
\hat{g}_B^{+}(\omega, \mu)&=&
{1 \over 4} \, \int_{\omega}^{\infty} \, d \rho \,
\bigg \{ (\rho - \omega)  \,
\left [\phi_B^{-}(\rho, \mu) - \phi_B^{+}(\rho, \mu)  \right ]
- 2 \, (\bar \Lambda - \rho) \, \phi_B^{+}(\rho, \mu) \bigg \} \,,
\nonumber \\
\hat{g}_B^{-}(\omega, \mu)&=&
{1 \over 4} \, \int_{\omega}^{\infty} \, d \rho \,
\bigg \{ (\rho - \omega)  \,
\left [\phi_B^{+}(\rho, \mu) - \phi_B^{-}(\rho, \mu)  \right ]
- 2 \, (\bar \Lambda - \rho) \, \phi_B^{-}(\rho, \mu) \bigg \} \,.
\label{results of gBpmhat}
\end{eqnarray}
We are then led to an equivalent form of the obtained factorization formula
(\ref{2PHT factorization formula: original})
\begin{eqnarray}
T_{\nu \mu}^{\rm 2PHT}(p, q) &=& - \frac{Q_{u} \, \tilde{f}_{B}(\mu) \, m_{B}}{n\cdot p} \,
\bigg \{  \left ( g_{\mu \nu}^{\perp} -  i \, \epsilon_{\mu \nu \rho \sigma} \, n^{\rho} \, v^{\sigma} \right ) \,
{\cal G}_{\rm NLP, \, L}^{\rm 2PHT}
+ \, \bar n_{\mu} \, \bar n_{\nu}  \,\,
{\cal G}_{{\rm NLP}, \, \bar n \, \bar n}^{\rm 2PHT}   \bigg \}  \,,
\label{2PHT factorization formula: new}
\end{eqnarray}
where the newly introduced invariant functions are defined as follows
\begin{eqnarray}
{\cal G}_{\rm NLP, \, L}^{\rm 2PHT} &=&
2 \, \int^{\infty}_{0} \, d \omega \, \frac{ \hat{g}_B^{+}(\omega, \mu)}{(\bar{n}\cdot p-\omega)^{2}}
- \int^{\infty}_{0} \, d\omega_1  \int^{\infty}_{0} \, d \omega_2 \,
\int^{1}_{0} du\, \frac{(1-u) \, \psi_{4} (\omega_1, \omega_2, \mu)}
{(\bar{n}\cdot p - \omega_1 -u \, \omega_2)^{2}}  \,,
\nonumber \\
{\cal G}_{{\rm NLP}, \, \bar n \, \bar n}^{\rm 2PHT} &=&
- 2 \, \int^{\infty}_{0} \, d \omega \, \frac{ \hat{g}_B^{-}(\omega, \mu)}{(\bar{n}\cdot p-\omega)^{2}}
+ \int^{\infty}_{0} \, d\omega_1  \int^{\infty}_{0} \, d \omega_2 \,
\int^{1}_{0} du\, \frac{(1-u) \, \psi_{5} (\omega_1, \omega_2, \mu)}
{(\bar{n}\cdot p - \omega_1 -u \, \omega_2)^{2}}  \,.
\hspace{0.5 cm}
\end{eqnarray}
Matching the tree-level SCET computation
of the correlation function $T_{\nu \mu}$ (\ref{2PHT factorization formula: new})
onto  the appropriate  hadronic representation  (\ref{New Lorentz decomposition}) yields
\begin{eqnarray}
F_{V, \,  {\rm NLP} }^{\rm 2PHT}
&=& \hat{F}_{A, \,  {\rm NLP} }^{\rm 2PHT}
= - \frac{2 \, Q_{u} \, \tilde{f}_{B}(\mu) \, m_{B}}{(n\cdot p)^2} \,
{\cal G}_{\rm NLP, \, L}^{\rm 2PHT}
+ {\cal O} \left(\alpha_s, \, {\Lambda / m_b} \right)  \,,
\nonumber \\
\hat{F}_{1, \,  {\rm NLP} }^{\rm 2PHT}
&=& {\cal O} \left(\alpha_s, \, {\Lambda / m_b} \right)  \,,
\nonumber \\
\hat{F}_{2, \,  {\rm NLP} }^{\rm 2PHT}
&=& {\cal O} \left(\alpha_s, \, {\Lambda / m_b} \right)  \,,
\nonumber \\
\hat{F}_{3, \,  {\rm NLP} }^{\rm 2PHT}
&=& {\cal O} \left(\alpha_s, \, {\Lambda / m_b} \right)  \,,
\nonumber \\
\hat{F}_{4, \,  {\rm NLP} }^{\rm 2PHT}
&=& - \frac{2 \, Q_{u} \, \tilde{f}_{B}(\mu) \, m_{B}}{(n\cdot p)^2} \,
{\cal G}_{{\rm NLP}, \, \bar n \, \bar n}^{\rm 2PHT}
+ {\cal O} \left(\alpha_s, \, {\Lambda / m_b} \right)  \,,
\label{2PHT NLP results}
\end{eqnarray}
where the factorized  expression of $F_{V, \,  {\rm NLP} }^{\rm 2PHT}$
($\hat{F}_{A, \,  {\rm NLP} }^{\rm 2PHT}$)
can be alternatively inferred from the two-particle subleading twist correction to
$B_{u}^{-} \to \gamma \, \ell \, \bar \nu_{\ell}$ \cite{Beneke:2018wjp}
and the established formula of $\hat{F}_{4, \,  {\rm NLP} }^{\rm 2PHT}$
is completely consistent with the counterpart contribution to the very correlation function
suitable for constructing the light-cone QCD sum rules for $B \to \pi, K$ form factors
\cite{Lu:2018cfc}.

The ``kinematic" power corrections to the radiative $B$-meson decay form factors
can be determined by employing an equivalent hadronic representation for the correlation function
$T_{\nu \mu}$  other than (\ref{Original Lorentz decomposition}) and  (\ref{New Lorentz decomposition})
\begin{eqnarray}
T_{\nu \mu}(p, q) &=& {n \cdot p  \over 2} \,
\left [ - i \, (1-\kappa_p) \epsilon_{\mu \nu \rho \sigma} \, n^{\rho} \, v^{\sigma} \, F_V
+ (1 + \kappa_p) \,  g_{\mu \nu}^{\perp} \, \hat{F}_A \right ]
\nonumber \\
&& + {n \cdot p  \over 4} \,  \left [ \left (\hat F_1 + \frac{\hat F_3}{2} \right )
+ \kappa_{p} \,  \left (\hat F_2+\frac{\hat F_3}{2}+2\,\hat F_4 \right ) \right ] \,
\bar{n}_\mu \, n_\nu
\nonumber \\
&& + {n \cdot p  \over 4} \, \left [ \left (\hat F_2 + \frac{\hat F_3}{2} \right )
+ \kappa_{p} \,  \left (\hat F_1 + \frac{\hat F_3}{2}+2\,\hat F_4 \right ) \right ] \,
n_\mu \, {\bar n}_\nu
\nonumber \\
&& + {n \cdot p  \over 4} \, \left [ \left (\hat F_1 + \hat F_2 + \frac{\hat F_3}{2}+ 2 \, \hat F_4 \right )
+ \kappa_{p} \,  \left (\frac{\hat F_3}{2} - 2 \, \hat F_4 \right ) \right ] \,
{\bar n}_\mu \, {\bar n}_\nu
\nonumber \\
&& +  {n \cdot p  \over 4} \, \left [ \frac{\hat F_3}{2}
+ \kappa_{p} \,  \left (\hat F_1 + \hat F_2 + \frac{\hat F_3}{2}  \right ) \right ] \,
n_\mu \, n_\nu
+ {\cal O} (\kappa_p^2) \,,
\label{Updated Lorentz decomposition}
\end{eqnarray}
where the dimensionless variable  $\kappa_p$  is explicitly defined by
\begin{eqnarray}
\kappa_{p} \equiv {p^2 \over (n\cdot p)^2} \sim {\cal O} \left ( {\Lambda_{\rm QCD} \over  m_b} \right ) \,.
\end{eqnarray}
Confronting (\ref{Updated Lorentz decomposition}) with the LP QCD expression
for the diagram \ref{fig: tree-level Feynman diagrams}(a) allows for the determination
of the subleading power  ``kinematic"  corrections at tree level
\begin{eqnarray}
F_{V, \,  {\rm NLP} }^{\rm hc, \, kin}
&=&  - \kappa_{p} \, \frac{Q_{u} \, \tilde{f}_{B}(\mu) \, m_{B}}{n\cdot p}\,
\int^{\infty}_{0} \, \frac{d\omega}{\bar{n}\cdot p-\omega} \,
\phi_B^{+}(\omega, \mu)
+ {\cal O} \left(\alpha_s, \, \kappa_p^2 \right) \,,
\nonumber \\
\hat{F}_{A, \,  {\rm NLP} }^{\rm hc, \, kin}
&=&   \kappa_{p} \, \frac{Q_{u} \, \tilde{f}_{B}(\mu) \, m_{B}}{n\cdot p}\,
\int^{\infty}_{0} \, \frac{d\omega}{\bar{n}\cdot p-\omega} \,
\phi_B^{+}(\omega, \mu)
+ {\cal O} \left(\alpha_s, \, \kappa_p^2 \right) \,,
\nonumber \\
\hat{F}_{1, \,  {\rm NLP} }^{\rm hc, \, kin}
&=&  - 2 \,  \kappa_{p} \, \frac{Q_{u} \, \tilde{f}_{B}(\mu) \, m_{B}}{n\cdot p}\,
\int^{\infty}_{0} \, \frac{d\omega}{\bar{n}\cdot p-\omega} \,
\phi_B^{-}(\omega, \mu)
+ {\cal O} \left(\alpha_s, \, \kappa_p^2 \right) \,,
\nonumber \\
\hat{F}_{2, \,  {\rm NLP} }^{\rm hc, \, kin}
&=&  - 2 \,  \kappa_{p} \, \frac{Q_{u} \, \tilde{f}_{B}(\mu) \, m_{B}}{n\cdot p}\,
\int^{\infty}_{0} \, \frac{d\omega}{\bar{n}\cdot p-\omega} \,
\phi_B^{-}(\omega, \mu)
+ {\cal O} \left(\alpha_s, \, \kappa_p^2 \right)  \,,
\nonumber \\
\hat{F}_{3, \,  {\rm NLP} }^{\rm hc, \, kin}
&=&  {\cal O} \left(\alpha_s, \, \kappa_p^2 \right) \,,
\nonumber \\
\hat{F}_{4, \,  {\rm NLP} }^{\rm hc, \, kin}
&=&  3 \,  \kappa_{p} \, \frac{Q_{u} \, \tilde{f}_{B}(\mu) \, m_{B}}{n\cdot p}\,
\int^{\infty}_{0} \, \frac{d\omega}{\bar{n}\cdot p-\omega} \,
\phi_B^{-}(\omega, \mu)
+ {\cal O} \left(\alpha_s, \, \kappa_p^2 \right)  \,.
\label{kinematical NLP results}
\end{eqnarray}

Furthermore, we  derive the power suppressed local contribution from the hard-collinear photon radiation
off the bottom quark as displayed in the diagram \ref{fig: tree-level Feynman diagrams}(b)
\begin{eqnarray}
F_{V, \, \rm NLP}^{Q_b, \, \rm loc} &=&
- \frac{\tilde{f}_{B} (\mu) \, m_{B}}{\overline{m}^{2}_{b}}\,
\frac{Q_{b}}{r_{3}-1} \,
+ {\cal O} \left(\alpha_s, \, {\Lambda / m_b} \right) \,,
\nonumber \\
\hat{F}_{A, \, \rm NLP}^{Q_b, \, \rm loc} &=&
 \frac{\tilde{f}_{B}(\mu) \, m_{B}}{\overline{m}^{2}_{b}}\,
\left ( 1 + 2 \, \frac{\overline{y_{B}}}{r_{1}} \right )\,
\frac{Q_{b}}{r_{3}-1}\,
+ {\cal O} \left(\alpha_s, \, {\Lambda / m_b} \right) \,,
\nonumber \\
\hat{F}_{1, \, \rm NLP}^{Q_b, \, \rm loc} &=&  \frac{4 \, \tilde{f}_{B}(\mu) \,  m_{B}}{\overline{m}^{2}_{b}}\,
\left ({r_2 \over r_1^2} + \frac{\overline{y_{B}}}{2 \, r_1} \right ) \, \frac{Q_{b}}{r_{3}-1}
+ {\cal O} \left(\alpha_s, \, {\Lambda / m_b} \right)  \,,
\nonumber \\
\hat{F}_{2, \, \rm NLP}^{Q_b, \, \rm loc} &=&  \frac{4 \, \tilde{f}_{B}(\mu) \,  m_{B}}{\overline{m}^{2}_{b}}\,
\left ({r_2 \over r_1^2} + \frac{\overline{y_{B}}}{2 \, r_1} \right ) \, \frac{Q_{b}}{r_{3}-1}
+ {\cal O} \left(\alpha_s, \, {\Lambda / m_b} \right)   \,,
\nonumber \\
\hat{F}_{3, \, \rm NLP}^{Q_b, \, \rm loc} &=&
\frac{4 \, \tilde{f}_{B}(\mu) \, m_{B}}{\overline{m}^{2}_{b}}\,
\frac{y_{B}}{r_{1}}\,
\frac{Q_{b}}{r_{3}-1}
+  {\cal O} \left(\alpha_s, \, {\Lambda / m_b} \right)  \,,
\nonumber \\
\hat{F}_{4, \, \rm NLP}^{Q_b, \, \rm loc} &=&
- \frac{\tilde{f}_{B}(\mu) \, m_{B}}{\overline{m}^{2}_{b}}\,
\left (1+2\,\frac{\overline{y_{B}}}{r_{1}} \right )\,
\frac{Q_{b}}{r_{3}-1}
+  {\cal O} \left(\alpha_s, \, {\Lambda / m_b} \right) \,,
\label{b-quark propagator NLP results}
\end{eqnarray}
by introducing further three dimensionless quantities 
\begin{eqnarray}
y_B=m_b / \overline{m_b} \,, \qquad \overline{y_B} = 1- y_B \,, 
\qquad r_3 \equiv q^2/\overline{m}_b^2=r_{2}-y_{B}\,r_{1}+y^{2}_{B} \,.
\end{eqnarray}
The obtained expressions for the vector and axial-vector form factors
are  compatible with the counterpart contributions to the radiative leptonic
$B_{u}^{-} \to \gamma \, \ell \, \bar \nu_{\ell}$ decay
by taking the on-shell photon limit $p^2 \to 0$ \cite{Beneke:2011nf}.

Collecting the individual NLP corrections discussed so far allows us
to write down the following  master formula
\begin{eqnarray}
{\cal F}_{i, \, \rm NLP} = {\cal F}_{i, \, \rm NLP}^{\rm hc, \, dyn}
+  {\cal F}_{i, \, \rm NLP}^{\rm 3PHT}
+  {\cal F}_{i, \, \rm NLP}^{\rm 2PHT}
+  {\cal F}_{i, \, \rm NLP}^{\rm hc, \, kin}
+  {\cal F}_{i, \, \rm NLP}^{Q_b, \, \rm loc} \,,
\qquad
({\cal F}_i = F_V,  \, \hat{F}_A, \, \hat{F}_{1,..., 4}) \,,
\label{the combined FFs in the hard-collinear p2 region}
\end{eqnarray}
where the detailed expressions of the separate terms appearing on the right-hand side
can be found in (\ref{dynamical NLP results}),  (\ref{3PHT NLP results}),  (\ref{2PHT NLP results}),
(\ref{kinematical NLP results}) and (\ref{b-quark propagator NLP results}).
Prior to concluding the explorations of factorization properties
for the generalized $B_{u}^{-} \to \gamma^{\ast} \,  W^{\ast}$ form factors
with a hard-collinear photon, we pause for a while to compare our computations
of the subleading power terms in the heavy quark expansion
with the previous theory analysis in the QCD framework
\cite{Bharucha:2021zay,Beneke:2021rjf,Ivanov:2021jsr}.

\begin{itemize}

\item{To facilitate an exploratory comparison with \cite{Beneke:2021rjf},
we first establish the conversion relation of their ${\it ``longitudinal"}$ form factor
$\hat{F}_{A _{\|}}$ and a complete set of the exclusive transition form factors
introduced in (\ref{Original Lorentz decomposition})
\begin{eqnarray}
\hat{F}_{A _{\|}} = - \left [  F_1 + { v \cdot p \over m_B} \, F_3 +  \hat{F}_A \right ]  \,,
\end{eqnarray}
which further implies another advantageous   representation with the aid of
our  equation (\ref{relations between Fi and Fihat}) as well as the two relevant identities
in (2.4) and (2.6) of \cite{Beneke:2021rjf}
\begin{eqnarray}
F_{A _{\|}} = - \left \{  \hat{F}_1 + { v \cdot p \over m_B} \, \hat{F}_3 +
\left [ 1 +  \frac{p^2} {(v \cdot p)^2 - p^2} \,
 \left ( 1 -  \frac{v \cdot p} {m_B}  \right )  \right ]  \,
\frac{Q_{\ell} \, f_B}{v \cdot p} \right \}  \,.
\label{FA-parr-Beneke}
\end{eqnarray}
Plugging the obtained NLP factorization formulae of $\hat{F}_1$ and  $\hat{F}_3$
at tree level into (\ref{FA-parr-Beneke}) and performing the (hard)-collinear
expansion up to the   ${\cal O}(\Lambda_{\rm QCD} / m_b)$ accuracy leads to
\begin{eqnarray}
F_{A _{\|}} &=&  - \frac{4 \, Q_u \,  \tilde{f}_{B}(\mu) \, m_{B}}{ (n \cdot p)^2 } \,
{\bar n \cdot p \over \lambda_B^{-}(\bar n \cdot p, \mu)}
+ { 2 \, \tilde{f}_{B}(\mu) \over n \cdot p} \, (Q_b -Q_u - Q_{\ell})
+  {\cal O} \left(\alpha_s, \, {\Lambda_{\rm QCD} \over m_b} \right)  \,,
\hspace{1.0 cm}
\label{our result of FA-parr}
\end{eqnarray}
where the second term  vanishes evidently due to the electric-charge conservation
and the inverse moment of the twist-three distribution amplitude
$\phi_B^{-}(\omega, \mu)$ is defined by \cite{Beneke:2001at}
\begin{eqnarray}
\frac{1} {\lambda_B^{-}(\bar n \cdot p, \mu)}
= \int_0^{\infty} \, d \omega \,
{\phi_B^{-}(\omega, \mu) \over \omega - \bar n \cdot p - i \, 0} \,.
\label{generalized lambdam moment}
\end{eqnarray}
Interestingly, the power suppressed three-particle contribution
from the ``genuine" twist-three distribution amplitude
$\phi_{3} (\omega_1, \omega_2, \mu)$ disappears at LO in the strong coupling constant,
owing to the complete cancellation of the distinct dynamical mechanisms
entering (\ref{dynamical NLP results}) and  (\ref{3PHT NLP results}).
It is straightforward to verify that  the yielding expression (\ref{our result of FA-parr})
for the form factor $F_{A _{\|}}$ reproduces the obtained result of \cite{Beneke:2021rjf}
adopting  the Wandzura-Wilczek approximation.
Inspecting the soft-collinear factorization formulae for the ``dynamical" power corrections
(\ref{dynamical NLP results}) reveals that the first term in the curly brackets of
(\ref{expansion of the hc propagator}) will generate the non-vanishing contributions
to the form factors $\hat{F}_1$ and $\hat{F}_3$ starting at next-to-next-to-leading-power (NNLP)
in an expansion in powers of $\Lambda_{\rm QCD} / m_b$, thus supporting the proposed ansatz for the non-perturbative
form factor $\xi^{\prime}(p^2, \bar n \cdot p)$ \cite{Beneke:2021rjf} analytically.
}

\item{We leave out the subleading power contributions to the  exclusive
$B_{u}^{-} \to \gamma^{\ast} \, \ell \, \bar \nu_{\ell}$ decay form factors
with a hard-collinear photon from the light-meson resonances (for instance $\rho$ and $\omega$)
discussed in \cite{Bharucha:2021zay,Beneke:2021rjf},
on account of (i) their insignificant numerical impacts in the kinematical regions
satisfying the constraints $p^2 \geq 1.5 \, {\rm GeV^2}$ and $n \cdot p \geq 3.0 \, {\rm GeV}$
as already observed in \cite{Beneke:2021rjf},
(ii) particularly a lack of the rigorous and systematic formalism to address the resonance
contributions. In addition, the OPE-controlled dispersion technique for
evaluating the soft NLP contributions to the $\gamma^{\ast} \gamma \to \pi^{0}$ form factor
\cite{Khodjamirian:1997tk,Wang:2017ijn,Gao:2021iqq} and the on-shell $B \to \gamma$  form factors
cannot be straightforwardly applied to the analogous computation for the generalized
$B_{u}^{-}(p_B) \to \gamma^{\ast}(p)  \,  W^{\ast}(q)$ form factors
in the time-like  regime of $p^2 \sim {\cal O}(\Lambda_{\rm QCD} \, m_b)$,
due to the yielding divergent dispersion integrals in the vicinity of $p^2=s_0$
with $s_0$ representing the threshold parameter in the $\rho$-meson channel
\begin{eqnarray}
{1 \over \pi} \, \int_{s_0}^{\infty} \, ds  \,\,
{{\rm Im}_{s} \, \mathcal{F}_i(s, n \cdot p) \over s - p^2 - i \, 0} \,,
\qquad
({\cal F}_i = F_V,  \, \hat{F}_A, \, \hat{F}_{1,..., 4})  \,,
\end{eqnarray}
which is precisely the argument to motivate an implementation of the phenomenological ansatz
for the soft form factor in the so-called B-type contribution
to $B_{d, \, s} \to \gamma \ell \bar \ell$ \cite{Beneke:2020fot}. }

\item{Applying the principle of gauge invariance of the QED interaction,
model-independent constraints on the radiative $B_{u}^{-} \to \gamma^{\ast} \, W^{\ast}$
form factors and the resulting phenomenological implications on
the differential distributions for $B_{u}^{-} \to  \ell^{\prime} \, \bar \ell^{\prime} \, \ell \, \bar \nu_{\ell}$
have been recently explored in \cite{Ivanov:2021jsr}, reaching the major observation
of the vanishing form factor $F_{2 A}(p^2, n \cdot p)$ in the on-shell photon limit
based upon their form-factor parametrization scheme.
Switching to our form-factor convention instead implies the following relation
\begin{eqnarray}
\lim_{p^2 \to 0} \, \left [  F_{A _{\|}}
+  \frac{p^2} {(v \cdot p)^2 - p^2} \,
\left ( 1 -  \frac{v \cdot p} {m_B}  \right )   \, F_A  \right ] = 0 \,,
\end{eqnarray}
which can be readily validated by employing the established result (\ref{our result of FA-parr})
for the transition form factor $F_{A _{\|}}$.
Actually, the vanishing longitudinal form factor in the on-shell photon limit
can be expected naturally from the very fact that there exists no longitudinal polarization
for an on-shell photon as already mentioned in \cite{Beneke:2021rjf}.
}

\end{itemize}

\section{QCD factorization for $B_{u}^{-} \to \gamma^{\ast} \, \ell \, \bar \nu_{\ell}$ with a hard photon}
\label{section: QCDF with a hard photon}

Now we turn to derive the perturbative factorization formulae for
a complete set of the $B_{u}^{-} \to \gamma^{\ast} \,  W^{\ast}$ form factors
with an off-shell photon state possessing the four-momentum $p_{\mu} \sim {\cal O}(m_b)$
by implementing the ${\rm QCD \rightarrow HQET}$ matching for
the $B$-meson-to-vacuum correlation function (\ref{definition: hadronic tensor}).
Evaluating the tree-level diagrams displayed in Figure \ref{fig: tree-level Feynman diagrams}
at leading power in the heavy quark expansion immediately leads to  the factorized expressions
\begin{eqnarray}
F_V^{\rm LO} &=&  - \frac{\tilde{f}_{B} (\mu) \, m_{B}}{\overline{m}^{2}_{b}}\,
\left [ \frac{Q_{u}}{r_{2}} +\frac{Q_{b}}{r_{3}-1} \right ]
+ {\cal O} \left(\alpha_s, \, {\Lambda_{\rm QCD} / m_b} \right) \,,
\nonumber \\
\hat{F}_A^{\rm LO} &=& - \frac{\tilde{f}_{B}(\mu) \, m_{B}}{\overline{m}^{2}_{b}}\,
\left[ \frac{Q_{u}}{r_{2}}
- \left ( 1 + 2 \,\frac{\overline{y_{B}}}{r_{1}} \right )\,
\frac{Q_{b}}{r_{3}-1}\, \right]
+ {\cal O} \left(\alpha_s, \, {\Lambda_{\rm QCD} / m_b} \right)  \,,
\nonumber \\
\hat{F}_1^{\rm LO} &=&  - \frac{\tilde{f}_{B}(\mu) \,  m_{B}}{\overline{m}^{2}_{b}}\,
\frac{4 \, r_2}{r^{2}_{1}-4\,r_{2}}\,
\left[  \frac{Q_{u}}{r_{2}}  -
\left (1 + \frac{\overline{y_{B}} \,r_{1}}{2 \, r_2} \right ) \, \frac{Q_{b}}{r_{3}-1}\, \right ]
+ {\cal O} \left(\alpha_s, \, {\Lambda_{\rm QCD} / m_b} \right)  \,,
\nonumber \\
\hat{F}_2^{\rm LO} &=&  - \frac{\tilde{f}_{B}(\mu) \,  m_{B}}{\overline{m}^{2}_{b}}\,
\frac{4 \, r_2}{r^{2}_{1}-4\,r_{2}}\,
\left [ \frac{Q_{u}}{r_{2}}
- \left (1 + \frac{\overline{y_{B}}\,r_{1}}{2 \, r_2} \right )\, \frac{Q_{b}}{r_{3}-1}\, \right ]
+  {\cal O} \left(\alpha_s, \, {\Lambda_{\rm QCD} / m_b} \right) \,,
\nonumber \\
\hat{F}_3^{\rm LO} &=& \frac{\tilde{f}_{B}(\mu) \, m_{B}}{\overline{m}^{2}_{b}}\,
\frac{4 \, r_2}{r^{2}_{1}-4\,r_{2}}\,
\left [ \frac{Q_{u}}{r_{2}}
- \left (1 - {y_{B}\,r_{1} \over r_2}
+ \frac{2\,(1+y_{B})}{r_{1}} \right )\,\frac{Q_{b}}{r_{3}-1}  \right ]
+  {\cal O} \left(\alpha_s, \, {\Lambda_{\rm QCD} / m_b} \right)  \,,
\nonumber \\
\hat{F}_4^{\rm LO} &=&   \frac{\tilde{f}_{B}(\mu) \, m_{B}}{\overline{m}^{2}_{b}}\,
\frac{r^{2}_{1}}{r^{2}_{1}-4\,r_{2}}\,
\left [ \frac{Q_{u}}{r_{2}}
- \left (1+2\,\frac{\overline{y_{B}}}{r_{1}} \right )\,
\frac{Q_{b}}{r_{3}-1}  \right ]
+  {\cal O} \left(\alpha_s, \, {\Lambda_{\rm QCD} / m_b} \right) \,.
\end{eqnarray}
It is evident that the large-recoil symmetry relation between the vector and axial-vector form factors
at $p^2 \sim {\cal O}(m_b \, \Lambda_{\rm QCD})$ is no longer valid for large $p^2$ of order $m_B^2$
even at ${\cal O}(\alpha_s^0)$, due to the emerged leading-power contribution from the hard photon radiation
off the heavy bottom-quark.
In addition, it is straightforward to verify that the resulting expressions of the four longitudinal form factors
$\hat{F}_{1,...,4}$ satisfy the obtained Ward-Takahashi identities (\ref{New WT relations}).
In contrast to the SCET factorization formulae (\ref{SCET factorization of form factors: hard-colinear region})
for the $B_{u}^{-} \to \gamma^{\ast} \, \ell \, \bar \nu_{\ell}$ form factors with a hard-collinear photon,
the yielding results of $\hat{F}_1$ and $\hat{F}_3$ are observed to be free of
the $\Lambda_{\rm QCD} /m_b$ suppression  compared with the remaining transition form factors
at $p^2 \sim {\cal O}(m_b^2)$.

\begin{figure}
\begin{center}
\includegraphics[width=1.0  \columnwidth]{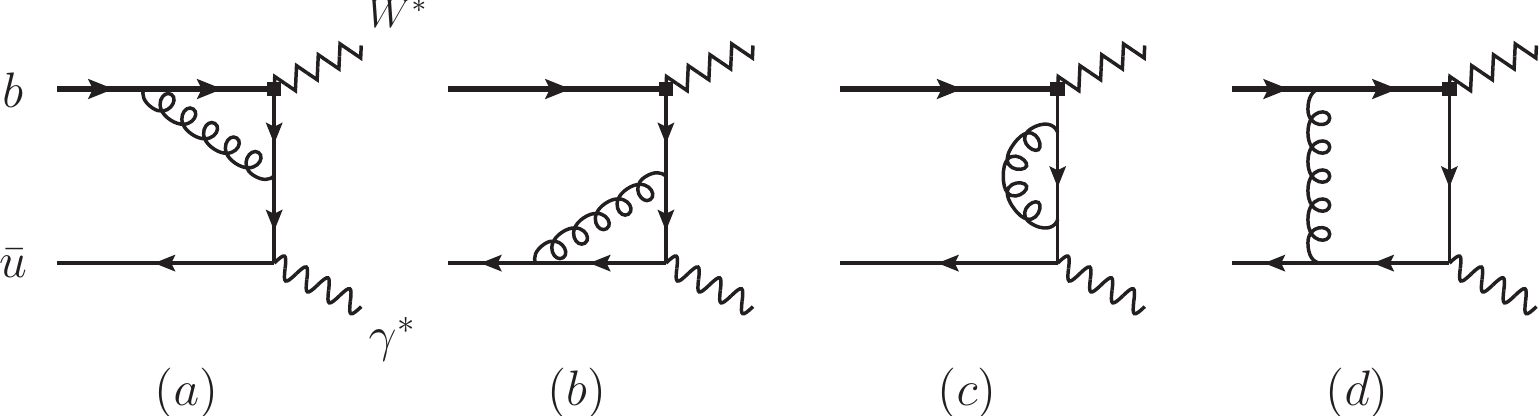}
\vspace*{-0.2 cm}
\caption{Diagrammatical representation of the NLO correction to the QCD correlation function
$T_{\nu \mu}(p_B, q)$ due to the hard photon radiation from the light anti-quark. }
\label{fig: 1-loop-diagrams-u-quark}
\end{center}
\end{figure}

We are now in a position to perform the NLO computation of the non-local matrix element
$T_{\nu \mu} (p_B, p)$ in the hard $p^2$ region
by applying the standard perturbative matching program,
which is somewhat more sophisticated than the ${\rm QCD \rightarrow HQET}$ matching for
the heavy-to-light currents at one loop \cite{Manohar:2000dt,Neubert:1993mb}.
It is apparent that the gluonic corrections to the short-distance  Wilson coefficients
can be conveniently split into two pieces with  the distinct electric charges
in the following
\begin{eqnarray}
{\cal F}_i^{\rm NLO} = {\cal F}_i^{\rm LO} + {\alpha_s(\mu)  \, C_F \over 2 \, \pi} \,
\frac{\tilde{f}_{B} (\mu) \, m_{B}}{\overline{m}^{2}_{b}}\,
\left [ \left ( { Q_u \over r_2} \right ) \, {\cal H}_i^{u}
+ \left ( {Q_b \over r_3 - 1} \right ) \, {\cal H}_i^{b} \right ], \,\,
({\cal F}_i = F_V,  \, \hat{F}_A, \, \hat{F}_{1,..., 4}) \,.
\hspace{0.5 cm}
\label{NLO factorization formula in hard region}
\end{eqnarray}
The renormalized coefficient function ${\cal H}_i^{u}$ can be readily determined
by identifying the hard contributions of the one-loop QCD diagrams
presented in Figure \ref{fig: 1-loop-diagrams-u-quark}
(apart from the wavefunction renormaization of the external bottom quark
at ${\cal O}(\alpha_s)$ \cite{Descotes-Genon:2002crx})
\begin{eqnarray}
{\cal H}_V^{u} &=& \left ( \frac{3}{4}\,\ln\frac{\mu^2}{m^2_b} +2 \right )
+\frac{1}{r_1^2-4r_2} \, \bigg  \{\frac{r_1-4r_2}{2}\,r_2\ln(-r_2)
-\left[ r_1^3+r_2^2-r_1\,r_2\,(4+r_2) \right ] \, {\cal C}_{0, \, u}
\nonumber\\
&& + \frac{r_1\,(r_1 + 3 \, r_2) - 2 \,r_2\,(3 +2 \, r_2)}{2\,r_3} \,
(1-r_3) \, \ln(1-r_3) \bigg \} \,,
\\
{\cal \hat{H}}_A^{u} &=& {\cal H}_V^{u}
+ \frac{r_2}{r_1\,(r_1^2-4r_2)} \,
\bigg \{ 2 \left[r_1^2-r_2\,(2+r_2)\right] \, {\cal C}_{0, \,  u}
+ \left[ r_1 \,(r_3-3) + 4\,r_2 \right] \,  \ln \left (- r_2 \right )
\nonumber\\
&& -\frac{(r_1+2)\,(1-r_3)-4}{r_3}\,(1-r_3)\ln(1-r_3)
- r_1^2 + 4 \, r_2 \bigg \} \,,
\\
{\cal \hat{H}}_1^{u} &=&  \frac{4 \, r_{2}}{r^{2}_{1}-4r_{2}}\,
\bigg\{  \left ({3 \over 4} \,\ln\frac{\mu^{2}}{m^{2}_{b}} + 2 \right )
-  \, \Big[\frac{r_{2}}{r^{2}_{1}-4r_{2}}\,(2+r_{1}\,r_{2}-2r_{3})+r_{1}\Big] \, {\cal C}_{0, \,  u}
\nonumber \\
&& + \frac{r_{3}-1}{r_{3}} \,
\Big[ \frac{2}{r^{2}_{1}-4 \, r_{2}} \, (r_{2}-r^{2}_{2}+r_{1}\,r_{3})
-\frac{(r_{2}-1)(r_{3}-1)}{2 \, r_{3}}\Big]\,\ln(1-r_{3})
\nonumber \\
&& +\frac{r_{1}}{2 \, (r^{2}_{1}-4r_{2})}\,(2+r_{1}\,r_{2}-2r_{3}) \, \ln(-r_{2})
+ \frac{2 - r_{1} - r_{1}\,r_{3}}{2\, r_{3}}  \bigg\}  \,,
\\
{\cal \hat{H}}_2^{u} &=&   \left ( - {r_1^2 \over 4 \, r_2} \right ) \, {\cal \hat{H}}_3^{u}
-  \left ({3 \over 4} \,\ln\frac{\mu^{2}}{m^{2}_{b}} + 1 \right ) \,,
\label{NLO H2hat in hard region} \\
{\cal \hat{H}}_3^{u} &=&  - \frac{4 \, r_{2}}{r^{2}_{1}-4r_{2}}\,
\bigg\{  \left ( {3 \over 4}\, \ln\frac{\mu^{2}}{m^{2}_{b}} + 2 \right )
+ \frac{2 \, r_{2}}{r_{1}}\, \left [\frac{r_{2}}{r^{2}_{1}-4r_{2}}\,(-4+r_{1}-2r_{2})-2 \right ]
\, {\cal C}_{0, \,  u}
\nonumber \\
&& + \frac{1- r_{3}}{r_{1}\,r_{3}}\,  \left [\frac{2\,r_{2}}{r^{2}_{1}-4r_{2}} \,
(-4+2\,r_{1}+r_{1}\,r_{3}) +\frac{(r_{2}-1)(r_{3}-1)}{r_{3}} \right ]\,\ln(1-r_{3})
\nonumber \\
&& + \frac{r_{2}}{r_{1}} \,
\left [ \frac{2}{r^{2}_{1}-4r_{2}}\,(2+r_{1}\,r_{2}-2r_{3})-1 \right ] \, \ln(-r_{2})
- \frac{2 \, r_2 \, (1+r_{3}) - r_1 }{r_{1}\,r_{3}} \bigg\}  \,,
\\
{\cal \hat{H}}_4^{u} &=&   \left ( - {r_1^2 \over 4 \, r_2} \right ) \, {\cal \hat{H}}_1^{u} \,,
\end{eqnarray}
where we have defined the perturbative loop function
\begin{eqnarray}
{\cal C}_{0, \,  u}
&=& \frac{1}{\sqrt{\lambda}}  \,
\bigg\{
\text{Li}_2\left(\frac{(r_3+1) \sqrt{\lambda} + r_u} {(r_3-1)\sqrt{\lambda} + r_u}\right)
+\text{Li}_2\left(\frac{(1+r_3) \sqrt{\lambda} + r_u}{(1-r_3) \sqrt{\lambda} + r_u}\right)
-\text{Li}_2\left(\frac{(1-r_3) \sqrt{\lambda} + r_u}{(r_3-1) \sqrt{\lambda} + r_u}\right)
\nonumber \\
&& + 2 \,\text{Li}_2\left(\frac{r_1 - \sqrt{\lambda}}{r_1 + \sqrt{\lambda}}\right)
+2\,\text{Li}_2\left(1-\frac{r_2}{\sqrt{\lambda}}\right)
+\frac{1}{2} \, \ln^2 \left (- \frac{r_1 + \sqrt{\lambda}}{r_1 - \sqrt{\lambda}} \right )
+ \frac{\pi^2}{3} \bigg\} \,,
\end{eqnarray}
with
\begin{eqnarray}
&& \lambda  \equiv   \lambda(1, r_2, r_3) = 1 + r_2^2 + r_3^2 - 2 \, r_2  - 2 \, r_3 - 2 \, r_2 \, r_3
= r_1^2 - 4 \, r_2  \,,
\nonumber \\
&& r_u \equiv r_2 \,(1+r_3) - (1-r_3)^2 \,.
\end{eqnarray}
The appearance of the second term in the yielding expression
(\ref{NLO H2hat in hard region}) for ${\cal H}_2^{u}$ follows from
the exact Ward-Takahashi relations (\ref{New WT relations})
and the matching equation (\ref{matching relation of fB})
for the QCD and HQET $B$-meson decay constants.

\begin{figure}
\begin{center}
\includegraphics[width=1.0  \columnwidth]{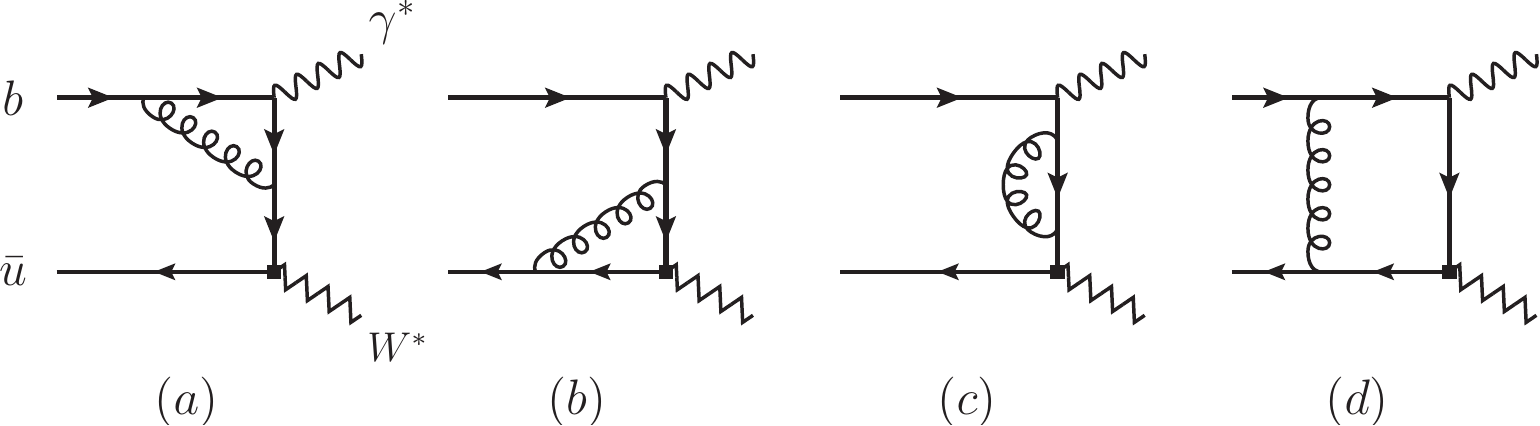}
\vspace*{-0.2 cm}
\caption{Diagrammatical representation of the NLO correction to the QCD correlation function
$T_{\nu \mu}(p_B, q)$ due to the hard photon radiation from the heavy bottom-quark. }
\label{fig: 1-loop-diagrams-b-quark}
\end{center}
\end{figure}

Along the same vein, we can derive the renormalized hard kernel ${\cal H}_i^{b}$
by extracting the perturbative contributions from the one-loop QCD diagrams displayed in
Figure \ref{fig: 1-loop-diagrams-b-quark}
\begin{eqnarray}
{\cal H}_V^{b} &=&  \left (\frac{4}{1-r_3} + 1 \right ) \,
\left ( {3 \over 4}\ln\frac{\mu^2}{m^2_b} + 1 \right )
+\left [ \frac{r_{2}}{r^{2}_{1}-4\,r_{2}}\,
\left (4+8\,r_{3}+r_2-r_{1} \, r_{2} \right ) + 2\,r_{2}+1 \right ] \,
{\cal C}_{0, \,  b}
\nonumber\\
&& +\frac{\ln(1-r_{3})}{2 \, r_{3}}\, \left \{ \frac{r_{2}}{r^{2}_{1}-4\,r_{2}}\,
\left [-2\,(7 + 2 \, r^{2}_{2} + 5 \,r_{3}) + r_{1} \,(8 + r_{2} + 10 \, r_{3}) \right ]
+ 3 \, r_{1}-2 \right \}\,
\nonumber\\
&& - \frac{r_b \, \left ( 8 + 4 \, r_3 - r_1 \right )} {2 \, (r^{2}_{1} -  4 \, r_{2})} \, \,
\ln\frac{2-r_2+r_b}{2} + 1  \,,
\\
{\cal \hat{H}}_A^{b} &=&  -{\cal H}_V^{b} + \frac{4}{r_{1}} \,
\left ( {3 \over 4} \, \ln\frac{\mu^{2}}{m^{2}_{b}} + 1 \right )
+ \left [\frac{r_{2}}{r^{2}_{1}-4 \, r_{2}}\,(8 +8 \, r_{3} - r_{1} \, r_{2})
+\frac{r_{2}}{r_1} \,(8+r_{2}) - 4 \right ]\, {{\cal C}_{0, \,  b} \over 2}
\nonumber\\
&& + \frac{\ln(1-r_{3})}{2\,r_{3}}\,\left \{\frac{r_{2}}{r^{2}_{1}-4r_{2}}\,
[3 \, r_{2} \, (r_1 - 2)  - 2 \, r_{3} \, (8 + r_2) ]
-\frac{r_{2}\,(r_{2}+8)}{r_{1}}  + 2 \, (2-r_2) \right \} \,
\nonumber \\
&&  - \frac{r_b \, (3+ r_3)}{r^{2}_{1}-4 \, r_{2}}\, \ln\frac{2-r_2+r_b}{2}
+ \frac{r_1-r_{2}}{r_{1}} \,,
\\
{\cal \hat{H}}_1^{b} &=&  \frac{4}{1-r_3}\,
\left[ \frac{r_{2}\,(r_{2}- 2 \, r_{1})}{r^{2}_{1} - 4 \, r_{2}}+1 \right] \,
\left ( {3 \over 4}  \, \ln\frac{\mu^2}{m^2_b} + 1 \right )
+ \frac{4 \, r_{2} \, (r_{1}-2)}{r_1^2 - 4 \, r_2}\,
\left [ \frac{r_{2}\, (r_{1}-r_{3}-3)}{r_1^2 - 4 \, r_2} -2 \right ] \,{\cal C}_{0, \,  b}
\nonumber\\
&& -\frac{2 \, r_{2}}{r_{3}} \,
\left \{\frac{4\,r_{2}}{(r_1^2 - 4 \, r_2)^{2}}\,
\left [ (2-r_1)^2 + r_3 \, (r_{3} - 2\, r_1)  \right ]
+\frac{5 - 4 \, r_1 + r_3 \, (2+r_3)}{r_1^2 - 4 \, r_2} \right \} \, \ln(1-r_{3})
\nonumber \\
&& + \frac{2 \, r_b}{r_1^2-4 \, r_2}\,
\left [ \frac{2 \, r_{2}\,(r_{1} - 2 \, r_{3}-2)}{r_1^2 - 4 \, r_2}-r_{3}-5 \right ] \,
\ln \frac{2-r_2+r_b}{2}
+ \frac{2 \, r_2}{1-r_{3}}\,
\frac{r_{1} \, (1+r_3) - 2 \, r_2}{r_1^2 - 4 \, r_2}  \,,
\nonumber \\
\\
{\cal \hat{H}}_2^{b} &=&   \left ( - {r_1^2 \over 4 \, r_2} \right ) \, {\cal \hat{H}}_3^{b}
+ \frac{r_{3}-1}{r_{2}} \, \left ({3 \over 4} \,\ln\frac{\mu^{2}}{m^{2}_{b}} + 1 \right ) \,,
\\
{\cal \hat{H}}_3^{b} &=&
\left ( -{4 \, r_{2} \over r_1^2} \right ) \,  {\cal \hat{H}}_1^{b}
+ {4 \,(r_{3}-5) \over r_1^2} \, \left ({3 \over 4} \,\ln\frac{\mu^{2}}{m^{2}_{b}} + 1 \right )
- {8 \, r_{2} \over r_1^2} \, \frac{(3- 2\, r_2 + r_3) \, (1-r_3)}{r_1^2-4r_2}  \,
{\cal C}_{0, \,  b}
\nonumber\\
&& + { 4\,r_{2} \over r_1^2} \, \left [ \frac{1-r_3}{r_3} \,
\frac{2 \, (1 + 3 \, r_3) - r_1 \, (1-r_3)}{r_1^2-4r_2} \right ] \, \ln(1-r_{3})
\nonumber \\
&& -\frac{4\,r_b}{r_1^2}\, {8 \, r_{2}+r_{1}\,(r_{3}-5) \over r_1^2-4r_2} \,
\ln \frac{2-r_2+r_b}{2}  - {4 \, r_2 \over r_1^2} \,,
\\
{\cal \hat{H}}_4^{b} &=&   \left ( - {r_1^2 \over 4 \, r_2} \right ) \, {\cal \hat{H}}_1^{b} \,,
\end{eqnarray}
where for brevity we have introduced the convention
\begin{eqnarray}
{\cal C}_{0, \,  b} &=&   \frac{1}{\sqrt{\lambda}} \,
\bigg \{
\text{Li}_2 \left(
\frac{\xi_1 - r_2 \, \sqrt{\lambda}}{\xi_1 - r_b \, \sqrt{\lambda}} \right )
+ \text{Li}_2 \left(
\frac{\xi_1 - r_2 \, \sqrt{\lambda}}{\xi_1 + r_b \, \sqrt{\lambda}} \right )
- \text{Li}_2 \left(
\frac{\xi_1 + r_2 \, \sqrt{\lambda}}{\xi_1 - r_b \, \sqrt{\lambda}} \right )
\nonumber \\
&& - \, \text{Li}_2 \left(
\frac{\xi_1 + r_2 \, \sqrt{\lambda}}{\xi_1 + r_b \, \sqrt{\lambda}} \right )
- \, \text{Li}_2 \left ( \frac{\xi_2}{(r_3-1)(\xi_3 - \sqrt{\lambda})} \right )
- \, \text{Li}_2 \left ( \frac{\xi_2}{(r_3-1)(\xi_3 + \sqrt{\lambda})} \right )
\nonumber \\
&& + \, \text{Li}_2 \left(\frac{\xi_3 - \sqrt{\lambda}}{\xi_3 + \sqrt{\lambda}} \right)
+ 2 \, \text{Li}_2 \left (1+\frac{\sqrt{\lambda}}{r_3-1} \right )
- \frac{\pi^2}{6}  \bigg \} \,,
\end{eqnarray}
with
\begin{eqnarray}
\xi_1 &=& r_2 \, (r_2-r_3-3) \,,
\qquad
\xi_2 = (r_3-1) \,(2-r_1) + (1+ r_3) \, \sqrt{\lambda} \,,
\nonumber \\
\xi_3 &=& 2 - r_1 \,,
\qquad
\hspace{1.6 cm}
r_b = \sqrt{r_2 \, (r_2-4) - i \, 0} \,.
\end{eqnarray}
Employing the RG evolution equations of the effective decay constant $\tilde{f}_B(\mu)$
as well as the bottom-quark mass $m_b$
\begin{eqnarray}
{d \tilde{f}_B(\mu) \over d \ln \mu} &=&
\gamma_{\rm hl}(\alpha_s) \, \tilde{f}_B(\mu)
= \left [ \sum_{k=0}^{\infty} \, \gamma_{k, \, \rm hl}  \,
\left ( {\alpha_s(\mu) \over 4 \, \pi} \right )^{k+1} \right ] \,
\tilde{f}_B(\mu) \,,
\nonumber \\
{d \, \overline{m}_b(\mu) \over d \ln \mu} &=&
\gamma_{m}(\alpha_s) \, \overline{m}_b(\mu)
= \left [ \sum_{k=0}^{\infty} \, \gamma_{k, \, m}  \,
\left ( {\alpha_s(\mu) \over 4 \, \pi} \right )^{k+1} \right ] \,
\overline{m}_b(\mu)  \,,
\end{eqnarray}
with the first two series coefficients given by \cite{Ji:1991pr,Broadhurst:1991fz}
\begin{eqnarray}
\gamma_{0, \, \rm hl} = 3 \, C_F \,, \qquad
\gamma_{1, \, \rm hl} = C_F \, \left [ \frac{127}{6} + \frac{14 \, \pi^2}{9} - \frac{5}{3} \, n_f \right ]   \,,
\end{eqnarray}
and \cite{Chetyrkin:1997dh,Vermaseren:1997fq}
\begin{eqnarray}
\gamma_{0, \, m} = 6 \, C_F \,, \qquad
\gamma_{1, \, m} = C_F \left [ 3 \, C_F + {97 \over 3}  \, C_A
- {20 \over 3} \, n_f \right ] \,,
\end{eqnarray}
it is then straightforward to demonstrate the factorization-scale independence of
the resulting expressions for  the exclusive $B_{u}^{-} \to \gamma^{\ast} \, \ell \, \bar \nu_{\ell}$ form factors
\begin{eqnarray}
{d  {\cal F}_i^{\rm NLO} \over d \ln \mu}
= {\cal O}(\alpha_s^2) \,,
\qquad
({\cal F}_i = F_V,  \, \hat{F}_A, \, \hat{F}_{1,..., 4})   \,.
\end{eqnarray}
In analogy to the SCET factorization for the radiative leptonic $B$-meson form factors
with a hard-collinear photon, the HQET decay constant $\tilde{f}_B(\mu)$ will be converted to
the QCD decay constant $f_B$ by means of the matching relation (\ref{matching relation of fB}).
Subsequently, the enhanced logarithms of order $\ln \left (m_b / \Lambda_{\rm QCD} \right )$
entering the hard function $K^{-1}(\mu)$ will be  summed at the NLL accuracy
according to the second identity in (\ref{Lee-Neubert solution}).
Importantly, the resulting NLO expressions (\ref{NLO factorization formula in hard region}) for
the four-body leptonic $B$-meson form factors  are observed to  comply with
the Ward-Takahashi constraints (\ref{New WT relations}) at the accuracy of ${\cal O}(\alpha_s)$,
thus providing a valuable check of our computation.

\section{Numerical results}
\label{section: numerical results}

We are now ready to explore the phenomenological implications of
the obtained factorization formulae for the exclusive
$B_{u}^{-} \to \gamma^{\ast} \,  W^{\ast}$ form factors
by applying a variety of  effective field theory approaches,
with an emphasis on the systematic computations of
the angular observables for the four-body decay process
$B_{u}^{-} \to \gamma^{\ast} ( \to  \ell^{\prime} \, \bar \ell^{\prime} ) \,  W^{\ast} (\to  \ell \, \bar \nu_{\ell})$
of experimental  importance.
To achieve this goal, we will proceed by specifying the different types of theory inputs
(the electroweak parameters, the bottom-quark mass, both the leading-twist and higher-twist $B$-meson
distribution amplitudes in HQET, and so on) entering  the factorized expressions of
the $B$-meson transition form factors emerged in the general decomposition of
the corresponding decay amplitude (\ref{general decay amplitude}).

\subsection{Theory inputs}
\label{subsection: Theory inputs}

In analogy to QCD factorization for the radiative
$B_{u}^{-} \to \gamma \, \ell \, \bar \nu_{\ell}$ decay \cite{Descotes-Genon:2002crx},
the twist-two $B$-meson distribution amplitude in HQET $\phi_B^{+}(\omega, \mu)$ apparently serves as the fundamental
non-perturbative ingredient appearing in the SCET factorization formulae of the exclusive
$B_{u}^{-} \to \gamma^{\ast} \, \ell \, \bar \nu_{\ell}$ form factors with a hard-collinear (off-shell) photon.
Following \cite{Beneke:2018wjp}, we will adopt the improved three-parameter ansatz for $\phi_B^{+}(\omega, \mu_0)$
with an attractive analytical behaviour under the RG evolution at the one-loop accuracy
\begin{eqnarray}
\phi_B^{+}(\omega, \mu_0) = {\Gamma(\beta) \over \Gamma(\alpha)}  \,
U \left (\beta-\alpha, 3-\alpha, {\omega \over \omega_0} \right ) \,
 {\omega \over \omega_0^2} \,\, {\rm exp} \left ( - {\omega \over \omega_0} \right ) \,,
\end{eqnarray}
where $U(a,b, z)$ stands for the confluent hypergeometric function of the second kind
possessing  an integral representation  for ${\rm Re} \, [a] >0$ and ${\rm Re} \, [z] >0$
\begin{eqnarray}
U(a,b, z) = {1 \over \Gamma(a)} \, \int_0^{\infty} d t \, e^{- z \, t} \,
t^{a-1} \, (t+1)^{b-a-1} \,.
\label{general-phiBplus-model}
\end{eqnarray}
In the numerical evaluation, we will adjust the  shape parameters
$\omega_0$, $\alpha$ and $\beta$  to cover the allowed ranges
for the inverse logarithmic moments  of the leading-twist
HQET distribution amplitude  displayed in Table \ref{table for the input parameters}
by applying the following identities \cite{Shen:2020hfq}
\begin{eqnarray}
\lambda_{B_u}(\mu_0) &=& \left ( {\alpha-1 \over \beta-1} \right ) \, \omega_0 \,,
\qquad
\widehat{\sigma}_{B_u}^{(1)} (\mu_0) =  \psi (\beta-1) - \psi (\alpha-1)
+ \ln  \left ( {\alpha-1 \over \beta-1} \right ) \,,
\nonumber \\
\widehat{\sigma}_{B_u}^{(2)} (\mu_0) &=&
\left [  \widehat{\sigma}_{B_u}^{(1)} (\mu_0)  \right ]^2 +
\psi^{(1)}(\alpha-1) - \psi^{(1)}(\beta-1)  + {\pi^2 \over 6} \,,
\end{eqnarray}
where the explicit definitions of $\lambda_{B_u}$,  $\widehat{\sigma}_{B_u}^{(1)}$
and $\widehat{\sigma}_{B_u}^{(2)}$ read \cite{Beneke:2018wjp}
\begin{eqnarray}
\lambda_{B_u}^{-1}(\mu) &=&   \int_0^{\infty} \, d \omega \,\,
{\phi_B^{+}(\omega, \mu)  \over \omega} \,,
\nonumber \\
\widehat{\sigma}_{B_u}^{(n)} (\mu) &=& \lambda_{B_u}(\mu) \,
\int_{0}^{\infty} \, {d \omega \over \omega} \,
\left [ \ln \left ( {\lambda_{B_u}(\mu) \over \omega} \right )
- \gamma_E \right ]^{n} \, \phi_{B}^{+}(\omega, \mu)  \,.
\label{definition of the inverse logarithmic moments}
\end{eqnarray}
Substituting (\ref{general-phiBplus-model}) into the constructed solution (\ref{Lee-Neubert solution})
to the Lange-Neubert evolution equation \cite{Lange:2003ff} results in
(see  \cite{Beneke:2018wjp} for the RG evolution function in coordinate space)
\begin{eqnarray}
&& \phi_B^{+}(\omega, \mu)
\nonumber \\
&& = {\rm exp} \bigg \{  - {\Gamma_{\rm cusp}^{(0)} \over 4 \, \beta_0^2}  \,
\bigg [ {4 \pi \over \alpha_s(\mu_0)} \, \left ( \ln {\alpha_s(\mu) \over \alpha_s(\mu_0)} - 1
+ {\alpha_s(\mu_0) \over \alpha_s(\mu)} \right )
- {\beta_1 \over 2 \, \beta_0} \, \ln^2 {\alpha_s(\mu) \over \alpha_s(\mu_0)}
+ \left ( {\Gamma_{\rm cusp}^{(1)} \over \Gamma_{\rm cusp}^{(0)}}
- {\beta_1 \over \beta_0}  \right )
\nonumber \\
&& \hspace{1.5 cm} \times \left ({\alpha_s(\mu) \over \alpha_s(\mu_0)} - 1
- \ln {\alpha_s(\mu) \over \alpha_s(\mu_0)} \right ) \bigg ] \bigg \} \,
\, \left ( {\alpha_s(\mu) \over \alpha_s(\mu_0)} \right )^{ \gamma_{\eta}^{(0)} / (2 \, \beta_0)} \,
\left ( {1 \over \omega_0} \right ) \,
\left ( {\mu_0 \, e^{2 \, \gamma_E} \over\omega_0 } \right )^{\kappa_s}
\nonumber  \\
&& \hspace{0.5 cm} \times \, \bigg \{  {\omega \over \omega_0} \,\,
\frac{\Gamma(\beta) \, \Gamma(2+\kappa_s) \, \Gamma(\alpha-\kappa_s-2)}
{\Gamma(\alpha) \,  \Gamma(\beta-\kappa_s-2)} \,\, {}_2F_2
\left ( \kappa_s+2, \kappa_s+3-\beta; 2, \kappa_s+3-\alpha, - {\omega \over \omega_0} \right )
\nonumber \\
&& \hspace{0.8 cm} + \left ( {\omega \over \omega_0} \right )^{\alpha-\kappa_s-1}
\,\, \frac{\Gamma(\beta) \, \Gamma(2+\kappa_s-\alpha)}
{\Gamma(\beta - \alpha) \,  \Gamma(\alpha-\kappa_s)}
\,\, {}_2F_2
\left ( \alpha, \alpha-\beta+1;  \alpha - \kappa_s -1, \alpha - \kappa_s, - {\omega \over \omega_0} \right )  \bigg \} \,,
\hspace{1.0 cm}
\end{eqnarray}
where the expansion coefficient $\kappa_s$ is explicitly defined by
\begin{eqnarray}
\kappa_s = {\Gamma_{\rm cusp}^{(0)}  \over 2 \, \beta_0}  \,
\ln {\alpha_s(\mu) \over \alpha_s(\mu_0)} \,.
\end{eqnarray}
In comparison with QCD factorization for the exclusive non-hadronic $B$-meson decays
$B_{u}^{-} \to \gamma \, \ell \, \bar \nu_{\ell}$ \cite{Beneke:2011nf,
Wang:2016qii,Wang:2018wfj,Beneke:2018wjp}
and $\bar{B}_{d, \, s} \to \gamma \gamma$ \cite{Descotes-Genon:2002lal,Bosch:2002bv,Shen:2020hfq}
in the heavy quark limit,
the LP contributions to the generalized $B_{u}^{-} \to \gamma^{\ast} \,  W^{\ast}$ form factors
presented in (\ref{SCET factorization of form factors: hard-colinear region}) are more sensitive to
the precise shape of the twist-two distribute  amplitude $\phi_B^{+}(\omega, \mu_0)$
rather than determined by the inverse moment $\lambda_B(\mu_0)$ completely at the one-loop accuracy.
Consequently, the four-body leptonic $B$-meson decays under discussion
are expected to  provide us with abundant opportunities
for probing the partonic landscape of the heavy-quark hadron system delicately.

\begin{table}
\centering
\renewcommand{\arraystretch}{2.0}
\resizebox{\columnwidth}{!}{
\begin{tabular}{|l|ll||l|ll|}
\hline
\hline
  Parameter
& Value
& Ref.
&  Parameter
& Value
& Ref.
\\
\hline
\hline
  $G_F$                                         & $1.166379 \times 10^{-5} \,\, {\rm GeV}^{-2} $
                                                                                                  & \cite{ParticleDataGroup:2020ssz} 
& $|V_{ub}|$                                    & $(3.70 \pm 0.10 \pm 0.12) \times 10^{-3}$       & \cite{ParticleDataGroup:2020ssz} 
\\
  $\alpha_s^{(5)}(m_Z)$                         & $0.1188 \pm 0.0017$                             & \cite{ParticleDataGroup:2020ssz} 
& $\alpha_{\rm em}^{(5)}(m_Z)^{-1}$             & $127.952 \pm 0.009$                             & \cite{ParticleDataGroup:2020ssz} 
\\
   $m_{e}$                                         &  $0.511$ MeV                                    &  \cite{ParticleDataGroup:2020ssz}
&  $m_{\mu}$                                       &  $105.658$ MeV                                  &  \cite{ParticleDataGroup:2020ssz}
\\
\hline
\hline
  $\overline{m}_b (\overline{m}_b)$                         & $4.198 \pm 0.012$  GeV                         &  \cite{ParticleDataGroup:2020ssz} 
& $m_b^{\rm PS} (2 \, {\rm GeV})$                & $4.532^{+0.013}_{-0.039}$  GeV                 & \cite{Beneke:2014pta} 
\\
  $m_{B_u}$                                     &  $5279.34 \pm 0.12$ MeV                         &  \cite{ParticleDataGroup:2020ssz}
& $\tau_{B_u}$                                  &  $(1.638 \pm 0.004)$ ps                         &  \cite{ParticleDataGroup:2020ssz}
\\
  $f_{B_u}|_{N_f = 2+1+1}$                      &  $190.0 \pm 1.3$ MeV                            &  \cite{Aoki:2019cca}
&                                  &                        &
\\
\hline
\hline
  $\lambda_{B_u}(\mu_0)$                        & $(350 \pm 150)$ MeV                             & \cite{Beneke:2020fot}  
 &                                              & $\{0.7, \, 6.0\}$                                &
\\
  $\lambda_E^2(\mu_0)/\lambda_H^2(\mu_0)$       & $0.50 \pm 0.10$                                & \cite{Beneke:2018wjp}
& $ \{\widehat{\sigma}_{B_u}^{(1)}(\mu_0), \,
\widehat{\sigma}_{B_u}^{(2)}(\mu_0)\}$          & $\{0.0, \, \pi^2/6\}$                            & \cite{Beneke:2020fot}
\\
  $2 \, \lambda_E^2(\mu_0) + \lambda_H^2(\mu_0)$       & $(0.25 \pm 0.15) \, {\rm GeV^2}$         & \cite{Beneke:2018wjp}
&                                                &  $\{-0.7, \, -6.0\}$                             &
\\
\hline
\hline
\end{tabular}
}
\renewcommand{\arraystretch}{1.0}
\caption{The numerical values of  the various input  parameters employed
in the  theory predictions for the four-body leptonic $B$-meson decays. }
\label{table for the input parameters}
\end{table}

Moreover, the two-particle and three-particle higher-twist $B$-meson distribution amplitudes in HQET
are evidently indispensable for evaluating the LP contributions to the two radiative form factors
$\hat{F}_{2(4), \, \rm LP}$ collected in (\ref{LP SCET factorization formula for F2hat})
and (\ref{SCET factorization of form factors: hard-colinear region})
as well as the subleading power corrections  calculable with the perturbative factorization technique
displayed in (\ref{dynamical NLP results}),  (\ref{3PHT NLP results}),  (\ref{2PHT NLP results})
and (\ref{kinematical NLP results}).
Following \cite{Beneke:2018wjp,Shen:2020hfq} we will adopt the concrete phenomenological models
fulfilling the classical equations of motion and the corresponding asymptotic behaviour
at small quark and gluon momenta from the conformal spin analysis \cite{Braun:1989iv}
(see \cite{Braun:2017liq,Wang:2018wfj} for the two sample choices of the higher-twist distribution
amplitudes at the twist-six accuracy)
\begin{eqnarray}
&& \phi_B^{-}(\omega, \mu_0) =  \phi_B^{-, \, \rm WW}(\omega, \mu_0)
+ \phi_B^{-, \, \rm tw3}(\omega, \mu_0)
\nonumber \\
&& = \left [  \int_{\omega}^{\infty} d \rho \, f(\rho) \right ]
+  {1 \over 6} \, \varkappa(\mu_0)  \,
\left [ \lambda_E^2 (\mu_0) - \lambda_H^2 (\mu_0) \right ] \,
\left [ \omega^2  \, f^{\prime}(\omega) + 4 \, \omega \, f(\omega)
- 2 \, \int_{\omega}^{\infty} d \rho \, f(\rho)  \right ],
\nonumber \\
&& \phi_3(\omega_1, \omega_2, \mu_0) =
- {1 \over 2} \, \varkappa(\mu_0) \, \left [ \lambda_E^2 (\mu_0) - \lambda_H^2 (\mu_0) \right ] \,
\omega_1 \, \omega_2^2 \, f^{\prime}(\omega_1 + \omega_2) \,,
\nonumber \\
&& \phi_4(\omega_1, \omega_2, \mu_0) =
{1 \over 2} \, \varkappa(\mu_0) \, \left [ \lambda_E^2 (\mu_0) + \lambda_H^2 (\mu_0) \right ] \,
\omega_2^2 \, f(\omega_1 + \omega_2) \,,
\nonumber \\
&& \psi_4(\omega_1, \omega_2, \mu_0) =
\varkappa(\mu_0) \,  \lambda_E^2 (\mu_0) \, \omega_1 \,  \omega_2 \, f(\omega_1 + \omega_2) \,,
\nonumber \\
&& \tilde{\psi}_4(\omega_1, \omega_2, \mu_0) =
\varkappa(\mu_0) \,  \lambda_H^2(\mu_0) \, \omega_1 \,  \omega_2 \, f(\omega_1 + \omega_2) \,,
\nonumber \\
&& \psi_5(\omega_1,\omega_2, \mu_0) =
\varkappa(\mu_0) \,  \lambda_E^2(\mu_0) \, \omega_2  \, \int_{\omega_1 + \omega_2}^\infty d \eta \, f(\eta),
\nonumber \\
&& \tilde{\psi}_5(\omega_1,\omega_2, \mu_0) =
\varkappa(\mu_0) \,  \lambda_H^2(\mu_0) \, \omega_2  \, \int_{\omega_1 + \omega_2}^\infty d \eta \, f(\eta),
\nonumber \\
&& \phi_6(\omega_1, \omega_2, \mu_0) =
\varkappa(\mu_0) \, \left [ \lambda_E^2 (\mu_0) - \lambda_H^2 (\mu_0) \right ] \,
\int_{\omega_1 + \omega_2}^\infty d \eta_1 \,  \int_{\eta_1}^{\infty} \, d \eta_2 \, f(\eta_2),
\end{eqnarray}
which further enable us to determine two-particle twist-four and twist-five distribution amplitudes
$g_B^{\pm}(\omega, \mu_0)$ by virtue of the obtained identities (\ref{results of gBpm}) and (\ref{results of gBpmhat}).
The non-perturbative profile function $f(\omega)$ and the normalization constant $\varkappa(\mu_0)$
are given by
\begin{eqnarray}
f(\omega) &=& {\Gamma(\beta) \over \Gamma(\alpha)}  \,
U \left (\beta-\alpha, 3-\alpha, {\omega \over \omega_0} \right ) \,
{1 \over \omega_0^2} \,\, {\rm exp} \left ( - {\omega \over \omega_0} \right ) \,,
\nonumber \\
\varkappa^{-1}(\mu_0) &=& {1 \over 2} \, \int_{0}^{\infty} d \omega \, \omega^3 \, f(\omega)
= \bar \Lambda^2 +  {1 \over 6} \, \left [ 2 \, \lambda_E^2 (\mu_0) + \lambda_H^2 (\mu_0) \right ] \,.
\end{eqnarray}

The appearing  HQET parameters $\lambda_E^2$ and $\lambda_H^2$ can be defined in terms of
the local effective matrix element of the dimension-five quark-gluon operator
\cite{Braun:2017liq,Neubert:1993mb}
\begin{eqnarray}
&& \langle 0 | \bar q (0) \, g_s \, G_{\mu \nu} \, \Gamma \, h_v(0)| \bar B_q(v) \rangle
\nonumber \\
&& = - {\tilde{f}_{B_q} \, m_{B_q} \over 6} \,
{\rm Tr} \left \{ \gamma_5 \, \Gamma \,\,
\left (  {1 + \slashed v \over 2} \right ) \,
\left [ \lambda_H^2  \, \left (  i \, \sigma_{\mu \nu} \right )
+ (\lambda_H^2 - \lambda_E^2) \,
\left ( v_{\mu} \, \gamma_{\nu} - v_{\nu} \, \gamma_{\mu} \right ) \right ]  \right \}   \,.
\end{eqnarray}
The RG evolution equations for  $\lambda_E^2$ and $\lambda_H^2$ at the one-loop accuracy read
\cite{Grozin:1996hk,Nishikawa:2011qk}
\begin{eqnarray}
{d \over d \ln \mu} \,
\left(
\begin{array}{c}
\lambda_E^2(\mu) \\
\lambda_H^2(\mu) \\
\end{array}
\right)
+  \gamma_{\rm EH}  \,
\left(
\begin{array}{c}
\lambda_E^2(\mu) \\
\lambda_H^2(\mu) \\
\end{array}
\right) = 0
\,,
\label{RGE of lambdaE and lamdbaH}
\end{eqnarray}
where the anomalous dimension matrix $\gamma_{\rm EH}$ takes the form
\begin{eqnarray}
\gamma_{\rm EH} =  {\alpha_s(\mu) \over 4 \, \pi}
\left(
                 \begin{array}{cc}
                   {8 \over 3} \, C_F + {3 \over 2}  \, N_c \,\,  &  \,\,  {4 \over 3} \, C_F - {3 \over 2}  \, N_c \\
                   {4 \over 3} \, C_F - {3 \over 2}  \, N_c \,\,  & \,\,  {8 \over 3} \, C_F + {5 \over 2}  \, N_c \\
                 \end{array}
               \right)
               + {\cal O}(\alpha_s^2)
\,.
\end{eqnarray}
The manifest solution to (\ref{RGE of lambdaE and lamdbaH}) can be readily constructed
by diagonalizing the achieved $2 \times 2$ mixing matrix $\gamma_{\rm EH}$
\cite{Grozin:1996hk,Gao:2019lta}.
The available predictions of these two HQET quantities from the method
of two-point QCD sum rules can be summarized as follows
\begin{eqnarray}
\left \{  \lambda_E^2(\mu_0),  \,\, \lambda_H^2(\mu_0) \right \}
= \left\{
\begin{array}{l}
\left \{  (0.11 \pm 0.06) \, {\rm GeV^2}, \,\, (0.18 \pm 0.07) \, {\rm GeV^2}  \right \}  \,,
\qquad \hspace{1.5 cm} \text{\cite{Grozin:1996pq}}
\vspace{0.5 cm} \\
\left \{  (0.03 \pm 0.02) \, {\rm GeV^2}, \,\, (0.06 \pm 0.03) \, {\rm GeV^2}  \right \}    \,,
 \qquad  \hspace{1.5 cm}
\text{\cite{Nishikawa:2011qk}}
\vspace{0.5 cm} \\
\left \{  (0.01 \pm 0.01) \, {\rm GeV^2}, \,\, (0.15 \pm 0.05) \, {\rm GeV^2}  \right \}    \,.
 \qquad  \hspace{1.5 cm}
\text{\cite{Rahimi:2020zzo}}
\end{array}
\hspace{0.8 cm} \right.
\label{QCDSR for lambdaE and lambdaH}
\end{eqnarray}
Apparently, the yielding results for the chromo-electric and chromo-magnetic matrix elements
deviate from each other significantly despite the implementations of the same calculational framework.
The dominating discrepancies for the numerical values displayed in \cite{Grozin:1996pq} and \cite{Nishikawa:2011qk}
can be attributed to the sizeable QCD radiative correction to the dimension-five quark-gluon condensate
and the yet higher-power contribution from the dimension-six vacuum condensate at tree level,
which have been included in the updated computation \cite{Nishikawa:2011qk} with a further improvement
on the (partial)-NLL resummation of the emerged large logarithms appearing in the HQET sum rules.
Instead of constructing the desired sum rules from the appropriate correlation functions with one two-body
and one three-body local current, the authors of \cite{Rahimi:2020zzo} suggested to employ the diagonal
Green functions for the sake of obtaining the alternative sum rules for  $\lambda_E^2$ and $\lambda_H^2$,
with the expectation that the parton-hadron duality approximation becomes more reliable
due to the positive definite property obviously.
Introducing the  higher dimensional correlation functions, however, turns out to worsen the OPE convergence
in contrast to the previously adopted non-diagonal ones, as already observed in \cite{Rahimi:2020zzo}.
In view of an absence of  the satisfactory evaluation for these two non-perturbative parameters,
we will follow closely the strategy of \cite{Beneke:2018wjp,Shen:2020hfq} such that
the numerical intervals of  the combinations $\lambda_E^2 / \lambda_H^2$  and  $2 \, \lambda_E^2 + \lambda_H^2$
collected in Table \ref{table for the input parameters} accommodate the obtained results from
the two-point sum rules \cite{Grozin:1996pq,Nishikawa:2011qk}
and particularly  lie within  the upper bounds for $\lambda_E^2$ and $\lambda_H^2$
derived from the diagonal correlation functions \cite{Rahimi:2020zzo}.
In addition,  the HQET parameter $\bar \Lambda$ entering the soft-collinear factorization formulae
(\ref{dynamical NLP results}) and (\ref{2PHT NLP results})
can be identified as the ``effective mass" of the heavy-meson state \cite{Neubert:1993mb,Manohar:2000dt}:
$\bar \Lambda = m_{B_u} - m_{b} + {\cal O}(\Lambda_{\rm QCD}^2/m_{b})$,
where we will take $m_b=(4.8 \pm 0.1) \, {\rm GeV}$ numerically
following the standard arguments presented in \cite{Beneke:2018wjp,Shen:2020hfq}.

Now we turn to discuss the practical implementations the interesting SM parameters
appeared in the exclusive $B_{u}^{-} \to \gamma^{\ast} \, \ell \, \bar \nu_{\ell}$
decay form factors. The strong coupling constant $\alpha_s(\mu)$ in the $\overline{\rm MS}$ scheme
will be computed from the initial condition  $\alpha_s^{(5)}(m_Z)$
summarized in Table \ref{table for the input parameters} with the
associated three-loop RG evolution equation, by further adopting the quark-flavour threshold scales
$\mu_{4} = 4.8 \, {\rm GeV}$ and $\mu_{3} = 1.2 \, {\rm GeV}$
for crossing $n_f=4$ and $n_f=3$, respectively.
Furthermore, the bottom-quark mass entering the perturbative hard matching functions
(\ref{one-loop hard functions in SCET}) is generally interpreted as the pole mass due
to the on-shell kinematics \cite{Beneke:2001at,Beneke:2004rc,Beneke:2020fot,Shen:2020hfq}.
However, the bottom-quark pole mass suffers from an intrinsic ambiguity of order $\Lambda_{\rm QCD}$
known as the infrared renormalon (see \cite{Beneke:1998ui} for an excellent review).
We will therefore employ the potential-subtracted (PS) renormalization scheme
for the bottom-quark mass \cite{Beneke:1998rk} and then convert the obtained expressions of the hard functions
from the pole scheme to the PS scheme for the mass parameter accordingly
(see \cite{Beneke:2021lkq} for an overview of the leading renormalon-free and short-distance mass definitions
for nearly on-shell heavy quarks).
Another important hadronic quantity governing the factorized result for the
four-body rare $B$-meson decay amplitude in the entire kinematic region
is the leptonic decay constant of the charged bottom-meson $f_{B_u}$,
whose interval collected in Table \ref{table for the input parameters}
is borrowed from the lattice-QCD computation with the number of dynamical
quark flavours  $N_f=2+1+1$ in the isospin symmetry limit \cite{Aoki:2019cca}
(see \cite{Bazavov:2017lyh} for the further discussion on the strong-isospin  violating effect
and \cite{Carrasco:2015xwa,Frezzotti:2020bfa,Beneke:2019slt} on the technical strategies to
address the more complicated electromagnetic correction).

Apart from the theory input parameters so far discussed,
we still need to specify the hard scales $\mu_{h1}$ and $\mu_{h2}$
entering the resummation improved  hard functions $C_V^{({\rm A0}), \, 1}$
and $K^{-1}$ presented in  (\ref{Lee-Neubert solution}),
which will be varied in the interval $\mu_{h1}=\mu_{h2} \in [m_b/2, \,\, 2 \, m_b]$
around the default value $m_b$.
Additionally, the factorization scale $\mu$ in the SCET expressions
(\ref{LP SCET factorization formula for FV}),
(\ref{LP SCET factorization formula for F1hat}),
(\ref{LP SCET factorization formula for F2hat}),
(\ref{LP SCET factorization formula for F3hat}),
(\ref{SCET factorization of form factors: hard-colinear region})
will be taken as $\mu \in [1.0, 2.0] \, {\rm GeV}$
with the central value $1.5 \, {\rm GeV}$.
By contrast, the allowed interval of the factorization scale $\mu$ appearing in the HQET expression
(\ref{NLO factorization formula in hard region}) will naturally read
$[m_b/2, \,\, 2 \, m_b]$ in that the typical short-distance fluctuation mode carries out
the four-momentum $p_{\mu} \sim {\cal O}(m_b)$.

\subsection{Theory predictions for the $B_{u}^{-} \to \gamma^{\ast} \, \ell \, \bar \nu_{\ell}$ form factors}

\begin{figure}
\begin{center}
\includegraphics[width=1.0  \columnwidth]{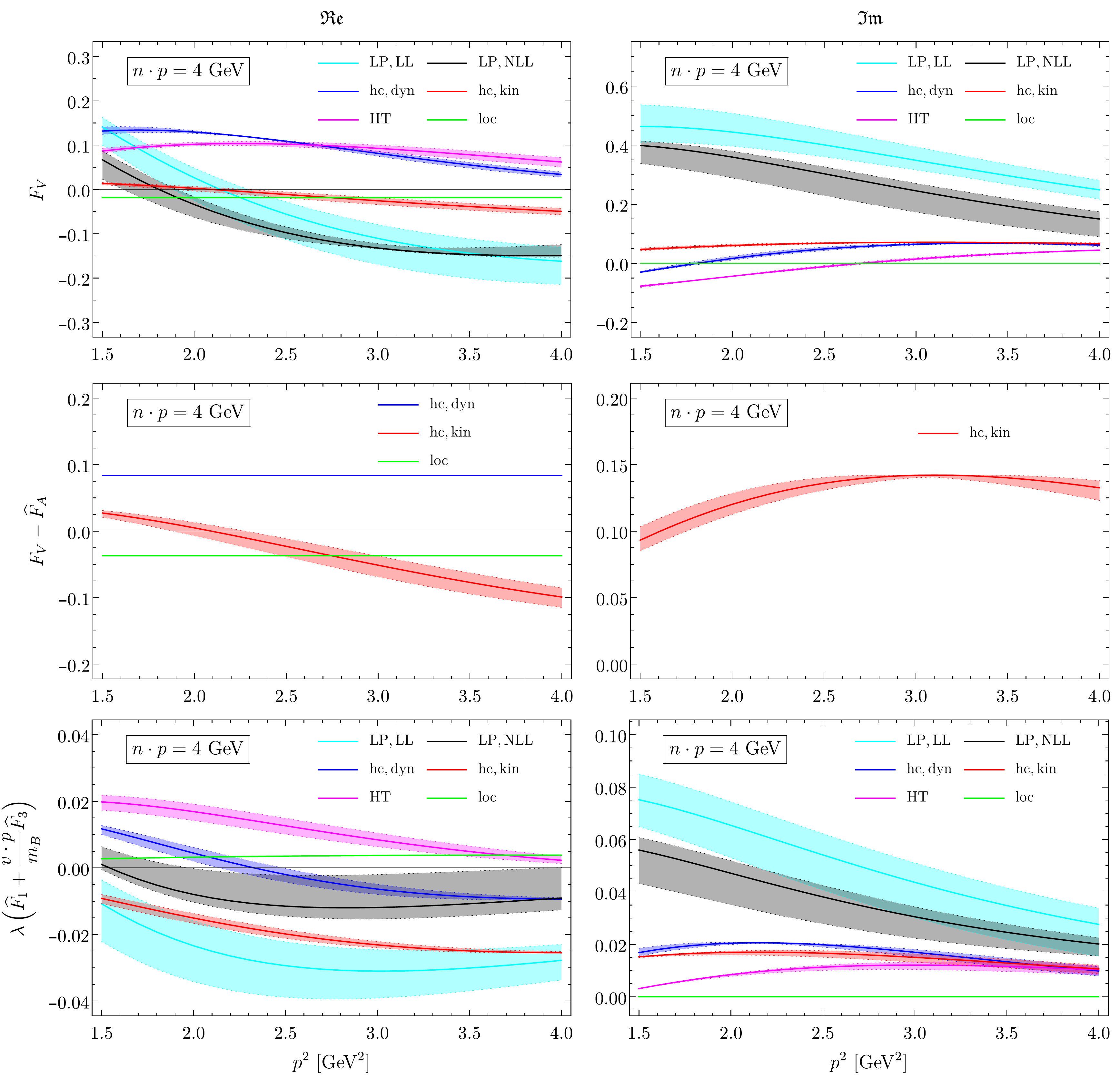}
\vspace*{-0.2 cm}
\caption{Breakdown of the distinct QCD mechanisms contributing to the transverse and longitudinal transition form factors
involved in the four-body leptonic decay amplitude (\ref{general decay amplitude})
in the kinematical region $p^2 \in [1.5, \, 4.0]  \, {\rm GeV^2}$,
with the theory uncertainties due to the variations of the hard and hard-collinear scales
indicated by the individual bands. The representative value of the large component $n \cdot p$
for the virtual photon momentum  is taken as $4.0 \, {\rm GeV}$.}
\label{fig: breakdown-FF-p2-hc}
\end{center}
\end{figure}

\begin{figure}
\begin{center}
\includegraphics[width=1.0  \columnwidth]{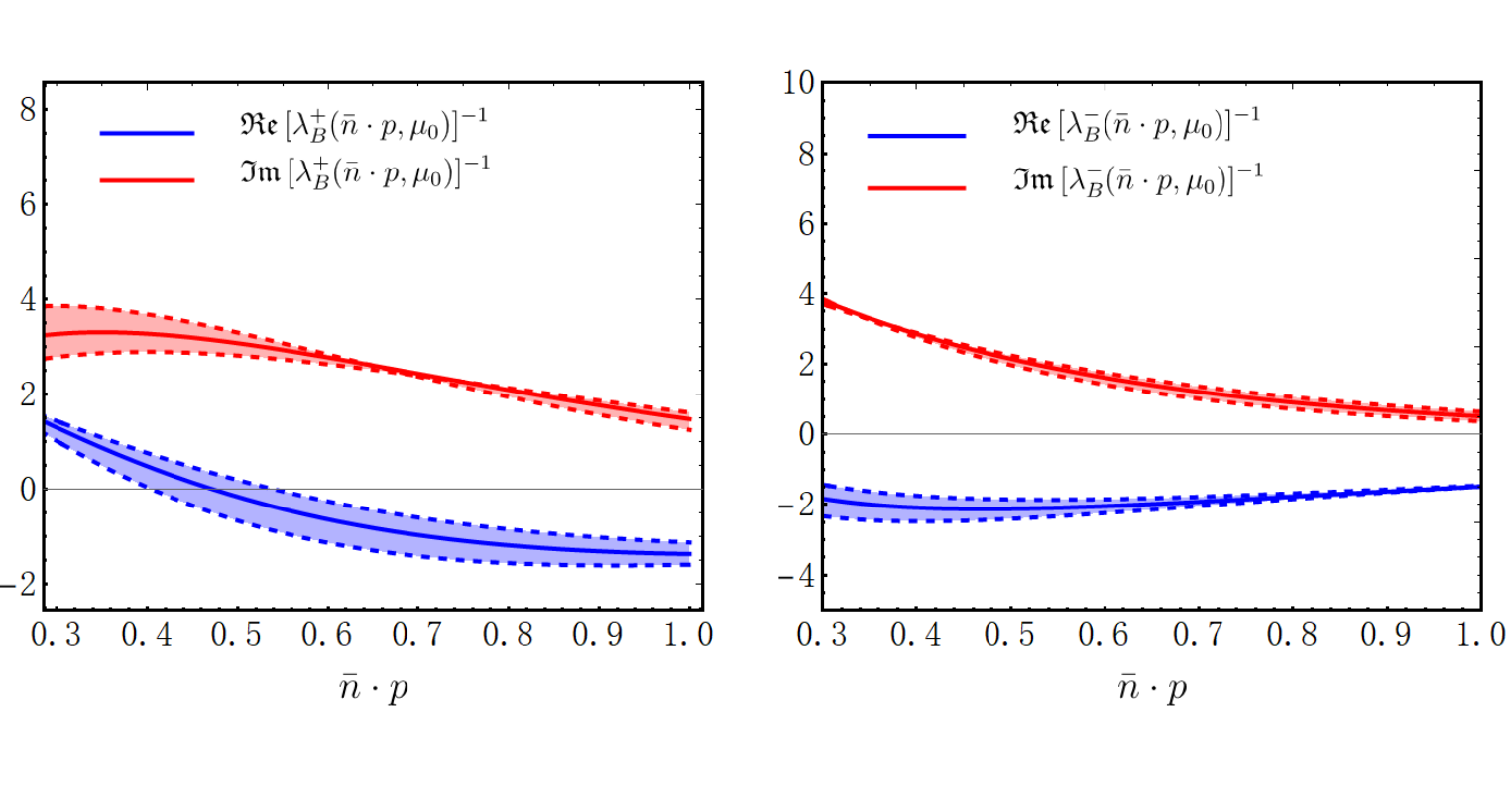}
\vspace*{-0.2 cm}
\caption{The photon-momentum dependence of the generalized inverse moments
$\lambda_B^{\pm}(\bar n \cdot p, \mu)$ for the two-particle $B$-meson distribution amplitudes
by employing the exponential model proposed in \cite{Grozin:1996pq} for illustration purposes.
The uncertainty bands arise from the variation of the non-perturbative shape parameter
$\omega_0 \in [300, 400] \, {\rm MeV}$.}
\label{fig: inverse-moment-general}
\end{center}
\end{figure}

We proceed to investigate the numerical impacts of the newly derived QCD radiative corrections
and the subleading power contributions to the three transition form factors
appearing in the four-body leptonic decay amplitude (\ref{general decay amplitude})
in the entire kinematic region.
To facilitate detailed explorations of the dynamical patterns dictating the exclusive
$B_{u}^{-} \to \gamma^{\ast} \, \ell \, \bar \nu_{\ell}$ form factors with a hard-collinear photon,
we first display the obtained leading-power contributions at the leading-logarithmic (LL) and NLL accuracy,
the power suppressed corrections from expanding the hard-collinear quark propagator
in the small parameter $\Lambda_{\rm QCD} / m_b$   beyond the LP approximation,
the higher-twist contributions from the two-particle and three-particle $B$-meson distribution amplitudes,
the  ``kinematic" power corrections presented in (\ref{kinematical NLP results}),
and the local NLP contributions due to the off-shell photon radiation from the heavy bottom quark
at $1.5 \,  {\rm GeV^2} \leq p^2 \leq 4.0 \, {\rm GeV^2}$  in Figure \ref{fig: breakdown-FF-p2-hc},
where the perturbative uncertainties from varying the hard scales
$\mu_{h1}$ and $\mu_{h2}$ as well as the factorization scale $\mu$ are further indicated by the yielding bands.
The resulting  uncertainties from the NLL resummation improved predictions
are evidently much less than the counterpart LL computations.
In particular, the achieved LL and NLL uncertainty bands for the imaginary part
of the vector form factor ${\rm Im} \, F_V$
turn out to be well separated in the majority of the hard-collinear $p^2$ regime.
The peculiar behaviours of the LP contributions to the two complex-valued form factors
$F_V$ and $\hat{F}_1 + {v \cdot p \over m_B} \, \hat{F}_3$ can be traced back to
the soft-collinear convolution integrals in the SCET factorization formulae
(\ref{LP SCET factorization formula for FV}),
(\ref{LP SCET factorization formula for F1hat}),
(\ref{LP SCET factorization formula for F2hat}),
(\ref{LP SCET factorization formula for F3hat}),
(\ref{SCET factorization of form factors: hard-colinear region})
which are effectively controlled by the generalized inverse moments  $\lambda_B^{\pm}(\bar n \cdot p, \mu)$
of the two-particle $B$-meson distribution amplitudes in HQET
\footnote{The explicit definition of the inverse moment $\lambda_B^{+}(\bar n \cdot p, \mu)$
is in analogy to (\ref{generalized lambdam moment}) with an obvious replacement
of the twist-three distribution amplitude: $\phi_B^{-}(\omega, \mu) \to \phi_B^{+}(\omega, \mu)$.}.
Applying the exponential model of $\phi_B^{\pm}(\omega, \mu_0)$ \cite{Grozin:1996pq} as an illustrative example,
the distinctive features of the two inverse moments  displayed in Figure \ref{fig: inverse-moment-general}
are indeed observed to  dictate the intricate  photon-momentum dependence of
the transverse and longitudinal decay form factors, respectively.
Furthermore, it is plainly not unexpected to discover from Figure \ref{fig: breakdown-FF-p2-hc}
the increasing significance of the ``kinematic power corrections"
to the two essential form factors $|F_V|$ and
$\left |\lambda \, \left ( \hat{F}_1 + {v \cdot p \over m_B} \, \hat{F}_3 \right ) \right |$,
when the off-shellness the hard-collinear photon moves towards higher values.
The very prominent subleading power contributions from the hard-collinear quark propagator
at tree level will consistently shift the LL predictions of $|F_V|$ and
$\left |\lambda \, \left ( \hat{F}_1 + {v \cdot p \over m_B} \, \hat{F}_3 \right ) \right |$
by an amount of approximately
$(20-30) \,  \%$ for $1.5 \,  {\rm GeV}^2 \leq p^2 \leq  \, 4.0 {\rm GeV}^2$.
In addition,  they constitute  an important source of generating the large-recoil symmetry violation
$|F_V- \hat{F}_A|$ between the vector and axial-vector form factors,
which is constructed to characterize the power suppressed terms
in the heavy quark expansion conveniently \cite{Beneke:2011nf}.
Our numerical studies of the subleading twist contributions to the radiative leptonic
$B$-meson form factors  summarized in (\ref{3PHT NLP results})  and (\ref{2PHT NLP results})
imply that the magnitudes of the yielding three-particle higher-twist corrections
are at least suppressed by  a factor of  twenty  when compared with the corresponding two-particle NLP effects
in virtue of the smallness of the two normalization constants $\lambda_E^2$ and $\lambda_H^2$.
Unsurprisingly, the local subleading power contributions  (\ref{b-quark propagator NLP results})
from the HQET formalism will  bring about insignificant impacts
on the exclusive non-hadronic $B_{u}^{-} \to \gamma^{\ast} \, W^{\ast}$ decay form factors:
$(4-6) \%$ for $|F_V|$ and  $(4-10) \%$ for
$\left |\lambda \, \left ( \hat{F}_1 + {v \cdot p \over m_B} \, \hat{F}_3 \right ) \right |$ numerically,
which are in accordance with the previous observation in the context of
$B \to \gamma \, \ell \, \bar \nu_{\ell}$ \cite{Wang:2018wfj}.

\begin{figure}
\begin{center}
\includegraphics[width=1.0  \columnwidth]{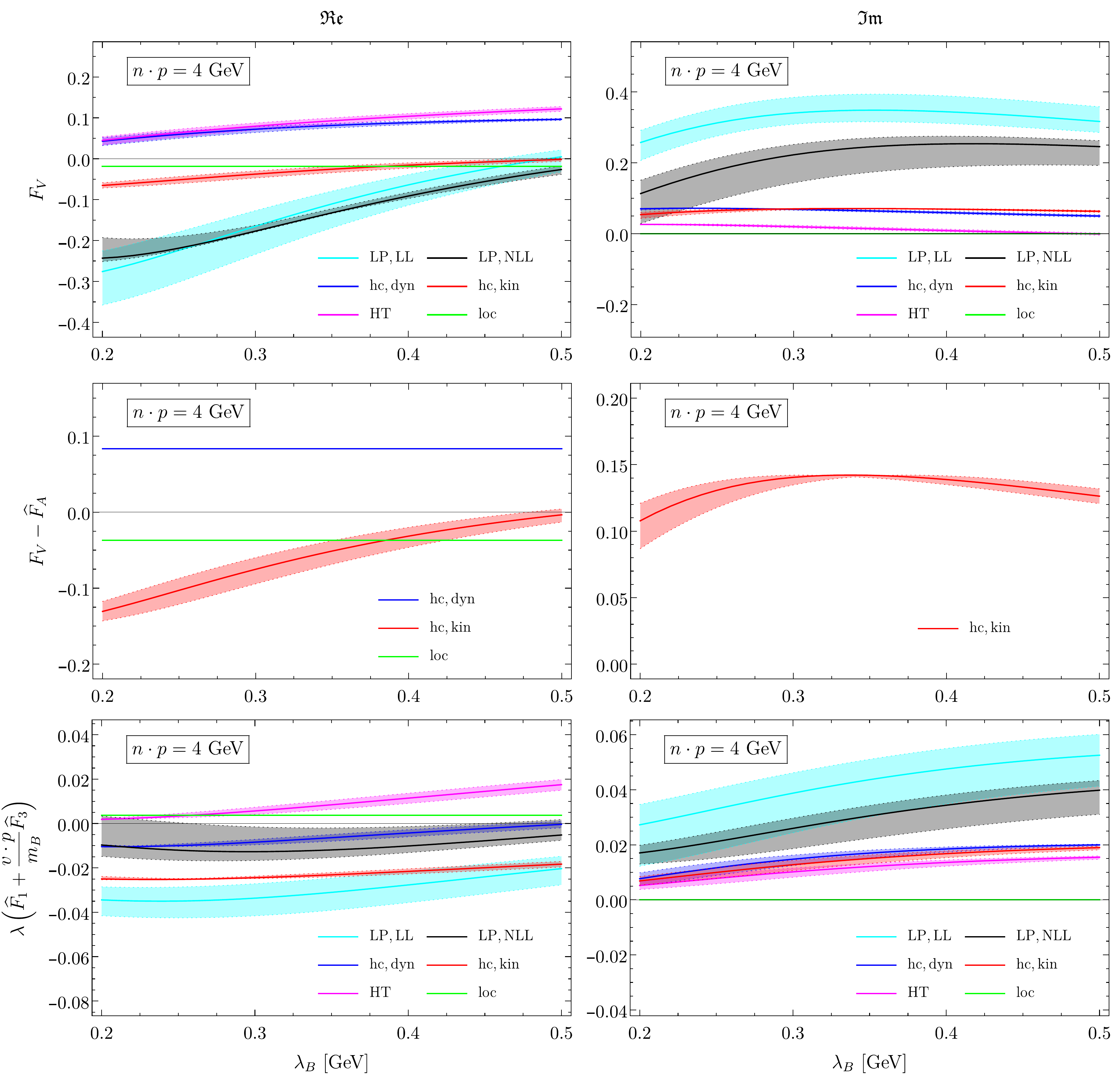}
\vspace*{-0.2 cm}
\caption{Theory predictions for the inverse-moment dependencies of all separate pieces
contributing to the radiative $B_{u}^{-} \to \gamma^{\ast} \, W^{\ast}$ form factors
in the decay amplitude (\ref{general decay amplitude})
in the interval $\lambda_{B_u} \in [200, 500] \, {\rm MeV}$,
with the uncertainty bands from varying the hard and hard-collinear scales.
The representative values of the large  and  small components for the virtual photon momentum  are taken as
$4.0 \, {\rm GeV}$ and $0.75 \, {\rm GeV}$.}
\label{fig: breakdown-FF-lambdaB-hc}
\end{center}
\end{figure}

We further present in Figure \ref{fig: breakdown-FF-lambdaB-hc}
the obtained theory predictions for the individual pieces
contributing to the three form factors of our interest, as the analytical functions of
the inverse moment $\lambda_{B_u}$, by adopting the input kinematic parameters
$n \cdot p= 4.0 \, {\rm GeV}$ and $\bar n \cdot p =0.75 \, {\rm GeV}$.
Interestingly,  the LP contribution of the vector form factor $|F_V|$ turns out to be significantly
more sensitive to $\lambda_{B_u}$ than that for the longitudinal form factor
$\left |\lambda \, \left ( \hat{F}_1 + {v \cdot p \over m_B} \, \hat{F}_3 \right ) \right |$,
stemming  from the different asymptotic behaviours of the very $B$-meson distribution amplitudes
at small quark and gluon momenta, which appear in the LP soft-collinear factorization formulae
(\ref{LP SCET factorization formula for FV}),
(\ref{LP SCET factorization formula for F1hat}),
(\ref{LP SCET factorization formula for F2hat}),
(\ref{LP SCET factorization formula for F3hat}),
(\ref{SCET factorization of form factors: hard-colinear region}).
It is further worth mentioning that the two-particle subleading twist corrections to
both form factors $|F_V|$ and $\left |\lambda \, \left ( \hat{F}_1 + {v \cdot p \over m_B} \, \hat{F}_3 \right ) \right |$
develop the yet stronger $\lambda_{B_u}$-dependence in comparison with the counterpart lower-twist contributions
from $\phi_B^{+}(\omega, \mu)$ and $\phi_B^{-}(\omega, \mu)$.
Additionally, the inverse-moment dependence of the helicity form factor $|F_V- \hat{F}_A|$
originates from the  ``kinematic" power corrections (\ref{kinematical NLP results}) entirely,
keeping in mind that the NLP dynamical contributions displayed in (\ref{dynamical NLP results})
merely generate the ``local" large-recoil symmetry breaking effects
independent of the higher-twist $B$-meson distribution amplitudes.

\begin{figure}
\begin{center}
\includegraphics[width=1.0  \columnwidth]{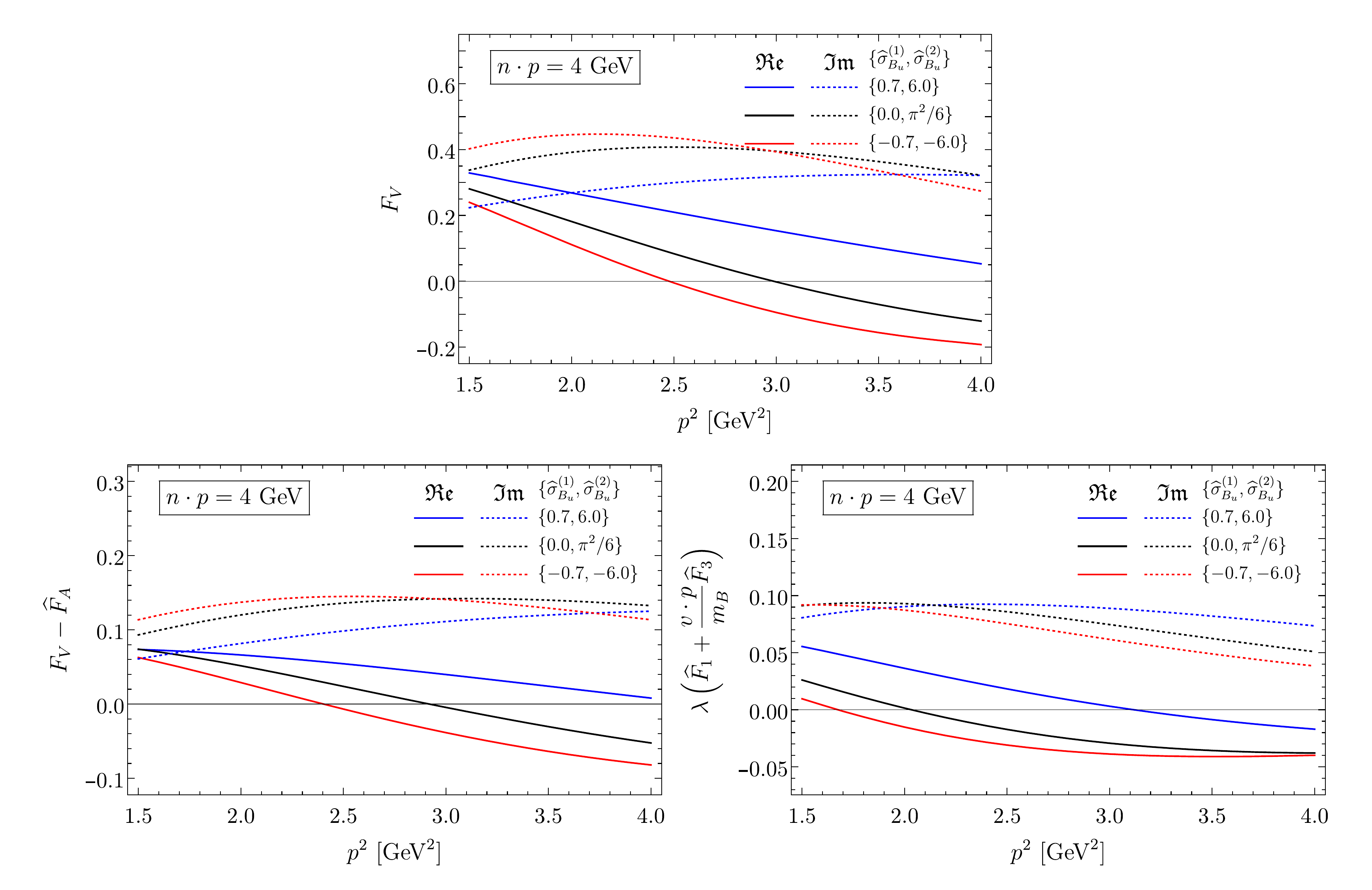}
\vspace*{-0.2 cm}
\caption{The predicted photon-momentum dependencies  of the three transition form factors governing the four-body leptonic
$B$-meson decay amplitude (\ref{general decay amplitude}) with three sample choices of the dimensionless shape parameters
$\widehat{\sigma}_{B_u}^{(1)}$ and $\widehat{\sigma}_{B_u}^{(2)}$, but with fixed $\lambda_{B_u}(\mu_0)=350 \, {\rm MeV}$,
characterizing the non-perturbative behaviours of the leading-twist distribution amplitude $\phi_B^{+}(\omega, \mu_0)$. }
\label{fig: shape-dependence-of-form-factors}
\end{center}
\end{figure}

In contrast to QCD factorization for $B \to \gamma \, \ell \, \bar \nu_{\ell}$ with an on-shell photon,
the LP contributions of the non-hadronic radiative $B_{u}^{-} \to \gamma^{\ast} \, \ell \, \bar \nu_{\ell}$ form factors
in the heavy quark expansion cannot be determined by the inverse moment $\lambda_{B_u}(\mu_0)$ completely
even without taking into account the higher-order gluonic corrections.
It is therefore  interesting to explore the actual dependencies  of the exclusive $B$-meson decay form factors
on the intricate shapes of the HQET distribution amplitudes,
along the lines of  \cite{DeFazio:2007hw,Wang:2015vgv,Wang:2015ndk,Beneke:2021rjf}.
To this end, we show in Figure \ref{fig: shape-dependence-of-form-factors}
the achieved predictions of  the exclusive $B_{u}^{-} \to \gamma^{\ast} \, W^{\ast}$ form factors
with three sample choices of the model parameters $\widehat{\sigma}_{B_u}^{(1)}$ and $\widehat{\sigma}_{B_u}^{(2)}$
for fixed $\lambda_{B_u}(\mu_0)=350 \, {\rm MeV}$ in the kinematic domain $ p^2 \in [ 1.5, \,  4.0] \, {\rm GeV^2}$.
It is not surprising to  observe the pronounced sensitivities of both the transverse and longitudinal form factors
under discussion to the precise shapes of the $B$-meson distribution amplitudes,
which are in accordance with the previous observations on the SCET sum-rule computations for
heavy-to-light $B$-meson decay form factors at large recoil \cite{Lu:2018cfc,Gao:2019lta}.
We are then led to conclude that the four-body leptonic bottom-meson decay processes allow us
to probe the partonic landscape of the heavy-hadron system diversely,
in a complementary manner to the semi-leptonic and electroweak penguin decays of $B$-mesons
\cite{Li:2012nk,Khodjamirian:2010vf,Khodjamirian:2012rm,Li:2012md},
with the aid of the upcoming sufficient experimental data.

\begin{figure}
\begin{center}
\includegraphics[width=1.0  \columnwidth]{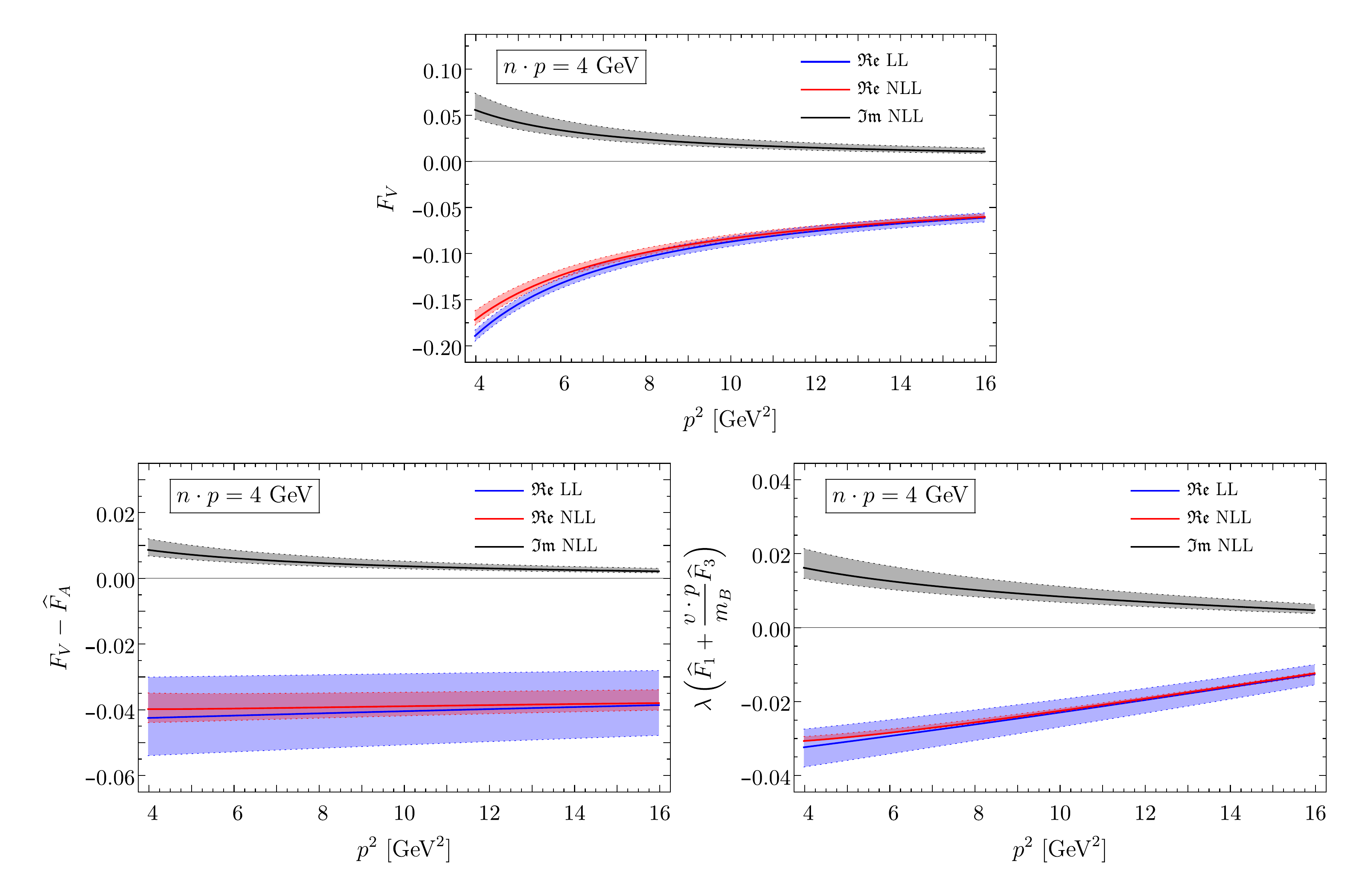}
\vspace*{-0.2 cm}
\caption{Theory predictions for the photon-momentum dependencies of the exclusive
$B_{u}^{-} \to \gamma^{\ast} \, \ell \, \bar \nu_{\ell}$ form factors with an off-shell photon
carrying the hard momentum $p_{\mu} \sim {\cal O}(m_b)$ in the LP approximation.
The uncertainty bands are obtained by varying the factorization scale in the allowed intervals
of $\mu  \in [m_b/2, \,\, 2 \, m_b]$.}
\label{fig: p2-dependence-of-form-factors-hard-region}
\end{center}
\end{figure}

We now turn to display the LP contributions to the radiative leptonic $B_{u}^{-} \to \gamma^{\ast} \, \ell \, \bar \nu_{\ell}$
form factors in the kinematic regime   $p^2 \in [4.0, \, 16.0]  \, {\rm GeV^2}$ at the LL and NLL accuracy
in Figure \ref{fig: p2-dependence-of-form-factors-hard-region}, where the residual perturbative uncertainties
from varying the factorization scale are further represented by the individual bands.
Unlike the factorized expressions for the exclusive $B_{u}^{-} \to \gamma^{\ast} \, \ell \, \bar \nu_{\ell}$
form factors with a hard-collinear photon,
the counterpart contributions of these form factors in the hard $p^2$ region
are apparently the real-valued functions at ${\cal O}(\alpha_s^0)$.
The emerged strong phases for $F_{V (A)}$  and
$\left ( \hat{F}_1 + {v \cdot p \over m_B} \, \hat{F}_3 \right )$ at the NLL accuracy
are generated perturbatively by the four one-loop diagrams shown in Figure \ref{fig: 1-loop-diagrams-u-quark},
which can be understood from the final-state rescattering mechanism
$B_{u}^{-} \to X_{u \, \bar u} \, \ell \, \bar \nu_{\ell} \to  \ell^{\prime} \, \bar \ell^{\prime} \, \ell \, \bar \nu_{\ell}$
at hadronic level with $X_{u \, \bar u}$ standing for the appropriate neutral light-hadron states.
Moreover, the higher-order QCD corrections to both the transverse and longitudinal form factors
at ${\cal O}(\alpha_s)$ will give rise to the very minor impacts on the corresponding tree-level predictions
at $4.0 \,  {\rm GeV^2} \leq p^2 \leq 16.0 \, {\rm GeV^2}$ (numerically less than $10 \%$ in magnitudes).
In particular,  the NLL resummation improved computations are highly beneficial for pining down the theory uncertainties
of the LL QCD predictions effectively. 

\subsection{Differential decay distribution for $B_{u}^{-} \to  \ell^{\prime} \, \bar \ell^{\prime} \, \ell \, \bar \nu_{\ell}$}

Having at our disposal the theory predictions for the exclusive radiative $B_{u}^{-} \to \gamma^{\ast} \, \ell \, \bar \nu_{\ell}$
form factors, we are now in a position to address the phenomenological aspects of the four-body leptonic $B$-meson decays
with an emphasis on the numerous decay observables of experimental interest.
In doing so, we begin to derive the five-fold differential decay width for
the process $B_{u}^{-} \to  \ell^{\prime} \, \bar \ell^{\prime} \, \ell \, \bar \nu_{\ell}$
with non-identical lepton flavours in terms of the two invariant masses $p^2$ and $q^2$  as well as the three angles
$\theta_1$, $\theta_2$ and $\phi$ (see Appendix \ref{appendix: Kinematics} for the detailed definitions)
\begin{eqnarray}
\frac{d^5 \Gamma(B_{u}^{-} \to  \ell^{\prime} \, \bar \ell^{\prime} \, \ell \, \bar \nu_{\ell}) }
{d p^2 \, dq^2 \, d \cos \theta_1 \, d \cos \theta_2 \, d \phi}
&=& \frac{G_F^2 \, \alpha_{\rm em}^2 \, |V_{ub}|^2}{2^{12} \, \pi^4 \, p^4} \,
m_{B}^3 \,\, \lambda^{1/2}(m_{B}^2, p^2, q^2) \, {\cal J}(p^2, \, q^2, \,  \theta_1, \,  \theta_2, \,  \phi) \,,
\nonumber \\
&=& {\cal N}(p^2, q^2) \, {\cal J}(p^2, \, q^2, \,  \theta_1, \,  \theta_2, \,  \phi) \,,
\label{full dfferential decay distribution}
\end{eqnarray}
by employing the explicit expression of the obtained decay amplitude (\ref{general decay amplitude})
and by further summing over spins of the final-state particles.
It is straightforward to decompose the angular distribution ${\cal J}$ into a set of the trigonometric functions
\begin{eqnarray}
{\cal J}(p^2, \, q^2, \,  \theta_1, \,  \theta_2, \,  \phi)
&=&  J_1 \, (1 + \cos^2 \theta_1) \, (1 + \cos^2 \theta_2)
+ J_2 \, \sin^2 \theta_1 \, \sin^2 \theta_2
+ J_3 \,   (1 + \cos^2 \theta_1) \,  \cos \theta_2
\nonumber \\
&& +  \left [ J_4 \, \sin \theta_2 +  J_5 \, \sin \, (2 \, \theta_2)  \right ] \,
\, \sin \, (2 \, \theta_1) \, \sin \phi
\nonumber \\
&& +  \left [ J_6 \, \sin \theta_2 +  J_7 \, \sin \, (2 \, \theta_2)  \right ] \,
\, \sin \, (2 \, \theta_1) \, \cos \phi
\nonumber \\
&& + J_8 \, \sin^2 \theta_1 \, \sin^2 \theta_2 \, \sin \,  (2 \, \phi)
+ J_9 \, \sin^2 \theta_1 \, \sin^2 \theta_2 \, \cos \,  (2 \, \phi)  \,,
\label{angular distribution for non-identical leptons}
\end{eqnarray}
where the nine independent coefficient functions $J_i \equiv J_i(p^2, q^2)$
($i=1,..., 9$) can be expressed by  the three radiative bottom-meson decay form factors
\begin{eqnarray}
J_1 &=&  \frac{1}{4}  \, \left [ | \widetilde{F}_A  |^2 +  | \widetilde{F}_V  |^2 \right ] \,,
\nonumber \\
J_2 &=&  \frac{1}{2}  \, \left [ | \widetilde{\mathbb{F}}_A  |^2 +  | \widetilde{F}_{\|}  |^2
+ 2 \, {\rm Re} \, \left (\widetilde{\mathbb{F}}_A \, \widetilde{F}_{\|}^{\ast} \right ) \right ] \,,
\nonumber \\
J_3 &=&   {\rm Re} \, \left (\widetilde{F}_A \, \widetilde{F}_{V}^{\ast} \right )  \,,
\nonumber \\
J_4 &=&  \frac{1}{2}  \, {\rm Im} \, \left (\widetilde{F}_A \, \widetilde{F}_{\|}^{\ast} \right )  \,,
\nonumber \\
J_5 &=&  \frac{1}{4}  \, {\rm Im} \, \left (\widetilde{F}_V \, \widetilde{\mathbb{F}}_A^{\ast}
+ \widetilde{F}_V \, \widetilde{F}_{\|}^{\ast}  \right )  \,,
\nonumber \\
J_6 &=&  - \frac{1}{2}  \, {\rm Re} \, \left (\widetilde{F}_V \, \widetilde{\mathbb{F}}_A^{\ast}
+ \widetilde{F}_V \, \widetilde{F}_{\|}^{\ast}  \right )  \,,
\nonumber \\
J_7 &=&  - \frac{1}{4}  \, \left [ | \widetilde{F}_A \, \widetilde{\mathbb{F}}_A |
+  {\rm Re} \, \left (\widetilde{F}_A \, \widetilde{F}_{\|}^{\ast} \right ) \right ] \,,
\nonumber \\
J_8 &=&  \frac{1}{2}  \, {\rm Im} \, \left (\widetilde{F}_A \, \widetilde{F}_{V}^{\ast} \right )  \,,
\nonumber \\
J_9 &=&  \frac{1}{4}  \, \left [ | \widetilde{F}_A  |^2 -   | \widetilde{F}_V  |^2 \right ] \,.
\end{eqnarray}
We have introduced the shorthand notations for the distinct combinations of
the transition  form factors  with the appropriate kinematic functions
\begin{eqnarray}
\widetilde{F}_V &=&  2 \, \sqrt{\hat{p}^2 \, \hat{q}^2 \,  \lambda(1, \hat{p}^2, \, \hat{q}^2)} \, F_V  \,,
\qquad
\widetilde{F}_A = 2 \, \sqrt{\hat{p}^2 \, \hat{q}^2 } \, (1 + \hat{p}^2  - \hat{q}^2 )  \, F_A   \,,
\nonumber \\
\widetilde{\mathbb{F}}_A  &=&   \left [ (1 - \hat{q}^2 )^2 -  \hat{p}^4 \right ] \, F_A  \,,
\qquad  \hspace{0.75 cm}
\widetilde{F}_{\|} = \lambda(1, \hat{p}^2, \, \hat{q}^2) \,
\left (  F_1 + { v \cdot p \over m_B} \, F_3  \right ) \,,
\end{eqnarray}
where  the two dimensionless hadronic variables are defined by  $\hat{p}^2 = p^2/m_{B}^2$ and $\hat{q}^2 = q^2/m_{B}^2$.
It remains important to point out that our expressions for the full angular distribution of
$B_{u}^{-} \to  \ell^{\prime} \, \bar \ell^{\prime} \, \ell \, \bar \nu_{\ell}$
coincide with Ref. \cite{Beneke:2021rjf} by  applying the  replacement rules
for the helicity angles and for the exclusive $B_{u}^{-} \to \gamma^{\ast} \,  W^{\ast}$ decay form factors
\begin{eqnarray}
\theta_{\gamma} \to \theta_1, \hspace{0.3 cm}
\theta_{W} \to \pi - \theta_2, \hspace{0.3 cm}
F_{A \perp} \to F_A, \hspace{0.3 cm}
F_{A_{\|}} \to - \left [  F_1 + { v \cdot p \over m_B} \, F_3
+  \frac{(1 - \hat{q}^2 )^2 -  \hat{p}^4}{\lambda(1, \hat{p}^2, \, \hat{q}^2)} \, F_A  \right ].
\hspace{0.5 cm}
\end{eqnarray}

\begin{figure}
\begin{center}
\includegraphics[width=0.95  \columnwidth]{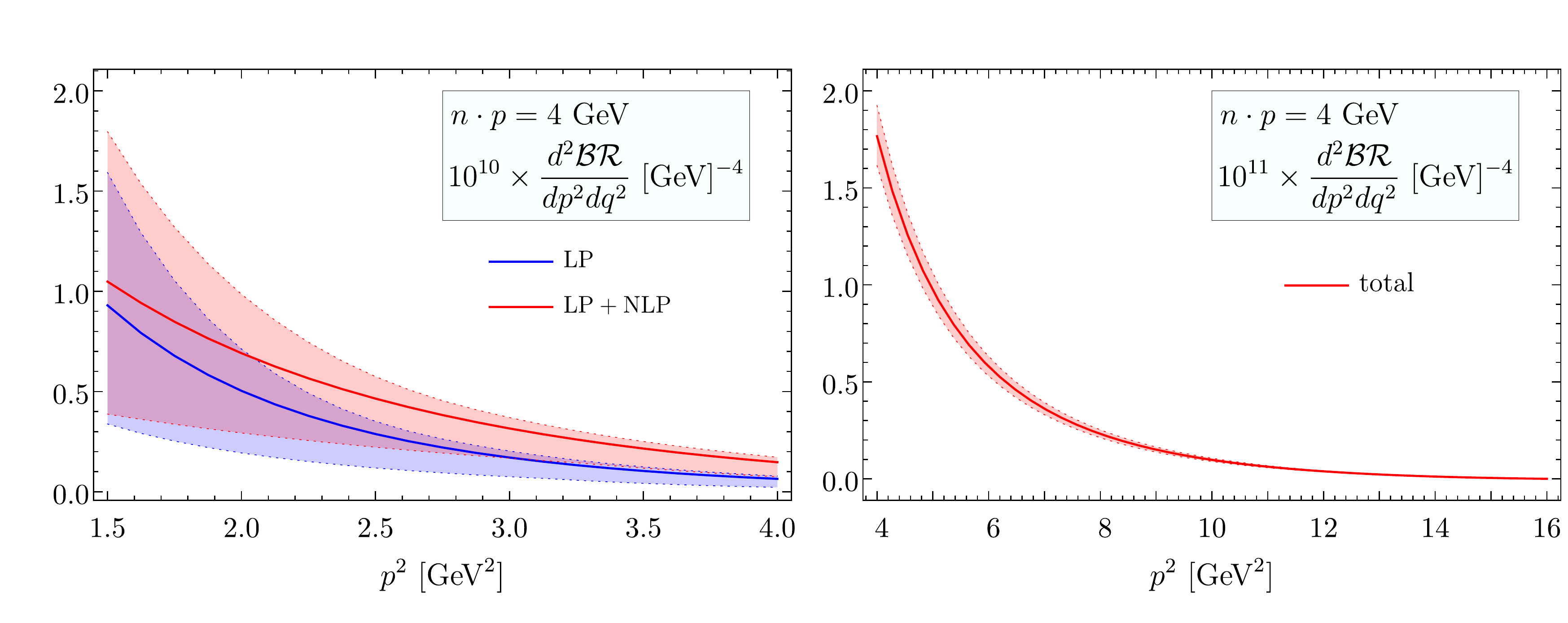}
\vspace*{0.5 cm}
\caption{Theory predictions for the double differential branching fractions
of the four-body leptonic $B$-meson decays with non-identical lepton flavours
in the hard-collinear $p^2$ region (left panel) and in the hard $p^2$ region (right panel),
where the numerical value of the large component of the virtual photon momentum is taken as $4 \, {\rm GeV}$.
The yielding uncertainty bands are obtained by adding the individual errors from
varying all the  theory  input parameters in quadrature.}
\label{fig: double differential BR}
\end{center}
\end{figure}

To facilitate the experimental explorations we proceed to construct the following
weighted angular integrals for  the five-fold differential decay width
(\ref{full dfferential decay distribution})
\begin{eqnarray}
X(p^2, q^2) = \int_{-1}^{1} d \cos \theta_1 \, \int_{-1}^{1} d \cos \theta_2  \,
\int_{0}^{2 \pi} d \phi \,
\frac{d^5 \Gamma(B_{u}^{-} \to  \ell^{\prime} \, \bar \ell^{\prime} \, \ell \, \bar \nu_{\ell}) }
{d p^2 \, dq^2 \, d \cos \theta_1 \, d \cos \theta_2 \, d \phi} \,
\omega_X(p^2, q^2, \theta_1, \theta_2, \phi),
\hspace{0.8 cm}
\end{eqnarray}
to obtain the double differential distributions in the invariant masses $p^2$ and $q^2$.
Adopting $\omega_X = 1$ immediately leads to the familiar differential decay rate
\begin{eqnarray}
\frac{d^2 \Gamma}{d p^2 \, dq^2}
= {1 \over  \tau_{B_u} } \, \frac{d^2 {\cal BR}}{d p^2 \, dq^2}
= {32 \, \pi \over 9} \,   {\cal N}(p^2, q^2) \, \left (4 \, J_1 + J_2  \right ) \,.
\end{eqnarray}
Inspecting the yielding predictions for the double differential branching fractions
of the four-body leptonic $B$-meson decay processes  displayed in
Figure \ref{fig: double differential BR} implies that the newly obtained subleading power corrections
computable  in the perturbative factorization framework appear to enhance the counterpart LP contribution
at the NLL accuracy significantly  in the hard-collinear $p^2$ region
(as large as ${\cal O} (40 \, \%)$ enhancement at $p^2=2.5 \, {\rm GeV^2}$ numerically)
for the fixed value $n \cdot p = 4 \, {\rm GeV}$,
where the substantial uncertainties represented by the blue and pink bands
are due to the poorly constrained shape parameters of the two-particle
and three-particle $B$-meson distribution amplitudes.
By contrast, the resulting uncertainties for the differential branching fractions in the hard $p^2$  region
turn out to be insignificant numerically (at the level of  ${\cal O} (10 \, \%)$),
which can be traced back to the very independence  of the HQET factorization formula
(\ref{NLO factorization formula in hard region}) on the non-local hadronic quantities.
Additionally, we predict the rapidly decreasing branching fractions in the kinematic regime
$4.0 \,  {\rm GeV}^2 \leq p^2 \leq  \, 16.0 \,  {\rm GeV}^2$
when the invariant mass of the $\ell^{\prime} \bar \ell^{\prime}$ pair moves towards
the higher values, which provides an explicit confirmation of the expected numerical features
dictated by our power counting scheme (see \cite{Beneke:2021rjf} for an earlier discussion).

Along the same vein, we can readily define the non-vanishing angular asymmetries
normalized to the differential decay width by taking the appropriate weight functions
\begin{eqnarray}
{\cal A}_{c2 \theta_1} &=&  \left [ \frac{d^2 \Gamma}{d p^2 \, dq^2}  \right ]^{-1} \,
\int_{-1}^{1} d \cos \theta_1 \,\,  {\rm sgn} \left (\cos (2 \, \theta_1) \right)  \,\,
\frac{d^3 \Gamma(B_{u}^{-} \to  \ell^{\prime} \, \bar \ell^{\prime} \, \ell \, \bar \nu_{\ell}) }
{d p^2 \, dq^2 \, d \cos \theta_1}
\nonumber \\
&=&  1- \frac{5} {2 \, \sqrt{2}} + \frac{3} {\sqrt{2}} \, \frac{J_1}{4\,J_1+J_2}   \,,
\nonumber \\
{\cal A}_{c \theta_2} &=&  \left [ \frac{d^2 \Gamma}{d p^2 \, dq^2}  \right ]^{-1} \,
 \int_{-1}^{1} d \cos \theta_2 \,\,  {\rm sgn} \left( \cos (\theta_2) \right)  \,\,
\frac{d^3 \Gamma(B_{u}^{-} \to  \ell^{\prime} \, \bar \ell^{\prime} \, \ell \, \bar \nu_{\ell}) }
{d p^2 \, dq^2 \, d \cos \theta_2}
= \frac{3}{2}\,\frac{J_3}{4\,J_1+J_2} \,,
\nonumber \\
{\cal A}_{s2 \phi} &=&  \left [ \frac{d^2 \Gamma}{d p^2 \, dq^2}  \right ]^{-1} \,
\int_{0}^{2 \pi} \,\,  {\rm sgn} \left( \sin (2 \, \phi) \right )  \,\,
\frac{d^3 \Gamma(B_{u}^{-} \to  \ell^{\prime} \, \bar \ell^{\prime} \, \ell \, \bar \nu_{\ell}) }
{d p^2 \, dq^2 \, d \phi}
= \frac{2}{\pi} \, \frac{J_8}{4\,J_1+J_2} \,,
\nonumber \\
{\cal A}_{c2 \phi} &=&  \left [ \frac{d^2 \Gamma}{d p^2 \, dq^2}  \right ]^{-1} \,
\int_{0}^{2 \pi} d \phi \,\,  {\rm sgn} \left( \cos (2 \, \phi) \right )  \,\,
\frac{d^3 \Gamma(B_{u}^{-} \to  \ell^{\prime} \, \bar \ell^{\prime} \, \ell \, \bar \nu_{\ell}) }
{d p^2 \, dq^2 \, d \phi}
= \frac{2}{\pi} \, \frac{J_9}{4\,J_1+J_2} \,,
\nonumber \\
{\cal A}_{s \phi, c \theta_1} &=&  \left [ \frac{d^2 \Gamma}{d p^2 \, dq^2}  \right ]^{-1} \,
 \int_{-1}^{1} d \cos \theta_1 \, \int_{0}^{2 \pi} d \phi \,\,
 {\rm sgn} \left( \sin (\phi) \right )  \,
{\rm sgn} \left( \cos(\theta_1) \right )  \,\,
\frac{d^4 \Gamma(B_{u}^{-} \to  \ell^{\prime} \, \bar \ell^{\prime} \, \ell \, \bar \nu_{\ell}) }
{d p^2 \, dq^2 \, d \cos \theta_1 \, d \phi}
\nonumber \\
&=& \frac{3}{4} \, \frac{J_4}{4\,J_1+J_2} \,,
\nonumber \\
{\cal A}_{c \phi, c \theta_1} &=&  \left [ \frac{d^2 \Gamma}{d p^2 \, dq^2}  \right ]^{-1} \,
 \int_{-1}^{1} d \cos \theta_1 \, \int_{0}^{2 \pi} d \phi \,\,
 {\rm sgn} \left( \cos (\phi) \right )  \,
{\rm sgn} \left( \cos(\theta_1) \right )  \,\,
\frac{d^4 \Gamma(B_{u}^{-} \to  \ell^{\prime} \, \bar \ell^{\prime} \, \ell \, \bar \nu_{\ell}) }
{d p^2 \, dq^2 \, d \cos \theta_1 \, d \phi}
\nonumber \\
&=& \frac{3}{4} \, \frac{J_6}{4\,J_1+J_2} \,,
\nonumber \\
{\cal A}_{s \phi, c \theta_1, c \theta_2} &=&  \left [ \frac{d^2 \Gamma}{d p^2 \, dq^2}  \right ]^{-1} \,
\int_{-1}^{1} d \cos \theta_1 \, \int_{-1}^{1} d \cos \theta_2 \, \int_{0}^{2 \pi} d \phi \,\,
\nonumber \\
&& \times \, {\rm sgn} \left( \sin (\phi) \right )  \,
{\rm sgn} \left( \cos(\theta_1) \right ) \,
{\rm sgn} \left( \cos(\theta_2) \right ) \,\,
\frac{d^5 \Gamma(B_{u}^{-} \to  \ell^{\prime} \, \bar \ell^{\prime} \, \ell \, \bar \nu_{\ell}) }
{d p^2 \, dq^2 \, d \cos \theta_1 \, d \cos \theta_2 \, d \phi}
\nonumber \\
&=& \frac{2}{\pi} \, \frac{J_5}{4\,J_1+J_2} \,,
\nonumber \\
{\cal A}_{c \phi, c \theta_1, c \theta_2} &=&  \left [ \frac{d^2 \Gamma}{d p^2 \, dq^2}  \right ]^{-1} \,
\int_{-1}^{1} d \cos \theta_1 \, \int_{-1}^{1} d \cos \theta_2 \, \int_{0}^{2 \pi} d \phi \,\,
\nonumber \\
&& \times \, {\rm sgn} \left( \cos (\phi) \right )  \,
{\rm sgn} \left( \cos(\theta_1) \right ) \,
{\rm sgn} \left( \cos(\theta_2) \right ) \,\,
\frac{d^5 \Gamma(B_{u}^{-} \to  \ell^{\prime} \, \bar \ell^{\prime} \, \ell \, \bar \nu_{\ell}) }
{d p^2 \, dq^2 \, d \cos \theta_1 \, d \cos \theta_2 \, d \phi}
\nonumber \\
&=& \frac{2}{\pi} \, \frac{J_7}{4\,J_1+J_2} \,,
\end{eqnarray}
where the sign function reads ${\rm sgn} (\pm |x|) = \pm 1$ for any non-zero real number $x$.
In contrast to the established angular observable ${\cal A}_{c \theta_2}$,
we observe the vanishing single forward-backward asymmetry in the angle $\theta_1$
by taking advantage of the derived differential decay distribution
(\ref{full dfferential decay distribution}) immediately.
In order to determine the underlying mechanism for such an interesting discrepancy,
we recall the well-known expression of the differential forward-backward asymmetry
for the electroweak penguin $B \to K^{\ast} \ell \bar \ell$ decay process
\cite{Altmannshofer:2008dz}
\begin{eqnarray}
{\cal A}_{\rm FB} (B \to K^{\ast} \ell \bar \ell) \propto
{\rm Re} \left ( A_{\|, \,  L} \, A_{\perp, \,  L}^{\ast} \right )
- \left ( L \rightarrow  R \right )  \,,
\label{AFB for B to Kstar ll}
\end{eqnarray}
where the four transversity amplitudes in the factorization approximation are given by
\begin{eqnarray}
A_{\|, \,  L(R)}  \propto  \left ( C_9^{\rm eff} \mp C_{10} \right ) \,
\frac{V(q^2)}{m_B + m_{K^{\ast}}} \,,
\qquad
A_{\perp, \,  L(R)}  \propto  \left ( C_9^{\rm eff} \mp C_{10} \right ) \,
\frac{A_1(q^2)}{m_B - m_{K^{\ast}}} \,.
\end{eqnarray}
Confronting further the exclusive three-body $B \to  W^{\ast} \, \ell^{\prime} \, \bar \ell^{\prime}$ decay amplitude
\begin{eqnarray}
 {\cal M} (B \to  W^{\ast} \, \ell^{\prime} \, \bar \ell^{\prime})
& \propto &  \left [ \bar \ell^{\prime}(p_1) \, \gamma^{\nu} \,  \ell^{\prime}(p_2) \right ] \,
\bigg  \{  i \, \epsilon_{\mu \nu p v} \,
\epsilon_{W^{\ast}}^{\mu} (q) \,  F_V(p^2,  n \cdot p)
+  \epsilon_{W^{\ast} \nu}(q) \, v \cdot p \,   F_A(p^2,  n \cdot p)
\nonumber \\
&& + \,  q_{\nu} \,  \frac{ p \cdot   \epsilon_{W^{\ast}}(q) } {m_B} \,
\left [  F_1(p^2,  n \cdot p) + { v \cdot p \over m_B} \, F_3(p^2,  n \cdot p)   \right ]  \bigg  \}   \,,
\label{B to Wstar ll decay amplitude}
\end{eqnarray}
with the analogous formula for the semileptonic $B \to K^{\ast} \ell \bar \ell$ decay amplitude
\begin{eqnarray}
 {\cal M} (B \to  K^{\ast} \, \ell \, \bar \ell)
& \propto &  \left [ \bar \ell(q_1) \, \gamma^{\mu} \,
\left ( C_9^{\rm eff} + C_{10} \, \gamma_5 \right )  \ell(q_2) \right ] \,
\bigg  \{ i \, \epsilon_{\mu \nu p v} \,
\epsilon_{K^{\ast}}^{\nu} (p) \,
\left [ \frac{2 \, m_B}{m_B + m_{K^{\ast}}} V(q^2) \, \right ]
\nonumber \\
&& + \, \left [ \epsilon_{K^{\ast} \mu}(p)  - \frac{q \cdot \epsilon_{K^{\ast}}(p)}{q^2}  \, q_{\mu} \right ] \,
\left (m_B + m_{K^{\ast}} \right )  \,  A_1(q^2)
+ ... \bigg  \}   \,,
\label{B to Kstar ll decay amplitude}
\end{eqnarray}
we are then allowed to express the differential forward-backward asymmetry of
$B \to  W^{\ast} \, \ell^{\prime} \, \bar \ell^{\prime}$ from (\ref{AFB for B to Kstar ll})
analytically by implementing necessary replacements for the hadronic form factors
and especially for the short-distance Wilson coefficients
\begin{eqnarray}
C_9^{\rm eff} \to 1, \qquad  C_{10} \to 0 \,,
\end{eqnarray}
which leads to the vanishing result apparently due to an exact cancellation
between the left-handed and right-handed contributions.

\begin{figure}
\begin{center}
\includegraphics[width=1.0  \columnwidth]{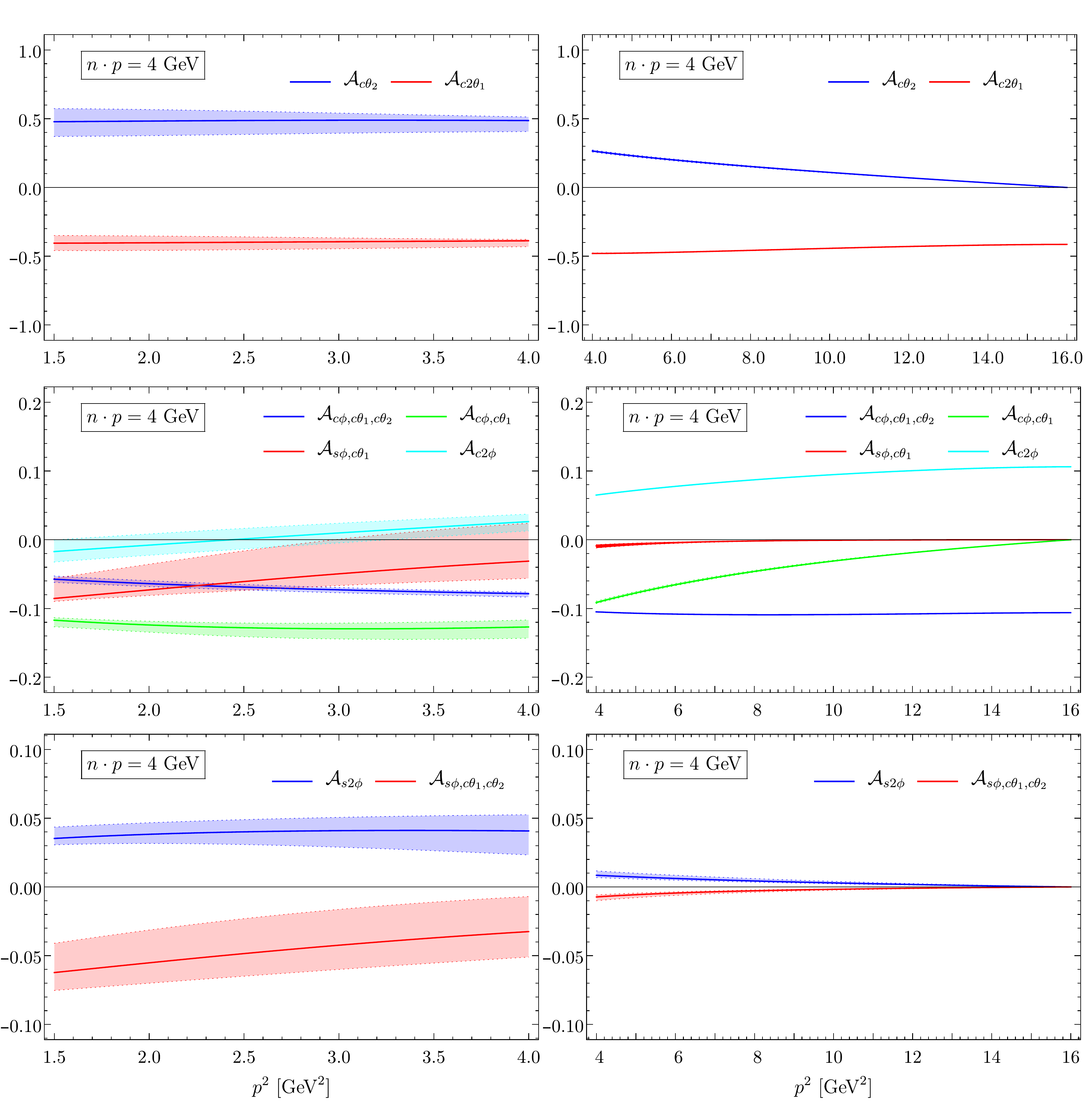}
\vspace*{0.2 cm}
\caption{Theory predictions for the various angular asymmetries of the four-body leptonic $B$-meson decays
with non-identical lepton flavours in the hard-collinear $p^2$ region (left panel) and in the hard $p^2$ region (right panel),
where the numerical value of the large component of the virtual photon momentum is taken as $4 \, {\rm GeV}$.
The yielding uncertainty bands are obtained by adding the individual errors from  varying all the theory input parameters
in quadrature. }
\label{fig: double_diff_Asymmetries}
\end{center}
\end{figure}

We now turn to present our predictions for the differential angular asymmetries  of the four-body leptonic
$B_{u}^{-} \to  \ell^{\prime} \, \bar \ell^{\prime} \, \ell \, \bar \nu_{\ell}$ decays
in the kinematic domain $1.5 \, {\rm GeV^2} \leq p^2 \leq 16.0 \, {\rm GeV^2}$
in Figure \ref{fig: double_diff_Asymmetries}, where the uncertainty bands  are obtained by adding all the separate
errors from the variations of the essential theory inputs in quadrature.
The yielding pronounced  results for the two  angular observables ${\cal A}_{c2 \theta_1}$
and ${\cal A}_{c \theta_2}$ in the hard-collinear $p^2$ region will evidently enable us to carry out the dedicated
measurements with the encouraging precision at the LHCb and Belle II experiments.
However, the four angular asymmetries ${\cal A}_{c 2 \phi}$, ${\cal A}_{s \phi, c \theta_1}$,
${\cal A}_{c \phi, c \theta_1}$ and ${\cal A}_{c \phi, c \theta_1, c \theta_2}$
can reach at most ${\cal O} (10 \, \%)$ numerically based upon our improved calculations
for the radiative $B_{u}^{-} \to \gamma^{\ast} \, \ell \, \bar \nu_{\ell}$ decay form factors.
In addition, the resulting predictions for the two remaining asymmetry observables ${\cal A}_{s 2 \phi}$
and ${\cal A}_{s \phi, c \theta_1, c \theta_2}$ appear to be merely at the level of  ${\cal O} (5 \, \%)$,
thus rendering their measurements considerably challenging  for the ongoing collision experiments.
It remains important to remark that the achieved uncertainties of the differential angular asymmetries
for $1.5 \, {\rm GeV^2} \leq p^2 \leq 4.0 \, {\rm GeV^2}$ are unsurprisingly improved,
when compared with the obtained results for the differential branching fractions presented in
Figure \ref{fig: double differential BR}, on account of the substantial cancellation
of the parametric uncertainties from the badly known $B$-meson distribution amplitudes in HQET.

\begin{table}
\centering
\renewcommand{\arraystretch}{2.0}
\resizebox{\columnwidth}{!}{
\begin{tabular}{|c||c|c|c||cccccc|}
\hline
\hline
\multirow{2}{*}{Observables}& \multirow{2}{*}{$[t_1, \,\, t_2]$ } & \multirow{2}{*}{LP} & \multirow{2}{*}{Total} & \multicolumn{6}{c|}{Uncertainties} \\
& $( {\rm GeV}^2 )$ & {(\rm NLL)}  & {(\rm LP + NLP)}  & $\mu$ &$\mu_{h1}$
& $\widehat{\sigma}_{B_u}^{(1, \, 2)}$ & $\lambda_{B_u}$  & $p^2_{\rm cut}$ & $|V_{ub}|$ \\
\hline
\multirow{4}{*}{$\langle {{\cal BR}\, [t_1, \, t_2] } \rangle \times 10^9$}& $\left.\text{[1.5, }m_B^2\right]$ & $0.88_{-0.37}^{+0.24}$ & $1.23_{-0.52}^{+0.30}$ & $\text{}_{-0.23}^{+0.06}$ & $\text{}_{-0.19}^{+0.07}$ & $\text{}_{-0.24}^{+0.21}$ & $\text{}_{-0.32}^{+0.13}$ & $\text{}_{-0.03}^{+0.05}$ & $\text{}_{-0.10}^{+0.11}$ \\
& $\left.\text{[2.0, }m_B^2\right]$ & $0.57_{-0.20}^{+0.16}$ & $0.83_{-0.29}^{+0.15}$ & $\text{}_{-0.14}^{+0.04}$ & $\text{}_{-0.11}^{+0.04}$ & $\text{}_{-0.11}^{+0.02}$ & $\text{}_{-0.17}^{+0.05}$ & $\text{}_{-0.03}^{+0.05}$ & $\text{}_{-0.07}^{+0.07}$ \\
& $\left.\text{[3.0, }m_B^2\right]$ & $0.32_{-0.09}^{+0.15}$ & $0.42_{-0.09}^{+0.07}$ & $\text{}_{-0.05}^{+0.02}$ & $\text{}_{-0.03}^{+0.01}$ & $\text{}_{-0.03}^{+0.00}$ & $\text{}_{-0.04}^{+0.00}$ & $\text{}_{-0.03}^{+0.05}$ & $\text{}_{-0.04}^{+0.04}$ \\
& $\left.\text{[4.0, }m_B^2\right]$ & $0.24_{-0.08}^{+0.02}$ & $0.24_{-0.04}^{+0.02}$ & $\text{}_{-0.00}^{+0.00}$ & $\text{}_{-0.00}^{+0.00}$ & $\text{}_{-0.00}^{+0.00}$ & $\text{}_{-0.00}^{+0.00}$ & $\text{}_{-0.03}^{+0.00}$ & $\text{}_{-0.02}^{+0.02}$ \\
\hline
\multirow{4}{*}{$\langle {{\cal A}_{c 2 \theta_1} \, [t_1, \, t_2] } \rangle$} & $\left.\text{[1.5, }m_B^2\right]$ & $-0.38_{-0.08}^{+0.06}$ & $-0.45_{-0.05}^{+0.04}$ & $\text{}_{-0.01}^{+0.00}$ & $\text{}_{-0.01}^{+0.00}$ & $\text{}_{-0.03}^{+0.04}$ & $\text{}_{-0.04}^{+0.01}$ & $\text{}_{-0.02}^{+0.01}$ & $-$ \\
& $\left.\text{[2.0, }m_B^2\right]$ & $-0.40_{-0.08}^{+0.05}$ & $-0.46_{-0.05}^{+0.04}$ & $\text{}_{-0.01}^{+0.00}$ & $\text{}_{-0.01}^{+0.00}$ & $\text{}_{-0.02}^{+0.02}$ & $\text{}_{-0.03}^{+0.01}$ & $\text{}_{-0.02}^{+0.01}$ & $-$ \\
& $\left.\text{[3.0, }m_B^2\right]$ & $-0.47_{-0.07}^{+0.05}$ & $-0.50_{-0.07}^{+0.06}$ & $\text{}_{-0.01}^{+0.00}$ & $\text{}_{-0.04}^{+0.00}$ & $\text{}_{-0.01}^{+0.05}$ & $\text{}_{-0.04}^{+0.01}$ & $\text{}_{-0.04}^{+0.02}$ & $-$ \\
& $\left.\text{[4.0, }m_B^2\right]$ & $-0.54_{-0.00}^{+0.05}$ & $-0.54_{-0.00}^{+0.01}$ & $\text{}_{-0.00}^{+0.00}$ & $\text{}_{-0.00}^{+0.00}$ & $\text{}_{-0.00}^{+0.00}$ & $\text{}_{-0.00}^{+0.00}$ & $\text{}_{-0.00}^{+0.01}$ & $-$ \\
\hline
\multirow{4}{*}{$\langle {{\cal A}_{c  \theta_2} \, [t_1, \, t_2] } \rangle$} & $\left.\text{[1.5, }m_B^2\right]$ & $0.48_{-0.13}^{+0.09}$ & $0.38_{-0.10}^{+0.07}$ & $\text{}_{-0.03}^{+0.00}$ & $\text{}_{-0.03}^{+0.01}$ & $\text{}_{-0.05}^{+0.06}$ & $\text{}_{-0.07}^{+0.03}$ & $\text{}_{-0.03}^{+0.02}$ & $-$ \\
& $\left.\text{[2.0, }m_B^2\right]$ & $0.43_{-0.13}^{+0.07}$ & $0.36_{-0.10}^{+0.06}$ & $\text{}_{-0.03}^{+0.00}$ & $\text{}_{-0.03}^{+0.01}$ & $\text{}_{-0.04}^{+0.04}$ & $\text{}_{-0.07}^{+0.02}$ & $\text{}_{-0.05}^{+0.03}$ & $-$ \\
& $\left.\text{[3.0, }m_B^2\right]$ & $0.28_{-0.10}^{+0.08}$ & $0.28_{-0.10}^{+0.05}$ & $\text{}_{-0.03}^{+0.01}$ & $\text{}_{-0.02}^{+0.01}$ & $\text{}_{-0.02}^{+0.00}$ & $\text{}_{-0.04}^{+0.01}$ & $\text{}_{-0.08}^{+0.05}$ & $-$ \\
& $\left.\text{[4.0, }m_B^2\right]$ & $0.17_{-0.01}^{+0.06}$ & $0.17_{-0.01}^{+0.07}$ & $\text{}_{-0.00}^{+0.00}$ & $\text{}_{-0.00}^{+0.00}$ & $\text{}_{-0.00}^{+0.00}$ & $\text{}_{-0.00}^{+0.00}$ & $\text{}_{-0.00}^{+0.07}$ & $-$ \\
\hline
\hline
\end{tabular}
}
\renewcommand{\arraystretch}{1.0}
\caption{Theory predictions for the binned distributions of the branching fraction
as well as the two angular asymmetries ${\cal A}_{c2 \theta_1}$ and ${\cal A}_{c \theta_2}$
for the four-body leptonic $B$-meson decays with non-identical lepton flavours,
where the numerically sizeable uncertainties from varying distinct input parameters
are further displayed for completeness.  }
\label{table for the binned observables with non-identical leptons}
\end{table}

We are now in a position to investigate the binned distributions for both the branching fraction and
the two promising angular  asymmetries with the required kinematic cut on  the large component
of the off-shell photon momentum
\begin{eqnarray}
\langle {{\cal BR}\, [t_1, \, t_2] } \rangle
&=& \tau_{B_u}  \, \int_{t_1}^{t_2} \, d p^2 \, \int_{0}^{(m_B - \sqrt{p^2})^2} \, d q^2 \,\,
\theta\left ( n \cdot p - 3 \, {\rm GeV} \right ) \,
\frac{d^2 \Gamma(B_{u}^{-} \to  \ell^{\prime} \, \bar \ell^{\prime} \, \ell \, \bar \nu_{\ell}) }
{d p^2 \, dq^2 }     \,,
\nonumber \\
\langle {{\cal A}_{c 2 \theta_1} \, [t_1, \, t_2] } \rangle
&=&  \frac{\tau_{B_u} }  {\langle {{\cal BR}\, [t_1, \, t_2] } \rangle} \,\,
\int_{t_1}^{t_2} \, d p^2 \, \int_{0}^{(m_B - \sqrt{p^2})^2} \, d q^2 \,
\int_{-1}^{1} d \cos \theta_1 \,\,  {\rm sgn} \left (\cos (2 \, \theta_1) \right)  \,
\nonumber \\
&& \times \,  \theta\left ( n \cdot p - 3 \, {\rm GeV} \right ) \,\,
\frac{d^3 \Gamma(B_{u}^{-} \to  \ell^{\prime} \, \bar \ell^{\prime} \, \ell \, \bar \nu_{\ell}) }
{d p^2 \, dq^2 \, d \cos \theta_1} \,,
\nonumber \\
\langle {{\cal A}_{c  \theta_2} \, [t_1, \, t_2] } \rangle
&=&  \frac{\tau_{B_u} }  {\langle {{\cal BR}\, [t_1, \, t_2] } \rangle} \,\,
\int_{t_1}^{t_2} \, d p^2 \, \int_{0}^{(m_B - \sqrt{p^2})^2} \, d q^2 \,
\int_{-1}^{1} d \cos \theta_2 \,\,  {\rm sgn} \left (\cos (\theta_2) \right)  \,
\nonumber \\
&& \times \, \theta\left ( n \cdot p - 3 \, {\rm GeV} \right ) \,\,
\frac{d^3 \Gamma(B_{u}^{-} \to  \ell^{\prime} \, \bar \ell^{\prime} \, \ell \, \bar \nu_{\ell}) }
{d p^2 \, dq^2 \, d \cos \theta_2}  \,,
\end{eqnarray}
where the conversion relations for the two scalar quantities $n \cdot p$
and  $\bar n \cdot p$  are given by
\begin{eqnarray}
n \cdot p = \frac{m_B^2 - q^2 + p^2 + \sqrt{\lambda(m_B^2, p^2, q^2)}} {2 \, m_B},
\qquad
\bar n \cdot p = \frac{m_B^2 - q^2 + p^2 - \sqrt{\lambda(m_B^2, p^2, q^2)}} {2 \, m_B} \,.
\hspace{0.8 cm}
\end{eqnarray}
We collect the numerical results for the LP and NLP computations of these binned observables in
Table \ref{table for the binned observables with non-identical leptons} subsequently
with the combined theory uncertainties by adding all the individual errors in quadrature.
Here the matching parameter $p^2_{\rm cut} = (4.0 \pm 1.0) \, {\rm GeV^2}$ has been introduced
to separate the hard and hard-collinear $p^2$ regimes such that the factorized expressions
of the non-hadronic $B_{u}^{-} \to \gamma^{\ast} \, \ell \, \bar \nu_{\ell}$ form factors
summarized in (\ref{the combined FFs in the hard-collinear p2 region})
are expected to be applicable for $p^2 \in [1.5 \, {\rm GeV^2}, \,  p^2_{\rm cut}]$,
whereas the obtained HQET expressions shown in (\ref{NLO factorization formula in hard region})
will be employed for evaluating the angular functions $J_i$
in the kinematic interval $p^2_{\rm cut} \leq p^2 \leq m_B^2$.
Comparing the numerical predictions for the first and last bins of $\langle {{\cal BR}\, [t_1, \, t_2] } \rangle$
indicates that the predicted hard-collinear contribution to the branching fraction
in the bin $p^2 \in [1.5 \, {\rm GeV^2}, \,  4.0 \, {\rm GeV^2}]$ is approximately
a factor of three larger than the counterpart effect from the hard $p^2$ bin
$[4.0 \, {\rm GeV^2}, \,  m_B^2]$ within the  sizeable theory uncertainties.
Importantly, the factorizable subleading power corrections to the binned branching fractions
of  $[1.5 \, {\rm GeV^2}, \,  2.0 \, {\rm GeV^2}]$
and $[1.5 \, {\rm GeV^2}, \,  4.0 \, {\rm GeV^2}]$ with the SCET factorization technique
will bring about  the notable enhancements for the corresponding LP predictions,
amounting to about ${\cal O}(30 \, \%)$ and ${\cal O}(50 \, \%)$ respectively.
In addition,  the yielding prediction of the binned decay rate
$\langle {{\cal BR}\, [1.5 \, {\rm GeV^2}, \, 2.0 \, {\rm GeV^2}] } \rangle$
deduced from Table \ref{table for the binned observables with non-identical leptons}
is observed to be compatible with the previous calculation  in the QCD factorization framework \cite{Beneke:2021rjf}.
Apparently, the newly derived subleading power corrections can  modify the corresponding LP predictions
for  the two binned asymmetries $\langle {{\cal A}_{c 2 \theta_1} \, [1.5 \, {\rm GeV^2}, \, m_B^2] } \rangle$
and  $\langle {{\cal A}_{c  \theta_2} \, [1.5 \, {\rm GeV^2}, \, m_B^2] } \rangle$
by an amount of approximately ${\cal O} (20 \, \%)$.
We further mention in passing that the fast-decreasing binned asymmetries
$\langle {{\cal A}_{c  \theta_2} \, [t_1, \, m_B^2] } \rangle$ with the growing value of $t_1$
can be actually understood from the distinctive feature of the very differential angular asymmetry
presented in Figure \ref{fig: double_diff_Asymmetries}.

Finally we turn to explore the phenomenological opportunities for the four-body charged-current
bottom-meson decays $B_{u}^{-} \to  \ell^{\prime} \, \bar \ell^{\prime} \, \ell \, \bar \nu_{\ell}$
with identical lepton flavours $\ell^{\prime} = \ell$, which will become quite challenging experimentally
due to the very appearance of the two indistinguishable like-sign leptons in the final state
and especially the practical implementation of the essential cut on the virtual photon momentum
for the sake of adopting the perturbative QCD factorization theorems.
To this end, we will begin with the manifest expression of the full decay amplitude
to  lowest non-vanishing  order in the electromagnetic interaction
\begin{eqnarray}
{\cal A}_{\rm tot} (B_{u}^{-} \to  \ell \, \bar \ell \, \ell \, \bar \nu_{\ell})
&=& {\cal A} (B_{u}^{-} (p_B) \to \gamma^{\ast}(p) (\to \ell(p_1) \,\, \bar \ell(p_2))
\,\, W^{\ast}(q) (\to \ell(q_1) \, \bar \nu_{\ell}(q_2)))
\nonumber \\
&& - \, {\cal A} (B_{u}^{-} (p_B) \to \gamma^{\ast}(\tilde{p}) (\to \ell(q_1) \,\, \bar \ell(p_2))
\,\, W^{\ast}(\tilde{q}) (\to \ell(p_1) \, \bar \nu_{\ell}(q_2)))
\nonumber \\
&\equiv& {\cal A}_{\rm dir}(B_{u}^{-} \to  \ell \, \bar \ell \, \ell \, \bar \nu_{\ell})
-  {\cal \tilde{A}}_{\rm exc}(B_{u}^{-} \to  \ell \, \bar \ell \, \ell \, \bar \nu_{\ell}) \,,
\label{total amplitude for the 4-body decay with identical leptons}
\end{eqnarray}
where the relative minus sign evidently stems from the Fermi-Dirac statistic for leptons.
It is then straightforward to write down the desired differential decay width
for the exclusive  transition process   $B_{u}^{-} \to  \ell \, \bar \ell \, \ell \, \bar \nu_{\ell}$
\begin{eqnarray}
d \Gamma (B_{u}^{-} \to  \ell \, \bar \ell \, \ell \, \bar \nu_{\ell})
= \left( {1 \over 2} \right )  \, \frac{(2 \, \pi)^4}{2 \, m_B}  \,
\left [ \left |{\cal A}_{\rm dir} \right |^2 + |{\cal \tilde{A}}_{\rm exc}|^2
- 2  \, {\rm Re} \left ({\cal A}_{\rm dir}^{\ast}  \, {\cal \tilde{A}}_{\rm exc} \right ) \right ] \,
d \Phi_{4, \, \rm PS} \,,
\end{eqnarray}
where we have introduced the degeneracy factor $1/2$ to prevent the double counting of the identical particles
in the final state  and $d \Phi_{4, \, \rm PS}$ stands for an element of relativistically invariant four-body phase space
\begin{eqnarray}
d \Phi_{4, \, \rm PS} &=& \frac{\lambda^{1/2}(m_{B}^2, p^2, q^2)}{2^8  \,\, (2 \, \pi)^{10} \, m_{B}^2} \, \,\,
d p^2 \, d q^2 \, d \cos \theta_1 \, d \cos \theta_2 \, d \phi
\nonumber \\
&=& \frac{\lambda^{1/2}(m_{B}^2, \tilde{p}^2, \tilde{q}^2)}{2^8  \,\, (2 \, \pi)^{10} \, m_{B}^2} \, \,\,
d \tilde{p}^2 \, d \tilde{q}^2 \, d \cos \tilde{\theta}_1 \, d \cos \tilde{\theta}_2 \, d \tilde{\phi}  \,.
\end{eqnarray}
The explicit definitions of the two invariant masses $\tilde{p}^2$ and $\tilde{q}^2$
together with the three helicity angles  $\tilde{\theta}_1$, $\tilde{\theta}_2$ and $\tilde{\phi}$
bear resemblance to the  counterpart  kinematic variables without a tilde symbol
(see Appendix \ref{appendix: Kinematics} for more details).
The yielding full differential decay rate for the four-body leptonic
$B$-meson decay with identical lepton flavours can be expressed as
\begin{eqnarray}
\frac{d^5 \Gamma(B_{u}^{-} \to  \ell \, \bar \ell \, \ell \, \bar \nu_{\ell}) }
{d p^2 \, dq^2 \, d \cos \theta_1 \, d \cos \theta_2 \, d \phi}
=  \frac{G_F^2 \, \alpha_{\rm em}^2 \, |V_{ub}|^2}{2^{13} \, \pi^4} \, m_{B}^3  \,\,
\lambda^{1/2}(m_{B}^2, p^2, q^2) \,
\bigg \{ \frac{1}{ p^4} \,
{\cal J}(p^2, \, q^2, \,  \theta_1, \,  \theta_2, \,  \phi) \,
&& \nonumber \\
+ \,  \frac{1}{\tilde{p}^4} \,
{\cal J}(\tilde{p}^2, \, \tilde{q}^2, \,  \tilde{\theta}_1, \,  \tilde{\theta}_2, \,  \tilde{\phi})
- {2 \over p^2 \, \tilde{p}^2} \,
{\cal J}_{\rm int}(p^2, \,  \tilde{p}^2, \,  q^2, \,  \tilde{q}^2, \,
\theta_1, \, \tilde{\theta}_1, \,  \theta_2, \, \tilde{\theta}_2, \, \phi, \, \tilde{\phi})  \bigg \} \,,
&&  \hspace{1.0 cm}
\label{full dfferential decay distribution with identical leptons}
\end{eqnarray}
where the angular distribution ${\cal J}$ of our interest has been previously derived in
(\ref{angular distribution for non-identical leptons})
and the emerged interference term ${\cal J}_{\rm int}$ remains invariant
under the following transformation
\begin{eqnarray}
p^2 \leftrightarrow \tilde{p}^2 \,,  \qquad
q^2 \leftrightarrow \tilde{q}^2 \,,   \qquad
\theta_1  \leftrightarrow  \tilde{\theta}_1 \,, \qquad
\theta_2  \leftrightarrow  \tilde{\theta}_2 \,, \qquad
\phi \leftrightarrow  \tilde{\phi} \,.
\end{eqnarray}
We can readily identify the translation rules between  the two complete sets of variables
with and without a tilde for later convenience
\begin{eqnarray}
\tilde{p}^2  &=&  \left ( {1 \over 4} \right )  \, \bigg [ - \lambda^{1/2}(m_B^2, p^2, q^2) \,
\left (\cos \theta_1 - \cos \theta_2 \right )
+ (m_B^2 - p^2 - q^2) \, \left ( 1 -  \cos \theta_1 \cos \theta_2  \right )
\nonumber \\
&& \hspace{1.3 cm} + \, 2 \, \sqrt{p^2 \, q^2} \, |\sin \theta_1 \sin \theta_2| \, \cos \phi  \bigg  ]  \,,
\nonumber \\
\tilde{q}^2  &=&  \left ( {1 \over 4} \right )  \, \bigg [ \lambda^{1/2}(m_B^2, p^2, q^2) \,
\left (\cos \theta_1 - \cos \theta_2 \right )
+ (m_B^2 - p^2 - q^2) \, \left ( 1 -  \cos \theta_1 \cos \theta_2  \right )
\nonumber \\
&& \hspace{1.3 cm} + \, 2 \, \sqrt{p^2 \, q^2} \, |\sin \theta_1 \sin \theta_2| \, \cos \phi  \bigg  ]  \,,
\nonumber \\
\cos \tilde{\theta}_1 &=&  \frac{1} {\lambda^{1/2}(m_B^2, \tilde{p}^2, \tilde{q}^2)} \,
\left [ \frac{ \lambda^{1/2}(m_B^2, p^2, q^2)}{2}  \, \left (\cos \theta_1 + \cos \theta_2 \right )
- p^2 + q^2 \right ] \,,
\nonumber \\
\cos \tilde{\theta}_2 &=&  \frac{1} {\lambda^{1/2}(m_B^2, \tilde{p}^2, \tilde{q}^2)} \,
\left [ \frac{ \lambda^{1/2}(m_B^2, p^2, q^2)}{2}  \, \left (\cos \theta_1 + \cos \theta_2 \right )
+ p^2 - q^2 \right ] \,,
\nonumber \\
\sin \tilde{\phi}  &=& - \frac{\lambda^{1/2}(m_B^2, p^2, q^2)}{\lambda^{1/2}(m_B^2, \tilde{p}^2, \tilde{q}^2)} \,
\frac{\sqrt{p^2 \, q^2}}{\sqrt{\tilde{p}^2 \, \tilde{q}^2}} \,
\left | \frac{\sin \theta_1 \sin \theta_2}{\sin \tilde{\theta}_1 \sin \tilde{\theta}_2} \right |  \,
\sin \phi \,,
\nonumber \\
\cos \tilde{\phi}  &=&  \frac{2 \, (p^2 + q^2) - (m_B^2 -\tilde{p}^2 -\tilde{q}^2)
\, (1- \cos \tilde{\theta}_1 \, \cos \tilde{\theta}_2)}
{2 \, \sqrt{\tilde{p}^2 \, \tilde{q}^2} \, \left |\sin \tilde{\theta}_1 \sin \tilde{\theta}_2 \right | } \,,
\end{eqnarray}
which will further enable us to derive the analytical form of  ${\cal J}_{\rm int}$
in terms of the five independent variables $(p^2, q^2, \theta_1, \theta_2, \phi)$
collected in Appendix \ref{appendix: interference angular function}.

As already pointed out in  \cite{Beneke:2021rjf}, the two resultant four-momenta $p = p_1+p_2$
and $\tilde{p} = q_1+p_2$ in this case cannot be distinguished experimentally.
We will therefore  implement both kinematic cuts on  the  light-cone components
$n \cdot p$ and $n \cdot \tilde{p}$ simultaneously  for constructing the accessible
binned distributions in order to validate the established factorization formulae
for the transverse and longitudinal $B_{u}^{-} \to \gamma^{\ast} \,  W^{\ast}$ form factors.
Moreover, the precise correspondence  between the invariant mass of the off-shell photon
and the kinematic variable $p^2$ does not hold anymore for the case of identical lepton flavours.
Consequently, we propose to define the double-binned observables for the branching fraction
and the angular asymmetries with the necessary kinematic constraints
\begin{eqnarray}
\langle {{\cal BR}\, [t_1, \, t_2; \, \tilde{t}_1, \, \tilde{t}_2] } \rangle
&=& \tau_{B_u}  \, \int_{0}^{m_B^2} \, d p^2 \, \int_{0}^{(m_B - \sqrt{p^2})^2} \, d q^2 \,
\int_{-1}^{1} d \cos \theta_1 \, \int_{-1}^{1} d \cos \theta_2  \,
\int_{0}^{2 \pi} d \phi \,\,
\nonumber \\
&& \Theta_{\rm meas} (p^2, \tilde{p}^2, t_1, t_2, \tilde{t}_1, \tilde{t}_2) \,\,
\frac{d^5 \Gamma(B_{u}^{-} \to  \ell \, \bar \ell \, \ell \, \bar \nu_{\ell})}
{d p^2 \, dq^2 \, d \cos \theta_1 \, d \cos \theta_2 \, d \phi} \,,
\nonumber \\
\langle {{\cal A}_{c  \theta_1 + c  \tilde{\theta}_1} \, [t_1, \, t_2; \, \tilde{t}_1, \, \tilde{t}_2] } \rangle
&=&  \frac{\tau_{B_u} }  {2 \, \langle {{\cal BR}\, [t_1, \, t_2; \, \tilde{t}_1, \, \tilde{t}_2] } \rangle} \,\,
\int_{0}^{m_B^2} \, d p^2 \, \int_{0}^{(m_B - \sqrt{p^2})^2} \, d q^2 \,
\int_{-1}^{1} d \cos \theta_1 \,
\nonumber \\
&& \int_{-1}^{1} d \cos \theta_2  \, \int_{0}^{2 \pi} d \phi  \,
\left [ {\rm sgn} \left (\cos (\theta_1) \right) +
{\rm sgn} \left (\cos (\tilde{\theta}_1) \right)  \right ] \,
\nonumber \\
&&  \times \,
\Theta_{\rm meas} (p^2, \tilde{p}^2, t_1, t_2, \tilde{t}_1, \tilde{t}_2) \,\,
\frac{d^5 \Gamma(B_{u}^{-} \to  \ell \, \bar \ell \, \ell \, \bar \nu_{\ell})}
{d p^2 \, dq^2 \, d \cos \theta_1 \, d \cos \theta_2 \, d \phi}   \,,
\nonumber \\
\langle {{\cal A}_{c  2 \theta_1 + c  2 \tilde{\theta}_1} \, [t_1, \, t_2; \, \tilde{t}_1, \, \tilde{t}_2] } \rangle
&=&  \frac{\tau_{B_u} }  {2 \, \langle {{\cal BR}\, [t_1, \, t_2; \, \tilde{t}_1, \, \tilde{t}_2] } \rangle} \,\,
\int_{0}^{m_B^2} \, d p^2 \, \int_{0}^{(m_B - \sqrt{p^2})^2} \, d q^2 \,
\int_{-1}^{1} d \cos \theta_1 \,\,
\nonumber \\
&& \int_{-1}^{1} d \cos \theta_2  \, \int_{0}^{2 \pi} d \phi  \,
\left [ {\rm sgn} \left (\cos (2 \theta_1) \right) +
{\rm sgn} \left (\cos (2 \tilde{\theta}_1) \right)  \right ] \,
\nonumber \\
&&  \times \,  \Theta_{\rm meas} (p^2, \tilde{p}^2, t_1, t_2, \tilde{t}_1, \tilde{t}_2) \,\,
\frac{d^5 \Gamma(B_{u}^{-} \to  \ell \, \bar \ell \, \ell \, \bar \nu_{\ell})}
{d p^2 \, dq^2 \, d \cos \theta_1 \, d \cos \theta_2 \, d \phi}  \,,
\nonumber \\
\langle {{\cal A}_{c  \theta_2 + c  \tilde{\theta}_2} \, [t_1, \, t_2; \, \tilde{t}_1, \, \tilde{t}_2] } \rangle
&=&  \frac{\tau_{B_u} }  {2 \, \langle {{\cal BR}\, [t_1, \, t_2; \, \tilde{t}_1, \, \tilde{t}_2] } \rangle} \,\,
\int_{0}^{m_B^2} \, d p^2 \, \int_{0}^{(m_B - \sqrt{p^2})^2} \, d q^2 \,
\int_{-1}^{1} d \cos \theta_1 \,\,
\nonumber \\
&& \int_{-1}^{1} d \cos \theta_2  \, \int_{0}^{2 \pi} d \phi  \,
 \left [ {\rm sgn} \left (\cos (\theta_2) \right) +
{\rm sgn} \left (\cos (\tilde{\theta}_2) \right)  \right ]
\nonumber \\
&&  \times \,  \Theta_{\rm meas} (p^2, \tilde{p}^2, t_1, t_2, \tilde{t}_1, \tilde{t}_2) \,\,
\frac{d^5 \Gamma(B_{u}^{-} \to  \ell \, \bar \ell \, \ell \, \bar \nu_{\ell})}
{d p^2 \, dq^2 \, d \cos \theta_1 \, d \cos \theta_2 \, d \phi}  \,,
\nonumber \\
\langle {{\cal A}_{c  \theta_1, c  \tilde{\theta}_1} \, [t_1, \, t_2; \, \tilde{t}_1, \, \tilde{t}_2] } \rangle
&=&  \frac{\tau_{B_u} }  {\langle {{\cal BR}\, [t_1, \, t_2; \, \tilde{t}_1, \, \tilde{t}_2] } \rangle} \,\,
\int_{0}^{m_B^2} \, d p^2 \, \int_{0}^{(m_B - \sqrt{p^2})^2} \, d q^2 \,
\int_{-1}^{1} d \cos \theta_1 \,\,
\nonumber \\
&& \int_{-1}^{1} d \cos \theta_2  \, \int_{0}^{2 \pi} d \phi  \,
 \left [ {\rm sgn} \left (\cos (\theta_1) \right)  \,
{\rm sgn} \left (\cos (\tilde{\theta}_1) \right)  \right ]
\nonumber \\
&&  \times \,  \Theta_{\rm meas} (p^2, \tilde{p}^2, t_1, t_2, \tilde{t}_1, \tilde{t}_2) \,\,
\frac{d^5 \Gamma(B_{u}^{-} \to  \ell \, \bar \ell \, \ell \, \bar \nu_{\ell})}
{d p^2 \, dq^2 \, d \cos \theta_1 \, d \cos \theta_2 \, d \phi} \,,
\nonumber \\
\langle {{\cal A}_{c  2 \theta_1, c  2 \tilde{\theta}_1} \, [t_1, \, t_2; \, \tilde{t}_1, \, \tilde{t}_2] } \rangle
&=&  \frac{\tau_{B_u} }  {\langle {{\cal BR}\, [t_1, \, t_2; \, \tilde{t}_1, \, \tilde{t}_2] } \rangle} \,\,
\int_{0}^{m_B^2} \, d p^2 \, \int_{0}^{(m_B - \sqrt{p^2})^2} \, d q^2 \,
\int_{-1}^{1} d \cos \theta_1 \,\,
\nonumber \\
&& \int_{-1}^{1} d \cos \theta_2  \, \int_{0}^{2 \pi} d \phi  \,
\left [ {\rm sgn} \left (\cos (2 \theta_1) \right)  \,
{\rm sgn} \left (\cos (2 \tilde{\theta}_1) \right)  \right ] \,
\nonumber \\
&&  \times \,  \Theta_{\rm meas} (p^2, \tilde{p}^2, t_1, t_2, \tilde{t}_1, \tilde{t}_2) \,\,
\frac{d^5 \Gamma(B_{u}^{-} \to  \ell \, \bar \ell \, \ell \, \bar \nu_{\ell})}
{d p^2 \, dq^2 \, d \cos \theta_1 \, d \cos \theta_2 \, d \phi} \,,
\nonumber \\
\langle {{\cal A}_{c  \theta_2, c  \tilde{\theta}_2} \, [t_1, \, t_2; \, \tilde{t}_1, \, \tilde{t}_2] } \rangle
&=&  \frac{\tau_{B_u} }  {\langle {{\cal BR}\, [t_1, \, t_2; \, \tilde{t}_1, \, \tilde{t}_2] } \rangle} \,\,
\int_{0}^{m_B^2} \, d p^2 \, \int_{0}^{(m_B - \sqrt{p^2})^2} \, d q^2 \,
\int_{-1}^{1} d \cos \theta_1 \,\,
\nonumber \\
&& \int_{-1}^{1} d \cos \theta_2  \, \int_{0}^{2 \pi} d \phi  \,  \left [ {\rm sgn} \left (\cos (\theta_2) \right)  \,
{\rm sgn} \left (\cos (\tilde{\theta}_2) \right)  \right ] \,
\nonumber \\
&&  \times \,  \Theta_{\rm meas} (p^2, \tilde{p}^2, t_1, t_2, \tilde{t}_1, \tilde{t}_2) \,\,
\frac{d^5 \Gamma(B_{u}^{-} \to  \ell \, \bar \ell \, \ell \, \bar \nu_{\ell})}
{d p^2 \, dq^2 \, d \cos \theta_1 \, d \cos \theta_2 \, d \phi} \,,
\label{observables for the 4-body decays with identical leptons}
\end{eqnarray}
where the newly defined measurement function is explicitly given by
\begin{eqnarray}
\Theta_{\rm meas}
&=& \bigg [ \theta(p^2 - t_1) \,  \theta(t_2 - p^2) \, \theta(\tilde{p}^2 - \tilde{t}_1) \, \theta(\tilde{t}_2 - \tilde{p}^2)
+ \, \theta(p^2 - \tilde{t}_1) \,  \theta(\tilde{t}_2 - p^2) \, \theta(\tilde{p}^2 - t_1) \, \theta(t_2 - \tilde{p}^2)
\nonumber \\
&& \hspace{0.3 cm} -  \, \theta(p^2 - t_{1,  \rm max}) \,  \theta(t_{2,  \rm min} - p^2) \,
\theta(\tilde{p}^2 - t_{1,  \rm max}) \,  \theta(t_{2,  \rm min} - \tilde{p}^2) \bigg ]
\nonumber \\
&& \times \, \bigg  [  \theta\left ( n \cdot p - 3 \, {\rm GeV} \right ) \,
\theta\left ( n \cdot \tilde{p} - 3 \, {\rm GeV} \right ) \bigg ] \,,
\nonumber \\
t_{1,  \rm max} &=& {\rm max} \left \{t_1,  \, \tilde{t}_1  \right \},
\qquad  t_{2,  \rm min}= {\rm min} \left \{t_2,  \, \tilde{t}_2  \right \}.
\end{eqnarray}
It is important to remark that our definitions for  the exclusive four-body
$B_{u}^{-} \to  \ell \, \bar \ell \, \ell \, \bar \nu_{\ell}$ decay observables
differ from the previous strategies suggested in \cite{Beneke:2021rjf}
on account of executing the essential kinematic cuts distinctly.
As displayed in Table \ref{table for the binned observables with identical leptons},
the yielding  double-binned branching fraction in the kinematic domain
$\{ p^2 \,, \tilde{p}^2 \} \in \left [1.5 \, {\rm GeV^2}, m_B^2 \right ]$
is predicted to be $(1.15^{+0.27}_{-0.49}) \times 10^{-9}$,
which lies well below the upper limit $1.6 \times 10^{-8}$
for ${\cal BR} (B_{u}^{-} \to  \mu^{+} \, \mu^{-} \, \mu^{-} \, \bar \nu_{\mu})$
reported by the LHCb Collaboration  with the lowest of the two $\mu^{+} \, \mu^{-}$
invariant masses below $0.96 \, {\rm GeV}^2$ \cite{LHCb:2018jvy}.
Despite of the different implementations of kinematic constraints,
our result for the partially integrated decay rate in the hard-collinear interval
$\{ p^2 \,, \tilde{p}^2 \} \in \left [1.5 \, {\rm GeV^2}, 2.0 \, {\rm GeV^2} \right ]$
turns out to be  comparable with the numerical value achieved in \cite{Beneke:2021rjf}.
In particular, the subleading power corrections to the binned branching fractions
of  $\{ p^2 \,, \tilde{p}^2 \} \in [1.5 \, {\rm GeV^2}, \,  2.0 \, {\rm GeV^2}]$
and $\{ p^2 \,, \tilde{p}^2 \} \in [1.5 \, {\rm GeV^2}, \,  4.0 \, {\rm GeV^2}]$
tend to enhance the counterpart LP predictions significantly.
Comparing the numerical predictions for $\langle {\cal BR} \rangle$  collected in Tables
\ref{table for the binned observables with non-identical leptons}
and \ref{table for the binned observables with identical leptons}
further leads us to conclude that the coherent interference of the direct and exchange
amplitudes merely generates the minor corrections to the double-binned
branching  fractions  of  $B_{u}^{-} \to  \ell \, \bar \ell \, \ell \, \bar \nu_{\ell}$
for all the hard-collinear $\{ p^2 \,, \tilde{p}^2 \}$-bins.
Furthermore, our numerical computation indicates that only three of the six asymmetry observables
$\langle {{\cal A}_{c  2 \theta_1 + c 2 \tilde{\theta}_1}  } \rangle$,
$\langle {{\cal A}_{c  \theta_1, \,  c  \tilde{\theta}_1}  } \rangle$
and $\langle {{\cal A}_{c  \theta_2, \,  c  \tilde{\theta}_2}  } \rangle$
shown in Table \ref{table for the binned observables with identical leptons}
can reach ${\cal O} (50 \, \%)$ in magnitudes and the yielding theory predictions
suffer from the relatively lower  uncertainties than those for $\langle {\cal BR} \rangle$,
mainly due to the anticipated cancellation of the non-perturbative uncertainties
from the HQET distribution amplitudes and of the perturbative uncertainties
from the residual factorization/renormalization scale dependencies.

\begin{table}
\centering
\renewcommand{\arraystretch}{2.0}
\resizebox{\columnwidth}{!}{
\begin{tabular}{|c||c|c|c||cccccc|}
\hline
\hline
\multirow{2}{*}{Observables}& \multirow{2}{*}{$[t_1, \,\, t_2]$ } & \multirow{2}{*}{LP} & \multirow{2}{*}{Total} & \multicolumn{6}{c|}{Uncertainties} \\
& $( {\rm GeV}^2 )$ & {(\rm NLL)}  & {(\rm LP + NLP)}  & $\mu$ &$\mu_{h1}$
& $\widehat{\sigma}_{B_u}^{(1, \, 2)}$ & $\lambda_{B_u}$  & $p^2_{\rm cut}$ & $|V_{ub}|$ \\
\hline
\multirow{4}{*}{$\langle {{\cal BR}\, [t_1, \, t_2; \, t_1, \, t_2] } \rangle \times 10^9$}& $\left.\text{[1.5, }m_B^2\right]$ & $0.76_{-0.33}^{+0.22}$ & $1.15_{-0.49}^{+0.27}$ & $\text{}_{-0.21}^{+0.05}$ & $\text{}_{-0.18}^{+0.07}$ & $\text{}_{-0.22}^{+0.19}$ & $\text{}_{-0.30}^{+0.12}$ & $\text{}_{-0.02}^{+0.05}$ & $\text{}_{-0.10}^{+0.10}$ \\
& $\left.\text{[2.0, }m_B^2\right]$ & $0.47_{-0.16}^{+0.16}$ & $0.74_{-0.26}^{+0.12}$ & $\text{}_{-0.12}^{+0.04}$ & $\text{}_{-0.10}^{+0.04}$ & $\text{}_{-0.10}^{+0.03}$ & $\text{}_{-0.16}^{+0.06}$ & $\text{}_{-0.02}^{+0.05}$ & $\text{}_{-0.07}^{+0.07}$ \\
& $\left.\text{[3.0, }m_B^2\right]$ & $0.23_{-0.08}^{+0.13}$ & $0.32_{-0.06}^{+0.06}$ & $\text{}_{-0.03}^{+0.01}$ & $\text{}_{-0.02}^{+0.01}$ & $\text{}_{-0.02}^{+0.00}$ & $\text{}_{-0.03}^{+0.00}$ & $\text{}_{-0.02}^{+0.05}$ & $\text{}_{-0.03}^{+0.03}$ \\
& $\left.\text{[4.0, }m_B^2\right]$ & $0.15_{-0.06}^{+0.02}$ & $0.15_{-0.03}^{+0.02}$ & $\text{}_{-0.00}^{+0.00}$ & $\text{}_{-0.00}^{+0.00}$ & $\text{}_{-0.00}^{+0.00}$ & $\text{}_{-0.00}^{+0.00}$ & $\text{}_{-0.02}^{+0.00}$ & $\text{}_{-0.02}^{+0.02}$ \\
\hline
\multirow{4}{*}{$\langle { {\cal A}_{c  2 \theta_1 + c 2 \tilde{\theta}_1} \, [t_1, \, t_2; \, t_1, \, t_2] } \rangle $}
& $\left.\text{[1.5, }m_B^2\right]$ & $-0.54^{+0.03}_{-0.09}$ & $-0.60^{+0.03}_{-0.05}$ & ${}^{+0.00}_{-0.02}$
& ${}^{+0.00}_{-0.01}$  & ${}^{+0.03}_{-0.02}$  & ${}^{+0.01}_{-0.03}$  & ${}^{+0.00}_{-0.01}$  & $-$ \\
& $\left.\text{[2.0, }m_B^2\right]$ & $-0.64^{+0.04}_{-0.05}$  & $-0.67^{+0.02}_{-0.04}$ & ${}^{+0.00}_{-0.01}$
& ${}^{+0.00}_{-0.01}$ & ${}^{+0.01}_{-0.02}$ & ${}^{+0.01}_{-0.03}$ & ${}^{+0.00}_{-0.01}$ & $-$ \\
& $\left.\text{[3.0, }m_B^2\right]$ & $-0.70^{+0.02}_{-0.05}$ & $-0.72^{+0.03}_{-0.01}$ & ${}^{+0.01}_{-0.01}$
& ${}^{+0.01}_{-0.00}$  & ${}^{+0.01}_{-0.00}$  & ${}^{+0.01}_{-0.00}$ & ${}^{+0.02}_{-0.00}$ & $-$ \\
& $\left.\text{[4.0, }m_B^2\right]$ & $-0.69^{+0.03}_{-0.02}$ & $-0.70^{+0.03}_{-0.00}$ & ${}^{+0.00}_{-0.00}$
& ${}^{+0.01}_{-0.00}$ & ${}^{+0.00}_{-0.00}$ & ${}^{+0.00}_{-0.00}$ & ${}^{+0.03}_{-0.00}$ & $-$ \\
\hline
\multirow{4}{*}{$\langle { {\cal A}_{c  \theta_1, \,  c  \tilde{\theta}_1} \, [t_1, \, t_2; \, t_1, \, t_2] } \rangle $}
& $\left.\text{[1.5, }m_B^2\right]$ & $0.25^{+0.09}_{-0.05}$  & $0.29^{+0.05}_{-0.04}$ & ${}^{+0.03}_{-0.01}$
& ${}^{+0.02}_{-0.01}$ & ${}^{+0.01}_{-0.03}$ & ${}^{+0.03}_{-0.02}$ & ${}^{+0.02}_{-0.00}$ & $-$ \\
& $\left.\text{[2.0, }m_B^2\right]$ & $0.31^{+0.07}_{-0.06}$ & $0.33^{+0.04}_{-0.06}$ & ${}^{+0.02}_{-0.01}$
& ${}^{+0.02}_{-0.01}$ & ${}^{+0.00}_{-0.01}$ & ${}^{+0.02}_{-0.01}$ & ${}^{+0.02}_{-0.00}$ & $-$ \\
& $\left.\text{[3.0, }m_B^2\right]$ & $0.42^{+0.03}_{-0.04}$ & $0.40^{+0.04}_{-0.03}$ & ${}^{+0.01}_{-0.01}$
& ${}^{+0.01}_{-0.01}$ & ${}^{+0.01}_{-0.00}$ & ${}^{+0.02}_{-0.00}$ & ${}^{+0.03}_{-0.02}$ & $-$ \\
& $\left.\text{[4.0, }m_B^2\right]$ & $0.50^{+0.02}_{-0.01}$ & $0.50^{+0.02}_{-0.01}$ & ${}^{+0.00}_{-0.00}$
& ${}^{+0.01}_{-0.00}$ & ${}^{+0.00}_{-0.00}$ & ${}^{+0.00}_{-0.00}$ & ${}^{+0.00}_{-0.01}$ & $-$ \\
\hline
\multirow{4}{*}{$\langle { {\cal A}_{c  \theta_2, \,  c  \tilde{\theta}_2} \, [t_1, \, t_2; \, t_1, \, t_2] } \rangle $}
& $\left.\text{[1.5, }m_B^2\right]$ & $0.24^{+0.10}_{-0.05}$ & $0.27^{+0.09}_{-0.02}$ & ${}^{+0.03}_{-0.00}$
& ${}^{+0.03}_{-0.00}$ & ${}^{+0.04}_{-0.01}$ & ${}^{+0.06}_{-0.01}$ & ${}^{+0.03}_{-0.02}$ & $-$ \\
& $\left.\text{[2.0, }m_B^2\right]$ & $0.32^{+0.06}_{-0.10}$ & $0.31^{+0.08}_{-0.02}$ & ${}^{+0.04}_{-0.00}$
& ${}^{+0.04}_{-0.00}$  & ${}^{+0.03}_{-0.00}$  & ${}^{+0.03}_{-0.00}$  & ${}^{+0.03}_{-0.02}$  & $-$ \\
& $\left.\text{[3.0, }m_B^2\right]$ & $0.44^{+0.04}_{-0.03}$ & $0.42^{+0.03}_{-0.02}$ & ${}^{+0.02}_{-0.01}$
& ${}^{+0.01}_{-0.01}$ & ${}^{+0.00}_{-0.00}$ & ${}^{+0.01}_{-0.00}$ & ${}^{+0.02}_{-0.02}$ & $-$ \\
& $\left.\text{[4.0, }m_B^2\right]$ & $0.52^{+0.04}_{-0.00}$ & $0.52^{+0.00}_{-0.02}$ & ${}^{+0.00}_{-0.00}$
& ${}^{+0.00}_{-0.01}$ & ${}^{+0.00}_{-0.00}$ & ${}^{+0.00}_{-0.00}$ & ${}^{+0.00}_{-0.00}$ & $-$ \\
\hline
\hline
\end{tabular}
}
\renewcommand{\arraystretch}{1.0}
\caption{Theory predictions for the binned distributions of the branching fraction
$\langle {\cal BR} \rangle$
as well as the three angular asymmetries
$\langle {{\cal A}_{c  2 \theta_1 + c 2 \tilde{\theta}_1}  } \rangle$,
$\langle {{\cal A}_{c  \theta_1, \,  c  \tilde{\theta}_1}  } \rangle$
and $\langle {{\cal A}_{c  \theta_2, \,  c  \tilde{\theta}_2}  } \rangle$
for the four-body leptonic $B$-meson decays with identical lepton flavours,
where the numerically sizeable uncertainties from varying distinct input parameters
are further displayed for completeness.}
\label{table for the binned observables with identical leptons}
\end{table}

\section{Summary and conclusions}
\label{section:summary}

In this paper we have performed the improved  QCD calculations of the exclusive radiative
$B_{u}^{-} \to \gamma^{\ast} \, \ell \, \bar \nu_{\ell}$  form factors with an off-shell photon
carrying either the hard-collinear momentum $p_{\mu} \sim m_b \, (1, \lambda^2, \lambda)$
or the hard momentum $p_{\mu} \sim m_b \, (1, 1, 1)$ by taking advantage of the modern SCET factorization program
and the traditional local OPE technique, respectively.
Applying further the renormalized jet functions in the factorized expressions of
the analogous $B$-meson-to-vacuum correlations for constructing the light-cone sum rules
of the semileptonic $B \to V$ transition form factors \cite{Gao:2019lta}
as well as  the hadronic photon correction to the on-shell $B \to \gamma$ form factors \cite{Wang:2018wfj},
we constructed explicitly the soft-collinear factorization formulae for both the transverse and longitudinal
$B \to \gamma^{\ast} \, \ell \, \bar \nu_{\ell}$ form factors at the NLL accuracy in the hard-collinear $p^2$-region
with the aid of the Ward-Takahashi identities due to the ${\rm U_{em}} (1)$ gauge symmetry
and  the standard momentum-space RG formalism.
Subsequently, we evaluated the distinct subleading-power contributions to the generalized
$B \to \gamma^{\ast} \, W^{\ast}$ form factors at $p^2 \sim {\cal O}(m_b \, \Lambda_{\rm QCD})$
from expanding the hard-collinear quark propagator beyond the LP approximation,
from the two-particle and three-particle higher-twist HQET distribution amplitudes,
from the ``kinematic" power corrections suppressed by the small (but non-vanishing)
component of the virtual photon momentum, and from the energetic photon radiation
off the bottom quark at LO in the strong coupling constant.
The yielding HQET factorization formulae of the non-hadronic $B \to \gamma^{\ast}$ form factors
were then derived in the NLL approximation for $p^2 \sim {\cal O}(m_b^2)$
with the effective decay constant $\tilde{f}_B$ encoding the non-perturbative dynamics
of the composite bottom-meson system.

Having at our disposal the desired expressions for the exclusive
$B \to \gamma^{\ast} \, \ell \, \bar \nu_{\ell}$ decay form factors,
we proceeded to explore their numerical implications with the three-parameter ansatz
for the essential $B$-meson distribution amplitudes
whose RG evolution behaviours can be determined analytically in terms of hypergeometric functions
at one-loop order \cite{Beneke:2018wjp}.
It has been observed that the resulting non-local power corrections from the subleading terms
in the expanded hard-collinear quark  propagator can shift the counterpart LP predictions
by an amount of approximately $(20-30) \%$  in magnitudes in the kinematic domain
$p^2 \in [1.5, \, 4.0] \, {\rm GeV^2}$.
In particular, the predicted LP contribution to the vector form factor $|F_V|$
appeared to develop the yet stronger sensitivity on the inverse moment $\lambda_{B_u}$
in comparison with the obtained longitudinal form factor
$\left |\lambda \, \left ( \hat{F}_1 + {v \cdot p \over m_B} \, \hat{F}_3 \right ) \right |$,
which can be traced back to the quite distinct asymptotic behaviours of
the two HQET distribution amplitudes $\phi_{B}^{\pm}(\omega, \mu)$
at small partonic momentum $\omega$.
Unsurprisingly, the generalized radiative $B \to \gamma^{\ast} \, W^{\ast}$ form factors
dependent on the invariant masses of both $\gamma^{\ast}$ and $W^{\ast}$
turned out to be rather sensitive to the precise shape of the $B$-meson distribution amplitudes.
By contrast, the resulting theory predictions for such non-hadronic decay form factors
in the hard $p^2$-regime suffered from the enormously reduced uncertainties
due to the apparent independence of the non-perturbative  light-ray distributions.
We then turned to investigate systematically the  angular observables
for the four-body leptonic decays $B_{u}^{-} \to  \ell^{\prime} \, \bar \ell^{\prime} \, \ell \, \bar \nu_{\ell}$
with non-identical lepton flavours $\ell \neq \ell^{\prime}$
and (more complicated) identical ones $\ell = \ell^{\prime}$.
Our numerical results indicated that the dominant contribution to the  branching fraction
of $B_{u}^{-} \to  \ell^{\prime} \, \bar \ell^{\prime} \, \ell \, \bar \nu_{\ell}$
indeed arises from the hard-collinear $p^2$-region rather than from the kinematic regime
of $p_{\mu} \sim m_b \, (1, 1, 1)$ as anticipated by the power-counting analysis.
It is also interesting to remark that the newly computed subleading power contributions
to the binned branching fractions resulted in  the sizeable enhancements of
the corresponding LP computations based upon the SCET factorization approach.
Additionally, the two promising asymmetry observables
$\langle {{\cal A}_{c 2 \theta_1} \, [t_1 \, t_2] } \rangle$
and  $\langle {{\cal A}_{c  \theta_2} \, [t_1 \, t_2] } \rangle$
for  non-identical lepton flavours have been predicted to be as large as
${\cal O} (20 - 50) \%$  for all the selected $p^2$-bins  with the substantially improved precision.
Furthermore, we quote our theory prediction for the double-binned branching fraction  of
$B_{u}^{-} \to  \ell \, \bar \ell  \, \ell \, \bar \nu_{\ell}$
as $(1.15^{+0.27}_{-0.49}) \times 10^{-9}$
in the kinematic interval $\{ p^2 \,, \tilde{p}^2 \} \in \left [1.5 \, {\rm GeV^2}, m_B^2 \right ]$
with the additional cuts on the large light-cone components of the two indistinguishable lepton-pair momenta
$n \cdot p \geq 3 \, {\rm GeV}$ and $n \cdot \tilde{p} \geq 3 \, {\rm GeV}$.

Confronting our theory predictions for the binned decay rates and  the angular asymmetries
with the upcoming  dedicated measurements  at the LHCb and Belle II experiments
will be evidently beneficial for advancing our knowledge of the poorly constrained $B$-meson
distribution amplitudes in a complementary manner to the previous determination
from the exclusive radiative $B \to \gamma \ell \bar \nu_{\ell}$ decays.
In this respect, it will be in high demand to carry out a global fit of all the key exclusive channels
including further a variety of semileptonic and nonleptonic bottom-meson decays
for the sake of obtaining the more stringent constraints on the fundamental HQET distribution amplitudes
(see also  \cite{Wang:2019msf} for an alternative strategy).
Further theoretical  investigations of  the non-hadronic $B \to \gamma^{\ast} \, W^{\ast}$ form factors
can be also pursued by constructing the soft-collinear factorization formulae for their subleading power
contributions systematically with the ${\rm QCD} \rightarrow \rm{SCET_I} \rightarrow \rm{SCET_{II}}$
matching procedure \cite{Beneke:2003pa}.

\subsection*{Acknowledgements}

We are grateful to Alexander Khodjamirian for illuminating discussions.
C.W is supported in part by the National Natural Science Foundation of China
with Grant No. 12105112 and  the Natural Science Foundation of
Jiangsu Education Committee with Grant No. 21KJB140027.
Y.M.W acknowledges support from
the  National Natural Science Foundation of China  with
Grant No. 11735010 and 12075125, and  the Natural Science Foundation of Tianjin
with Grant No. 19JCJQJC61100.
Y.B.W is supported in part by the Alexander-von-Humboldt Stiftung.


\appendix

\section{Kinematics  for the four-body  decays $B_{u}^{-} \to  \ell^{\prime} \, \bar \ell^{\prime} \, \ell \, \bar \nu_{\ell}$}
\label{appendix: Kinematics}

\begin{figure}
\begin{center}
\includegraphics[width=0.75  \columnwidth]{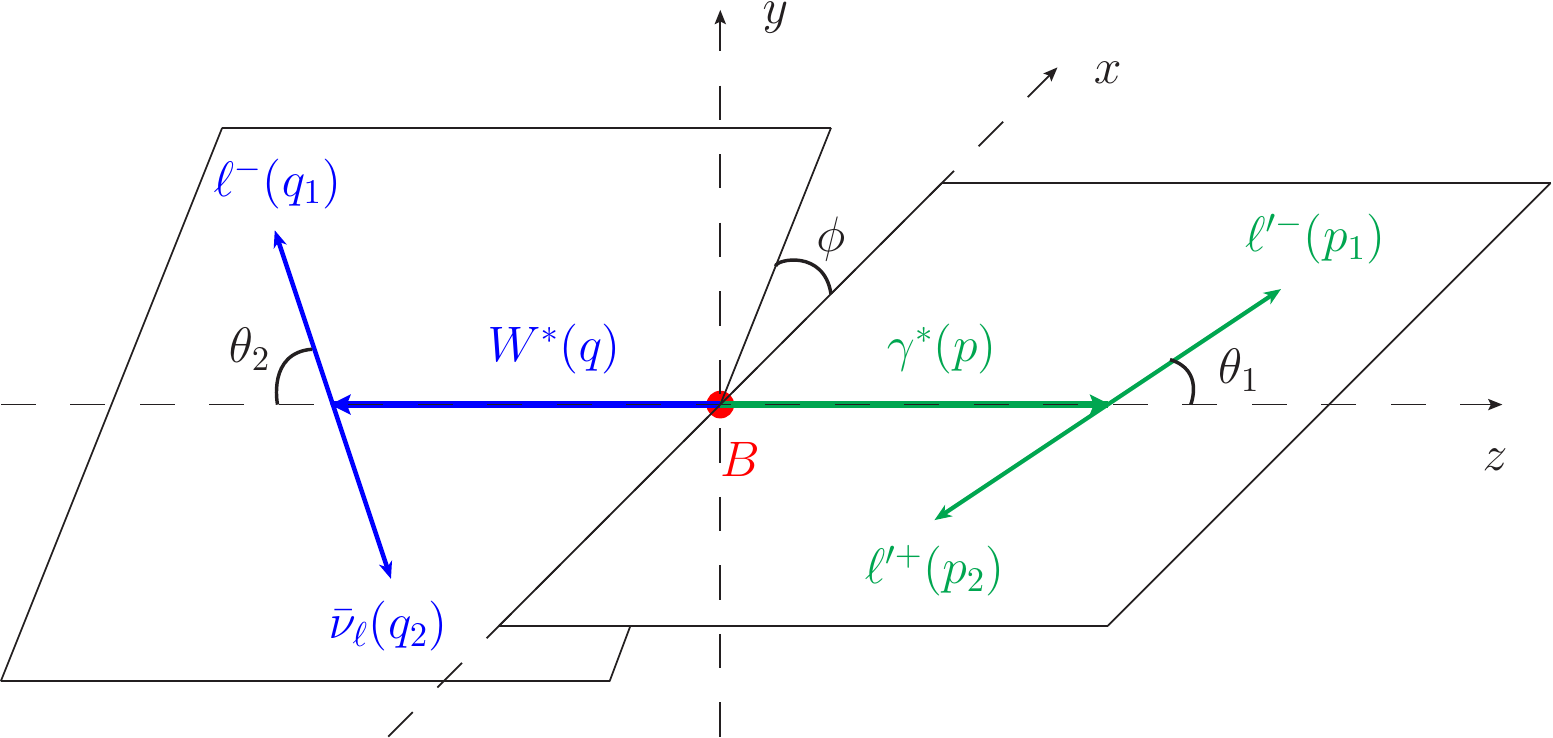}
\vspace*{0.2 cm}
\caption{Kinematics of the four-body leptonic decay
$B_{u}^{-} \to  \ell^{\prime} \, \bar \ell^{\prime} \, \ell \, \bar \nu_{\ell}$.}
\label{fig: kinematics-four-body-decays}
\end{center}
\end{figure}

In this appendix, we will begin with the essential kinematics 
for the four-body leptonic decays
$B_{u}^{-} (p_B) \to \gamma^{\ast}(p) (\to \ell^{\prime}(p_1) \,\, \bar \ell^{\prime}(p_2) )\,
W^{\ast}(q) (\to \ell(q_1) \, \bar \nu_{\ell}(q_2))$
with non-identical lepton flavours $\ell^{\prime} \neq \ell$.
Following the discussion for the exclusive $B \to K^{\ast} (\to K \pi) \, \ell \, \bar \ell$ decays
\cite{Altmannshofer:2008dz}, it is customary and convenient to express the full differential decay distribution
of  $B_{u}^{-} \to  \ell^{\prime} \, \bar \ell^{\prime} \, \ell \, \bar \nu_{\ell}$
in terms of the five kinematic variables:
the two invariant masses  $p^2$ and $q^2$, the helicity angles $\theta_1$ and $\theta_2$,
as well as the azimuthal angle $\phi$ (see Figure \ref{fig: kinematics-four-body-decays}).

More explicitly, we choose the $z$-axis along the flight direction of the  off-shell photon momentum $\bf{p}$
in the $B$-meson rest frame. The angle $\theta_1$ is then defined as the angle between the $\ell^{\prime -}$
direction of flight and the $z$-axis in the dilepton rest frame.
As a consequence, the leptonic momenta in the dilepton rest frame ($2 \, \ell^{\prime}$-RF)
in the massless limit are given by
\begin{eqnarray}
p_1 |_{2 \, \ell^{\prime}-{\rm RF}} &=& {\sqrt{p^2} \over 2} \, (1, \, \sin \theta_1, \, 0, \cos \theta_1) \,,
\nonumber \\
p_2 |_{2 \, \ell^{\prime}-{\rm RF}} &=& {\sqrt{p^2} \over 2} \, (1, \, - \sin \theta_1, \, 0, - \cos \theta_1) \,.
\end{eqnarray}
In addition, $\theta_2$ is the angle between the $\ell^{-}$-momentum $\bf{q_1}$
and the negative $z$ direction in the $W^{\ast}$-boson rest frame.
The azimuthal angle $\phi$ is defined by the relative angle
between the decay plane of the $\ell^{\prime} \bar {\ell^{\prime}}$  system
and the $\ell \, \bar \nu_{\ell}$ decay plane.
Accordingly,  the two momenta $q_1$ and $q_2$ in the  $\ell \, \bar \nu_{\ell}$ rest frame
($\ell \, \bar \nu_{\ell}$-RF) can be written as
\begin{eqnarray}
q_1 |_{\ell \, \bar \nu_{\ell}-{\rm RF}} &=& {\sqrt{q^2} \over 2} \,
(1, \, \sin \theta_2 \, \cos \phi, \, \sin \theta_2 \, \sin \phi, \,  - \cos \theta_2) \,,
\nonumber \\
q_2 |_{\ell \, \bar \nu_{\ell}-{\rm RF}} &=& {\sqrt{q^2} \over 2} \,
(1, \, -\sin \theta_2 \, \cos \phi, \, - \sin \theta_2 \, \sin \phi, \, \cos \theta_2)\,.
\end{eqnarray}

Now we proceed to investigate the more complicated kinematics for the exclusive four-body
bottom-meson decays   $B_{u}^{-} \to  \ell^{\prime} \, \bar \ell^{\prime} \, \ell \, \bar \nu_{\ell}$
with identical lepton flavours $\ell^{\prime} = \ell$.
In order to evaluate the so-called exchange amplitude ${\cal \tilde{A}}_{\rm exc}$ displayed in
(\ref{total amplitude for the 4-body decay with identical leptons}),
it proves more convenient to introduce further the two invariant masses in the following
\begin{eqnarray}
\tilde{p}^2 = (q_1 + p_2)^2 \,,  \qquad  \tilde{q}^2 = (p_1 + q_2)^2  \,,
\end{eqnarray}
as well as the three alternative helicity angles $\tilde{\theta}_1$, $\tilde{\theta}_2$ and $\tilde{\phi}$
such that
\begin{eqnarray}
q_1 |_{\ell \, \bar \ell^{\prime}-{\rm RF}} &=& {\sqrt{\tilde{p}^2} \over 2} \,
(1, \, \sin \tilde{\theta}_1, \, 0, \cos \tilde{\theta}_1) \,,
\nonumber \\
p_2 |_{\ell \, \bar \ell^{\prime}-{\rm RF}} &=& {\sqrt{\tilde{p}^2} \over 2} \,
(1, \, - \sin \tilde{\theta}_1, \, 0, - \cos \tilde{\theta}_1) \,,
\end{eqnarray}
correspond to the four-momenta of the  final-state leptons from the cascade decay process
$\gamma^{\ast}(\tilde{p}) \to  \ell (q_1) \, \bar \ell^{\prime}(p_2)$
in the dilepton rest frame ($\ell \, \bar \ell^{\prime}$-RF),
and
\begin{eqnarray}
p_1 |_{\ell^{\prime} \, \bar \nu_{\ell}-{\rm RF}} &=& {\sqrt{\tilde{q}^2} \over 2} \,
(1, \, \sin \tilde{\theta}_2 \, \cos \tilde{\phi}, \, \sin \tilde{\theta}_2 \, \sin \tilde{\phi}, \,  - \cos \tilde{\theta}_2) \,,
\nonumber \\
q_2 |_{\ell^{\prime} \, \bar \nu_{\ell}-{\rm RF}} &=& {\sqrt{\tilde{q}^2} \over 2} \,
(1, \, -\sin \tilde{\theta}_2 \, \cos \tilde{\phi}, \, - \sin \tilde{\theta}_2 \, \sin \tilde{\phi}, \, \cos \tilde{\theta}_2)\,,
\end{eqnarray}
coincide with the four-momenta of the final-state particles produced from the cascade  weak transition
$W^{\ast}(\tilde{q}) \to  \ell^{\prime} (p_1) \, \bar \nu_\ell(q_2)$
in the $\ell^{\prime} \, \bar \nu_\ell$ rest frame ($\ell^{\prime} \, \bar \nu_\ell$-RF), respectively.

\section{Explicit expressions for the angular function ${\cal J}_{\rm int}$}
\label{appendix: interference angular function}

Here we will present the detailed expressions for ${\cal J}_{\rm int}$ entering the differential decay distribution
of the exclusive four-body bottom-meson decay with identical lepton flavours
(\ref{full dfferential decay distribution with identical leptons})
due to  the coherent interference  of the direct and exchange amplitudes.
By analogy with the result (\ref{angular distribution for non-identical leptons})
for the non-hadronic process
$B_{u}^{-} \to  \ell^{\prime} \, \bar \ell^{\prime} \, \ell \, \bar \nu_{\ell}$
with $\ell^{\prime} \neq \ell$, the yielding decomposition for the intricate angular function
${\cal J}_{\rm int}$ can be cast in the form of
\begin{eqnarray}
{\cal J}_{\rm int} &=& \tilde{J}_1 \, \left (\cos \theta_1 + \cos \theta_2 \right )^2
+ \tilde{J}_2 \, \left (\cos \theta_1 + \cos \theta_2 \right ) \, \left (1 + \cos \theta_1 \, \cos \theta_2 \right )
\nonumber \\
&& + \,  \tilde{J}_3 \,  \left (\cos \theta_1 + \cos \theta_2 \right ) \,  \sin^2 \theta_1 \, \sin^2 \theta_2
+   \tilde{J}_4 \, \left (\cos \theta_1 + \cos \theta_2 \right ) \,
\left (\sin^2 \theta_1 +  \sin^2 \theta_2 \right )
\nonumber \\
&&  + \,  \tilde{J}_5 \, \left (1 + \cos \theta_1 \, \cos \theta_2 \right ) \,
\left (\sin^2 \theta_1 +  \sin^2 \theta_2 \right )
 + \,  \tilde{J}_6 \,  \sin^2 \theta_1 \, \sin^2 \theta_2
\nonumber \\
&&  + \, \tilde{J}_7 \, \left (\cos \theta_1 + \cos \theta_2 \right )  \,
\sin \theta_1 \, \sin \theta_2 \, \sin \phi
\nonumber \\
&&  + \, \tilde{J}_8 \, \left (\cos \theta_1 + \cos \theta_2 \right ) \,
\left (1 + \cos \theta_1 \, \cos \theta_2 \right ) \, \sin \theta_1 \, \sin \theta_2
\, \sin \phi
\nonumber \\
&&  + \, \tilde{J}_9 \,  \left (1 + \cos \theta_1 \, \cos \theta_2 \right ) \, \sin \theta_1 \, \sin \theta_2
\, \sin \phi
\nonumber \\
&&  + \, \tilde{J}_{10} \, \sin \theta_1 \, \sin \theta_2  \,
\left (\sin^2 \theta_1 +  \sin^2 \theta_2 \right ) \, \sin \phi
\nonumber \\
&&  + \, \tilde{J}_{11} \, \left (\cos \theta_1 + \cos \theta_2 \right ) \,
\sin \theta_1 \, \sin \theta_2 \, \cos \phi
\nonumber \\
&&  + \, \tilde{J}_{12} \, \left (1 + \cos \theta_1 \, \cos \theta_2 \right ) \, \sin \theta_1 \, \sin \theta_2
\, \cos \phi
\nonumber \\
&&  + \, \tilde{J}_{13} \, \left (\cos \theta_1 + \cos \theta_2 \right )^2 \, \sin \theta_1 \, \sin \theta_2
\, \cos \phi
\nonumber \\
&&  + \, \tilde{J}_{14} \, \left (\cos \theta_1 + \cos \theta_2 \right ) \,
\sin (2 \theta_1)  \, \sin (2 \theta_2) \, \cos \phi
\nonumber \\
&&  + \, \tilde{J}_{15} \, \sin^2 \theta_1 \, \sin^2 \theta_2 \, \sin (2 \phi)
+  \tilde{J}_{16} \, \left (\cos \theta_1 + \cos \theta_2 \right ) \,
\sin^2 \theta_1 \, \sin^2 \theta_2 \, \sin (2 \phi)
\nonumber \\
&&  + \, \tilde{J}_{17} \, \sin^2 \theta_1 \, \sin^2 \theta_2 \, \cos (2 \phi)
+  \tilde{J}_{18} \, \left (\cos \theta_1 + \cos \theta_2 \right ) \,
\sin^2 \theta_1 \, \sin^2 \theta_2 \, \cos (2 \phi) \,.
\label{explicit form of Jint}
\end{eqnarray}
The coefficient functions $\tilde{J}_{i}$ (with $i=1, 2,...18$) can be further expressed in terms of
the generalized $B_{u}^{-} \to \gamma^{\ast} \, W^{\ast}$ transition form factors
$F_{k} \equiv F_k(p^2, q^2)$ and ${\mathbb F}_k \equiv F_k(\tilde{p}^2, \tilde{q}^2)$
together with the suitable kinematic functions ($k=V, A,  \parallel$)
\begin{eqnarray}
\tilde{J}_1 &=& - \,\, \hat{p}^2 \, \hat{q}^2 \, (1 + \hat{p}^2 -\hat{q}^2 )  \,\,
{\rm Re} \left [ (1 + \hat{\tilde{p}}^2 -\hat{\tilde{q}}^2 )  \, F_A  \,
 {\mathbb F}_A^{\ast} + F_A  \, {\mathbb F}_{\|}^{\ast} \right ] \,,
\nonumber \\
\tilde{J}_2 &=& - \,\, \hat{p}^2 \, \hat{q}^2 \, \lambda^{1/2}(1, \hat{p}^2, \hat{q}^2 ) \,\,
{\rm Re} \left [ (1 + \hat{\tilde{p}}^2 -\hat{\tilde{q}}^2 )  \,
F_V  \, {\mathbb F}_A^{\ast} + F_V  \, {\mathbb F}_{\|}^{\ast} \right ] \,,
\nonumber \\
\tilde{J}_3 &=& - {1 \over 8} \, \, \lambda^{3/2}(1, \hat{p}^2, \hat{q}^2 ) \,
\bigg \{ \left [ 1 +  \frac{6 \,  \hat{p}^2 \, \hat{q}^2 } {\lambda (1, \hat{p}^2, \hat{q}^2 ) }  \right ]  \,
(1 + \hat{p}^2 -\hat{q}^2 ) \,
{\rm Re} \left ( F_A  \, {\mathbb F}_V^{\ast}  \right )
\nonumber \\
&&  +  \,  (1- \hat{p}^2 - \hat{q}^2) \,
\, {\rm Re} \left ( F_{\|}  \, {\mathbb F}_V^{\ast}  \right )  \bigg \} \,,
\nonumber \\
\tilde{J}_4 &=& {1 \over 2} \,\,  \hat{p}^2 \, \hat{q}^2 \, (1 + \hat{p}^2 -\hat{q}^2 ) \,
\lambda^{1/2}(1, \hat{p}^2, \hat{q}^2 ) \,\,
{\rm Re} \left ( F_A  \, {\mathbb F}_V^{\ast}  \right ) \,,
\nonumber \\
\tilde{J}_5 &=& {1 \over 2} \,\,  \hat{p}^2 \, \hat{q}^2 \, \lambda (1, \hat{p}^2, \hat{q}^2 ) \,\,
{\rm Re} \left ( F_V  \, {\mathbb F}_V^{\ast}  \right ) \,,
\nonumber \\
\tilde{J}_6 &=& \left [ {1 \over 2} - \frac{(1- \hat{p}^2 - \hat{q}^2)^2} {4 \, \hat{p}^2 \, \hat{q}^2 } \right ] \,
\tilde{J}_1 - \tilde{J}_5 - \frac{1- \hat{p}^2 - \hat{q}^2}{2 \, \sqrt{\hat{p}^2 \, \hat{q}^2}} \, \tilde{J}_{12}
+ { \lambda^{2}(1, \hat{p}^2, \hat{q}^2 )  \over 4} \, \, {\rm Re} \left ( F_{\|}  \, {\mathbb F}_{\|}^{\ast}  \right ) \,,
\nonumber \\
\tilde{J}_7 &=& {1 \over 2} \,\, \sqrt{\hat{p}^2 \, \hat{q}^2} \, \lambda (1, \hat{p}^2, \hat{q}^2 ) \,\,
{\rm Im} \left [ (1 + \hat{p}^2 -\hat{q}^2 ) \, F_{A}  \, {\mathbb F}_{\|}^{\ast}
+  (1 + \hat{\tilde{p}}^2 -\hat{\tilde{q}}^2 ) \, F_{\|}  \, {\mathbb F}_{A}^{\ast}
+  F_{\|}  \, {\mathbb F}_{\|}^{\ast}   \right ] \,,
\nonumber \\
\tilde{J}_8 &=& - {1 \over 4} \,\, \sqrt{\hat{p}^2 \, \hat{q}^2} \,  (1- \hat{p}^2 - \hat{q}^2) \,
\lambda (1, \hat{p}^2, \hat{q}^2 ) \,
{\rm Im} \left ( F_V  \, {\mathbb F}_V^{\ast}  \right ) \,,
\nonumber \\
\tilde{J}_9 &=&  - \,\,  \sqrt{\hat{p}^2 \, \hat{q}^2} \,
\left [ \frac{1- \hat{p}^2 - \hat{q}^2}{\hat{p}^2 \, \hat{q}^2}  \, \tilde{J}_{15}
- {\lambda^{3/2} (1, \hat{p}^2, \hat{q}^2 ) \over 2}\,
{\rm Im} \left ( F_V  \, {\mathbb F}_{\|}^{\ast}  \right )  \right ]  \,,
\nonumber \\
\tilde{J}_{10} &=&   - {1 \over 4} \,\, \sqrt{\hat{p}^2 \, \hat{q}^2} \,
\lambda^{3/2} (1, \hat{p}^2, \hat{q}^2 ) \,
{\rm Im} \left ( F_{\|}  \, {\mathbb F}_{V}^{\ast}  \right )   \,,
\nonumber \\
\tilde{J}_{11} &=& - {1 \over 2} \,\, \sqrt{\hat{p}^2 \, \hat{q}^2} \,
\left [  \frac{1- \hat{p}^2 - \hat{q}^2}{\hat{p}^2 \, \hat{q}^2} \, \tilde{J}_{2}
+ { \lambda^{3/2} (1, \hat{p}^2, \hat{q}^2 ) \over 2 }\,
{\rm Re} \left ( F_{\|}  \, {\mathbb F}_{V}^{\ast} + 2 \,  F_{V}  \, {\mathbb F}_{\|}^{\ast} \right )  \right ]  \,,
\nonumber \\
\tilde{J}_{12} &=& -  \,\,  \sqrt{\hat{p}^2 \, \hat{q}^2} \,
\bigg \{  { \lambda (1, \hat{p}^2, \hat{q}^2 ) \over 2} \,
{\rm Re} \left [ (1 + \hat{p}^2 -\hat{q}^2 ) \, F_{A}  \, {\mathbb F}_{\|}^{\ast}
- \, (1 + \hat{\tilde{p}}^2 -\hat{\tilde{q}}^2 ) \,  F_{\|}  \, {\mathbb F}_{A}^{\ast}
- \,  F_{\|}  \, {\mathbb F}_{\|}^{\ast}  \right ]
\nonumber \\
&& + \,  \frac{1- \hat{p}^2 - \hat{q}^2}{\hat{p}^2 \, \hat{q}^2} \, \tilde{J}_{1}  \bigg  \} \,,
\nonumber \\
\tilde{J}_{14} &=&  {1 \over 8} \,\, \sqrt{\hat{p}^2 \, \hat{q}^2} \,
\lambda^{3/2} (1, \hat{p}^2, \hat{q}^2 ) \,
\left [ \frac{(1- \hat{q}^2)^2 -  \hat{p}^4} {\lambda (1, \hat{p}^2, \hat{q}^2 ) } \,\,
{\rm Re} \left ( F_{A}  \, {\mathbb F}_{V}^{\ast}  \right )
+  \left ( {1 \over 2} \right ) \,\, {\rm Re} \left ( F_{\|}  \, {\mathbb F}_{V}^{\ast}  \right )  \right ]  \,,
\nonumber \\
\tilde{J}_{15} &=&  {1 \over 2} \,\, \hat{p}^2 \, \hat{q}^2 \,
\lambda^{1/2} (1, \hat{p}^2, \hat{q}^2 ) \,
{\rm Im} \left [ (1 + \hat{p}^2 -\hat{q}^2 )  \, F_{A}  \, {\mathbb F}_{V}^{\ast}
+ \, (1 + \hat{\tilde{p}}^2 -\hat{\tilde{q}}^2 )  \,  F_{V}  \, {\mathbb F}_{A}^{\ast}
+ \,  F_{V}  \, {\mathbb F}_{\|}^{\ast}  \right ] \,,
\hspace{1.0 cm}
\label{independent Jtilde functions}
\end{eqnarray}
where for convenience we have introduced the shorthand notations
\begin{eqnarray}
F_{\|} & \equiv & F_1(p^2, q^2) + {v \cdot p \over m_B} \, F_3(p^2, q^2) \,,
 \qquad
{\mathbb F}_{\|} \equiv F_1(\tilde{p}^2, \tilde{q}^2) + {v \cdot \tilde{p} \over m_B}  \, F_3(\tilde{p}^2, \tilde{q}^2) \,,
\nonumber \\
\hat{\tilde{p}}^2 &=& \tilde{p}^2 /m_B^2 \,,
 \qquad
\hat{\tilde{q}}^2 = \tilde{q}^2 /m_B^2 \,.
\end{eqnarray}
The remaining four angular coefficients appearing in (\ref{explicit form of Jint}) turn out to be
linearly dependent of those already derived in (\ref{independent Jtilde functions}) by virtue of the following relations
\begin{eqnarray}
\tilde{J}_{13} &=& \frac{1- \hat{p}^2 - \hat{q}^2}{2 \, \sqrt{\hat{p}^2 \, \hat{q}^2}} \, \tilde{J}_{5} \,,
\qquad
\tilde{J}_{16} =  - \frac{\sqrt{\hat{p}^2 \, \hat{q}^2}}{1- \hat{p}^2 - \hat{q}^2}\, \tilde{J}_{8} \,,
\nonumber \\
\tilde{J}_{17} &=&  {1 \over 2} \, \tilde{J}_{1} -   \tilde{J}_{5} \,,
\qquad
\hspace{0.8 cm} \tilde{J}_{18} =  - {1 \over 2} \, \tilde{J}_{4} \,.
\end{eqnarray}


\end{document}